\documentclass[iop]{emulateapj}
 
\usepackage{wasysym}
\usepackage{natbib}
\usepackage{longtable}
\usepackage{graphicx}
\usepackage{fancyref}
\usepackage{gensymb}
\usepackage{adjustbox}
\usepackage{booktabs}
\usepackage{scrextend}

\newcommand{\ra}[1]{\renewcommand{\arraystretch}{#1}}

\shortauthors{Fischer et al.}

\begin{document}

\title{Hubble Space Telescope Observations of Extended [O III] $\lambda$5007 Emission in Nearby QSO2s: New Constraints On AGN / Host Galaxy Interaction}

\author{Travis C. Fischer\altaffilmark{1}\altaffilmark{\textdagger},
S. B. Kraemer\altaffilmark{2},
H. R. Schmitt\altaffilmark{3},
L.F. Longo Micchi\altaffilmark{2},
D. M. Crenshaw\altaffilmark{4},
M. Revalski\altaffilmark{4}\altaffilmark{$*$},
M. Vestergaard\altaffilmark{5,6},
M. Elvis\altaffilmark{7},
C. M. Gaskell\altaffilmark{8},
F. Hamann\altaffilmark{9},
L. C. Ho\altaffilmark{10},
J. Hutchings\altaffilmark{11},
R. Mushotsky\altaffilmark{12},
H. Netzer\altaffilmark{13},
T. Storchi-Bergmann\altaffilmark{14},
A. Straughn\altaffilmark{1},
T. J. Turner\altaffilmark{15},
M. J. Ward\altaffilmark{16}}

\altaffiltext{1}{Astrophysics Science Division, Goddard Space Flight Center,
Code 665, Greenbelt, MD 20771, USA}

\altaffiltext{2}{Institute for Astrophysics and Computational Sciences,
Department of Physics, The Catholic University of America, Washington, DC
20064, USA}

\altaffiltext{3}{Naval Research Laboratory, Washington, DC 20375, USA}

\altaffiltext{4}{Department of Physics and Astronomy, Georgia State 
University, Astronomy Offices, 25 Park Place, Suite 600,
Atlanta, GA 30303, USA}

\altaffiltext{5}{Dark Cosmology Centre, Niels Bohr Institute, University of Copenhagen, Juliane Maries Vej 30, DK-2100 Copenhagen {\O}, Denmark}

\altaffiltext{6}{Steward Observatory and Department of Astronomy, University of Arizona, 933 N. Cherry Avenue, Tucson AZ 85721}

\altaffiltext{7}{Harvard-Smithsonian Center for Astrophysics, 60 Garden St., Cambridge, MA 02138, USA}

\altaffiltext{8}{Department of Astronomy and Astrophysics, University of California, Santa Cruz, CA 95064, USA}

\altaffiltext{9}{Department of Physics and Astronomy, University of California, Riverside, CA 92507, USA}

\altaffiltext{10}{Kavli Institute for Astronomy and Astrophysics, Peking University; School of Physics, Department of Astronomy, Peking University; Beijing 100871, China}

\altaffiltext{11}{Dominion Astrophysical Observatory, NRC Herzberg Institute 
of Astrophysics, 5071 West Saanich Road, Victoria, BC, V9E 2E7, Canada}

\altaffiltext{12}{Department of Astronomy, University of Maryland, College Park, MD 20742, USA}

\altaffiltext{13}{School of Physics and Astronomy, Tel Aviv University, Tel Aviv 69978, Israel}

\altaffiltext{14}{Departamento de Astronomia, Universidade Federal do Rio Grande do Sul, IF, CP 15051, 91501-970 Porto Alegre, RS, Brazil}

\altaffiltext{15}{Department of Physics, University of Maryland Baltimore County, 1000 Hilltop Circle, Baltimore, MD 21250, USA}

\altaffiltext{16}{Centre for Extragalactic Astronomy, Department of Physics, University of Durham, South Road, Durham DH1 3LE, UK}

\altaffiltext{\textdagger}{James Webb Space Telescope NASA Postdoctoral Program Fellow; travis.c.fischer@nasa.gov}

\altaffiltext{$*$}{National Science Foundation Graduate Research Fellow}

\begin{abstract}
We present a {\it Hubble Space Telescope} ({\it HST}) survey of extended [O III] $\lambda$5007 emission for a sample of 12 nearby
(z $<$ 0.12), luminous Type 2 quasars (QSO2s), which we use to measure the extent and kinematics of their
AGN-ionized gas. We find the size of the observed [O~III] regions scale with luminosity in comparison to 
nearby, less luminous Seyfert galaxies and radially outflowing kinematics to exist in all targets. 
We report an average maximum outflow radius of $\sim$600 pc, with gas continuing to be kinematically influenced by the 
central AGN out to an average radius of $\sim$ 1130 pc. These findings question the effectiveness of AGN being 
capable of clearing material from their host bulge in the nearby universe and suggest that disruption of gas 
by AGN activity may prevent star formation without requiring evacuation. Additionally, we find a dichotomy in 
our targets when comparing [O~III] radial extent and nuclear FWHM, where QSO2s with compact [O~III] morphologies 
typically possess broader nuclear emission-lines. 

\end{abstract}


~~~~~

\section{Introduction}

Active galactic nuclei (AGN) are powered by the accretion of matter onto a supermassive black hole (SMBH), 
which generates massive amounts of radiation within a very small volume. Mass outflows of ionized gas, 
a spatially-resolved component of AGN feedback, are also generated and have been thought to be critical 
in the chemical enrichment of the intergalactic medium \citep{Kha08}, and the self-regulation of SMBH growth \citep{Hop05}.
The relationship between the SMBH mass and the stellar velocity of its galaxy bulge, i.e. the 
M$_{BH} - \sigma$ relation (\citealt{Kor13} and references therein), is credited to ignition of the 
AGN quenching star formation and evacuating gas from the bulge. 

AGN feedback exists in two forms: radio jets and $``$AGN winds". Jets are powerful and clearly impact 
their host galaxies and extragalactic environments. However, they are highly focused, and strong jets 
occur in only 5 - 10\% of the AGN population \citep{Raf09}. Alternatively, winds are prevalent in most 
AGN \citep{Mul13,Gen14,Woo16}, including nearby, moderate luminosity Seyfert galaxies. 
AGN winds are often observed as UV and X-ray absorption lines blueshifted with respect to their host galaxies, 
traveling at velocities up to $\sim$ 2500 km s$^{-1}$ within tens of parsecs from the central SMBH \citep{Cre03,Vei05,Cre12,Kin15},
or emission-line gas in AGN narrow-line regions (NLRs) on larger scales  (100s - 1000s of pc), with 
outflow velocities up to $\sim$ 2000 km s$^{-1}$ \citep{Cre05,Cre10b,Mul11,Fis13,Fis14,Bae16,Nev16}. 
This is likely to be the region where interaction between AGN and nuclear star formation occurs.
One aspect not addressed in this work is molecular gas outflows. Recent 
studies \citep{Vei13,Cic14} have found evidence for molecular outflows in AGN on large ($>$ 1 kpc) scales; 
however, most of the AGN in these works are in merged systems, whereas only one 
source, 2MASX-J0802593+2552551, shows evidence for an ongoing merger in our sample. 
Therefore, although we cannot rule out the presence of large-scale molecular outflows, 
the distribution of gas and dynamics in these AGN are likely much different than in 
merged systems and more consistent with typical low-z Seyfert galaxies \citep{Fis13}.

Recently, we used the Gemini Near-Infrared Integral Field Spectrograph (NIFS) to observe the 
NLR of the luminous, nearby Seyfert 2 galaxy Mrk 573 \citep{Fis17}. From this work, we found 
the NLR morphology to be consistent with an intersection between spiral arms in the host disk 
and ionizing radiation from the central engine, as dust lanes rotating in the host disk (also 
traced by molecular hydrogen emission) were found to connect to arcs of ionized gas inside the 
AGN ionizing bicone from outside of the field of ionizing radiation. Host disk gas inside the 
ionizing field at small radii ($r < 750$ pc) is radiatively driven away from the central engine 
and forms the NLR of the AGN, while ionized gas at greater radii follows the rotation kinematics 
of the host disk as part of the Extended NLR (ENLR; \citealt{Ung87}). Applying these findings to other, similar 
high-resolution kinematics studies of nearby, inclined Seyferts \citep{Das05,Das06,Cre10b,Sto10,Mul11,Fis13}, 
where projection effects on outflow distances are small, gas in the radially driven NLRs (i.e. 
gas with predominantly radial velocity structure) typically extend less than 1 kpc from the central engine. 
These observations question how successful AGN feedback is on galaxy-bulge scales that may 
be required in a bulge-quenching, negative feedback scenario. One aspect that may explain these findings 
is that Seyferts are relatively low-luminosity AGN. Therefore, outflows may not be powerful enough in 
these nearby AGN to drive gas out to bulge-radius distances. 

While one would expect that the power of these outflows would scale with luminosity, as suggested by 
\citet{Gan08}, ground-based studies of QSO2s \citep{Gre11,Liu13b,Har14,McE15} have 
found mostly chaotic, low velocity [O III] profiles attributed to AGN activity, with kpc-scale, high-velocity outflows, 
being exceptionally rare. Studies of SDSS J135646.11+102609.1 \citep{Gre11,Gre12}, for example, find outflows on scales of $\sim$10 
kpc, with deprojected velocities of $\sim$1000 km s$^{-1}$. Additionally, previous analysis of optical long-slit 
observations of the nearby QSO2 Mrk 34 \citep{Fis13}, measured bright, non-rotational kinematics
(i.e. outflows) at distances of $>$ 1 kpc. However, the absence of evidence for such outflows 
in a majority of observations raises the question of whether kpc-scale winds exist in most QSO2s. If they do not, 
it follows that outflows are not a critical component of quasar-mode feedback and the evolution of galaxy bulges, 
implying that the star formation is quenched in bulges by other means.
As such, we have obtained {\it HST} imaging and spectroscopy of 12 of the most luminous QSO2s within 
z = 0.12, through Hubble Program ID 13728 (PI: Kraemer) and archive observations of Mrk 34, to map 
[O III] velocities and widths as a function of radial distance and determine the extent of AGN-driven 
outflows in each system.

\section{Sample, Observations, and Measurements}
\label{obs_sec}

\subsection{Target Selection}

Our sample is listed in Table \ref{tab:obs} and includes 12 of the 15 most luminous targets from the \citet{Rey08} QSO2 
sample under z = 0.12. Our choice of distance limit was made to ensure our 
ability to map the structure of the [O~III] gas at scales of several 100 pcs for our most distant targets, while 
opting for the most luminous targets assures that we are studying AGN more luminous than the Seyfert galaxies in our 
previous kinematic studies. All targets have a log L$_{[O~III]} \ge$ 42.28, satisfying the conventional B-band absolute 
magnitude criterion of a "quasar", M$_B$ $<$ -23, where a corresponding L$_{[O~III]}$ is $>3 \times 10^8$L$_{\astrosun}$ \citep{Zak03}, 
and are among the top 25\% of QSO2s under z = 0.3 in the \citet{Rey08} sample. The majority of the sample are radio quiet, 
with a few being intermediately radio-loud, which indicates, in analogy to nearby Seyferts, that outflows are not likely to 
be jet driven. Minimum SMBH mass estimates required to produce the observed [O III] radiation range between 
10$^{7.7}$ - 10$^{8.3}$ M$_{\astrosun}$, calculated using the [O III] measurements from \citet{Rey08} and assuming an AGN radiating 
at Eddington with L$_{bol}$ = L$_{[O~III]} \times$ 3500 \citep{Hec04}. If these sources are heavily reddened, the bolometric 
correction and resulting black hole masses would be larger. In order to best sample extended NLR kinematics in each target, 
Type 2 QSOs were chosen rather than QSO1s, as AGN-ionized gas morphologies in QSO1s can be strongly foreshortened by projection effects.

\subsection{{\it HST} Observations}

Knowledge of the ionized-gas morphology of each AGN was required in order to properly place the STIS long-slit across its 
greatest extension in each AGN. Therefore, the observing program for our sample was performed in a two-step process: obtaining 
narrow-band images of each AGN to determine ideal STIS position angles and returning at a later date for the follow-up 
spectroscopic observation. Images were obtained for each QSO2 in our sample using FR505N and FR551N filters, chosen depending 
on redshift of each target to observe [O~III], with the Wide-Field Channel (WFC) of {\it HST} / ACS. Continuum observations 
were obtained using the FR647M filter, selected to provide a relatively broad continuum region free of strong emission lines, 
for continuum subtraction. Images were obtained between 2014 October 29 and 2015 July 8. 

Long-slit spectra were obtained using {\it HST}/STIS, with the CCD detector employing the $52'' \times 0.2''$ slit. Spectra 
were obtained between 2015 January 29 and 2016 March 3, using either the G430M or G750M medium-dispersion grating, also dependent 
on the redshift of each target to observe [O~III]. Spectral resolutions for the G430M and G750M gratings are 0.56 and 1.1 \AA~ (for 
bandwidths of 286 and 572\AA)respectively, with an angular resolution of 0.051$''$ per pixel in the cross-dispersion direction. Slit 
position angles for each observation were aligned within 5$\degree$ of the projected axis of the ionized-gas observed in the previously 
obtained {\it HST} imaging. Spectra were taken as a combination of three exposures of similar exposure lengths optimized to maximize 
available observing time in a single orbit. Observations were dithered by $\pm 0.25''$ along the slit with respect to the first spectrum 
to avoid problems due to hot pixels, and wavelength calibration lamp spectra were taken during Earth occultation. 

Our sample also includes archival imaging and spectroscopic observations (Proposal IDs: 10873 and 8253, respectively) of the QSO2 Mrk 34, 
as it matches the criteria of our sample and has similar {\it HST} imaging (WFPC2 for Mrk 34 versus ACS for the rest of the sample) and 
long-slit observations. Archival Mrk 34 observations were gathered from the Mikulski Archive for Space Telescopes (MAST). [O~III] imaging for 
this target is produced from F547M WFPC2 imaging with F467M continuum subtraction, as alignment between [O~III] and continuum imaging is required 
to properly map the STIS slit locations to the [O~III] imaging. Further details of all {\it HST} observations for our entire sample are 
listed in Table \ref{tab:obs}.

\begin{table*}[h!]
  \centering
  \ra{1.3}
  \caption{QSO2 Sample Characteristics and {\it HST} Observations Summary.}
  \label{tab:obs}
  \label{{bh}}
  \begin{adjustbox}{max width=\textwidth}
  \begin{tabular}{@{}lccccccrlrcc@{}}
    \toprule
    Target 						& R.A. 		& Dec		&	Redshift  	& Scale  	&  logL$_{[O~III]}$   	& Min. M$_{BH} $		 	& $S_{20cm}$ &Filter /	& Total Exp   	& Center $\lambda$ 	& PA$_{slit}$	  \\
    							&(2000)	  	& (2000) 		&	 		&(kpc/$''$)&  ergs s$^{-1}$ 	& log(M/M$_{\astrosun}$)		& (mJy)	&  Grating		&  (s)          	& ($\AA$)			& ($\degree$)	 \\
    \midrule
    2MASX J07594101+5050245	    & 07 59 40.9	& +50 50 23   	 &	0.054 	& 1.02 	& 42.35			& 7.8				& 45.55	& FR647M	& 200		& 5803.1	 		&  --			 \\
     							& 			& 		   	 &			& 		&				& 						& 	 	& FR551N		& 1956		& 5280.5			&  --			  \\
						    	& 			& 		   	 &			& 		&				& 						& 		& G430M		& 2270		& 5216.0				& 102.7 		 \\
    2MASX J08025293+2552551	    & 08 02 52.9	& +25 52 55 	 &	0.081 	& 1.48	& 42.35 			& 7.8						& 29.37	& FR647M	& 200		& 5949.6			&  --			  \\
         						& 			& 		   	 &			& 		&				& 						& 		& FR551N		& 1788		& 5413.7			&   --			 \\
							    & 			& 		   	 &			& 		&				& 						& 		& G430M		& 1997		& 5471.0				& -156.4 		\\							
    MRK 34					    & 10 34 08.6	& +60 01 52 	 &     	0.051	& 0.95	& 42.39			& 7.8						& 17.01 	& F467M		& 5200		& 4670.0			& --			\\					
						    	& 			& 		   	 &			& 		&				& 						& 		& F547M		& 7700		& 5483.0			&  --			  \\      						
						    	& 			& 		   	 &			& 		&				& 						& 		& G430M		& 1500		& 5216.0				& 152.6 		\\
    2MASX J11001238+0846157	    & 11 00 12.4	& +08 46 15       &	0.101 	& 1.80	& 42.69 			& 8.1						& 58.54 	& FR647M	& 200		& 6056.1			&  --			  \\    
     							& 			& 		   	 &			& 		&				& 						& 		& FR551N		& 1769		& 5510.1			&  --			  \\    
     							& 			& 		   	 &			& 		&				& 						& 		& G430M		& 1883		& 5471.0				& 160.6 		\\		
    SDSS J115245.66+101623.8 	& 11 52 45.7	& +10 16 23	 &	0.070 	& 1.30 	& 42.28 			& 7.7						& 3.56 	& FR647M	& 200		& 5887.8			&  --			  \\
     							& 			& 		   	 &			& 		&				& 						& 		& FR551N		& 1773		& 5358.0			&  --			  \\   
     							& 			& 		   	 &			& 		&				& 						& 		& G430M		& 1944		& 5471.0				& -169.9		\\ 
    FIRST J120041.4+314745 		& 12 00 41.4	& +31 47 46	 &	0.116 	& 2.04	& 42.89 			& 8.3						& 7.31	& FR647M	& 200		& 6138.5			&  --			  \\
     							& 			& 		   	 &			& 		&				& 						& 		& FR551N		& 1804		& 5585.6			&  --			  \\
     							& 			& 		   	 &			& 		&				& 						& 		& G750M		& 2015		& 5734.0				& -89.7		\\    
    2MASX J13003807+5454367	    & 13 00 38.1	& +54 54 36	 &	0.088 	& 1.59	& 42.47 			& 7.9 					& 2.19	& FR647M	& 200		& 5989.4			&  --			  \\
     							& 			& 		   	 &			& 		&				& 						& 		& FR551N		& 1956		& 5449.4			&  --			  \\
     							& 			& 		   	 &			& 		&				& 						& 		& G430M		& 2210		& 5471.0				& 164.4		\\    
    2MASX J14054117+4026326	    & 14 05 41.2	& +40 26 32	 &	0.081 	& 1.47 	& 42.29			& 7.7						& 16.81	& FR647M	& 200		& 5946.6			&  --			  \\
     							& 			& 		   	 &			& 		&				& 						& 		& FR551N		& 1863		& 5411.7			&  --			  \\
     							& 			& 		   	 &			& 		&				& 						& 		& G430M		& 2177		& 5471.0				& 137.1		\\    
    B2 1435+30					& 14 37 37.9	& +30 11 01	 &	0.092 	& 1.66      	& 42.38			& 7.8						& 63.91	& FR647M	& 200		& 6010.4			&  --			  \\
     							& 			& 		   	 &			& 		&				& 						& 		& FR551N		& 1804		& 5469.3			&  --			  \\
     							& 			& 		   	 &			& 		&				& 						& 		& G430M		& 2055		& 5471.0				& 80.1		\\    
    MRK 477					    & 14 40 38.1	& +53 30 16	 &	0.038 	& 0.72	& 42.30			& 7.7						& 57.59	& FR647M	& 200		& 5707.5			&  --			  \\
      							& 			& 		   	 &			& 		&				& 						& 		& FR505N		& 1912		& 5191.5			&  --			  \\
     							& 			& 		   	 &			& 		&				&						& 		& G430M		& 2224		& 5216.0				& 29.8		\\   
    2MASX J16531506+2349431	    & 16 53 15.1	& +23 49 42	 &	0.103 	& 1.83	& 42.54 			& 8.0						& 6.93	& FR647M	& 200		& 6072.0			&  --			  \\
     							& 			& 		   	 &			& 		&				& 						& 		& FR551N		& 1779		& 5525.0			&  --			  \\
     							& 			& 		   	 &			& 		&				& 						& 		& G430M		& 2169		& 5471.0				& -104.9		\\    
    2MASX J17135038+5729546	    & 17 13 50.3	& +57 29 54	 &	0.113 	& 1.97	& 42.53 			& 8.0						& 7.62	& FR647M	& 200		& 6122.7			&  --			  \\   
      							& 			& 		   	 &			& 		&				& 						& 		& FR551N		& 2017		& 5571.7			&  --			  \\
     							& 			& 		   	 &			& 		&				& 						& 		& G750M		& 2214		& 5734.0				& 27.8		\\   
    \bottomrule
  \end{tabular}
  
  \end{adjustbox}
 
\raggedright Columns 1 - 4 list each target and their coordinates and redshift. Column 5 lists the cosmology-corrected scale from the NASA Extragalactic Database. Column 6 lists the [O III] luminosity from \citet{Rey08}. Column 7 
lists the minimum SMBH mass required to produce the observed radiation, assuming an AGN radiating at Eddington with L$_{bol}$ = L$_{[O~III]} \times$ 3500 \citep{Hec04}. Column 8 lists radio fluxes from \citet{Rey08}. Columns 9 and 10 
list the filter/grating and exposure time of each data set. Column 11 lists central wavelengths for each observation. Column 12 lists spectroscopic long-slit position angles.

\end{table*}

\subsection{Data Reduction}

\begin{figure*}
\centering

\includegraphics[width=0.49\textwidth]{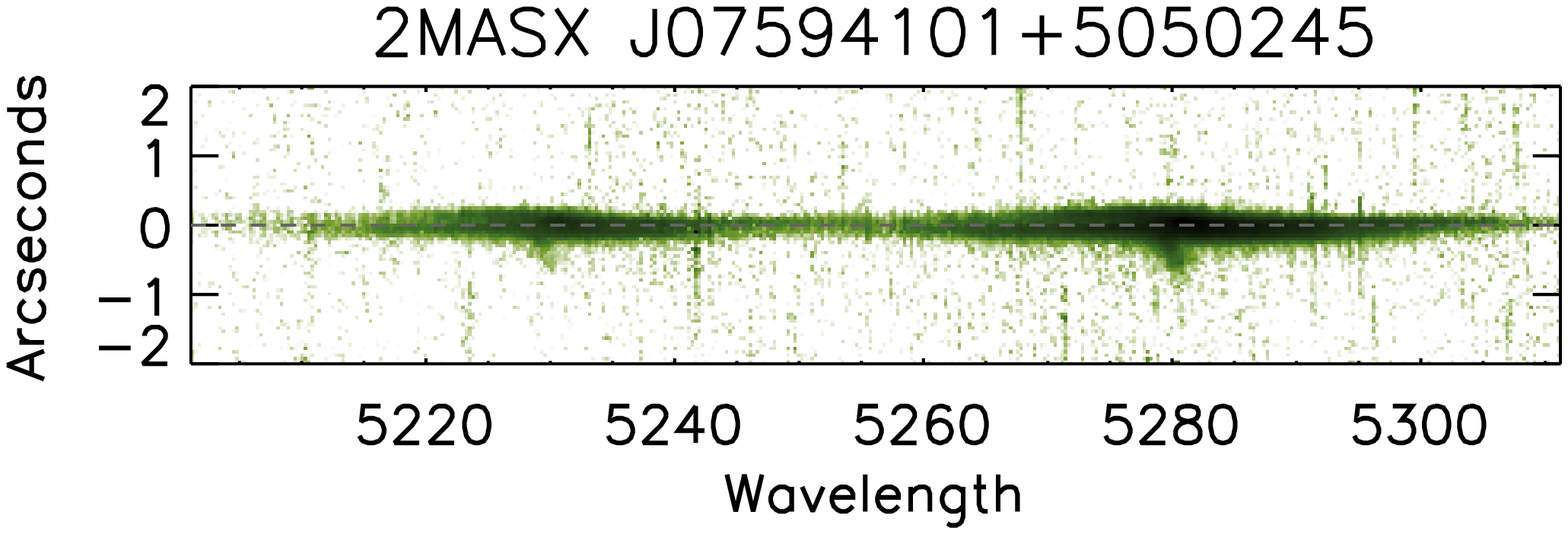}
\includegraphics[width=0.49\textwidth]{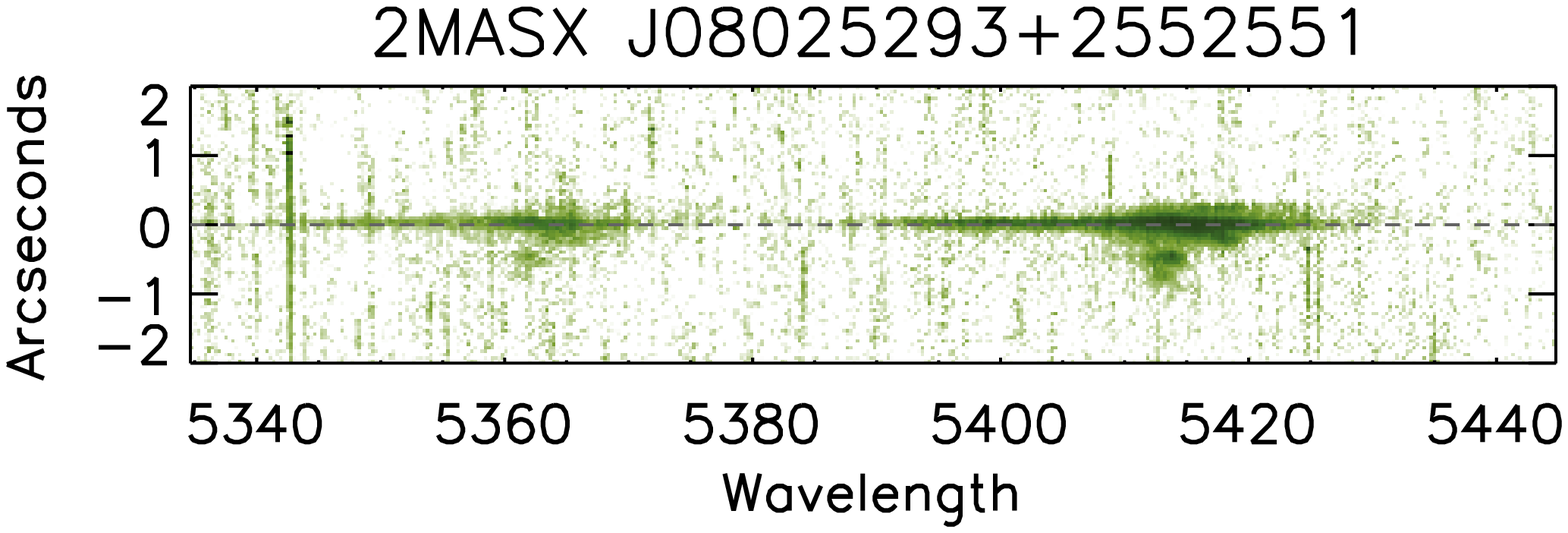}\\
\includegraphics[width=0.49\textwidth]{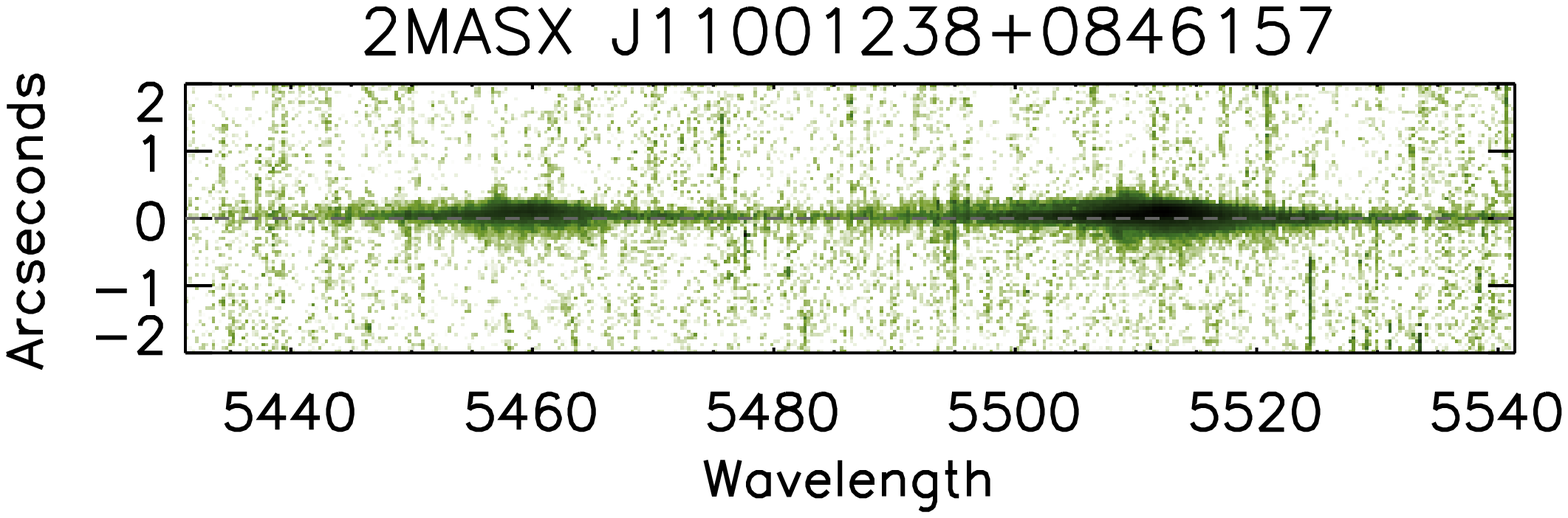}
\includegraphics[width=0.49\textwidth]{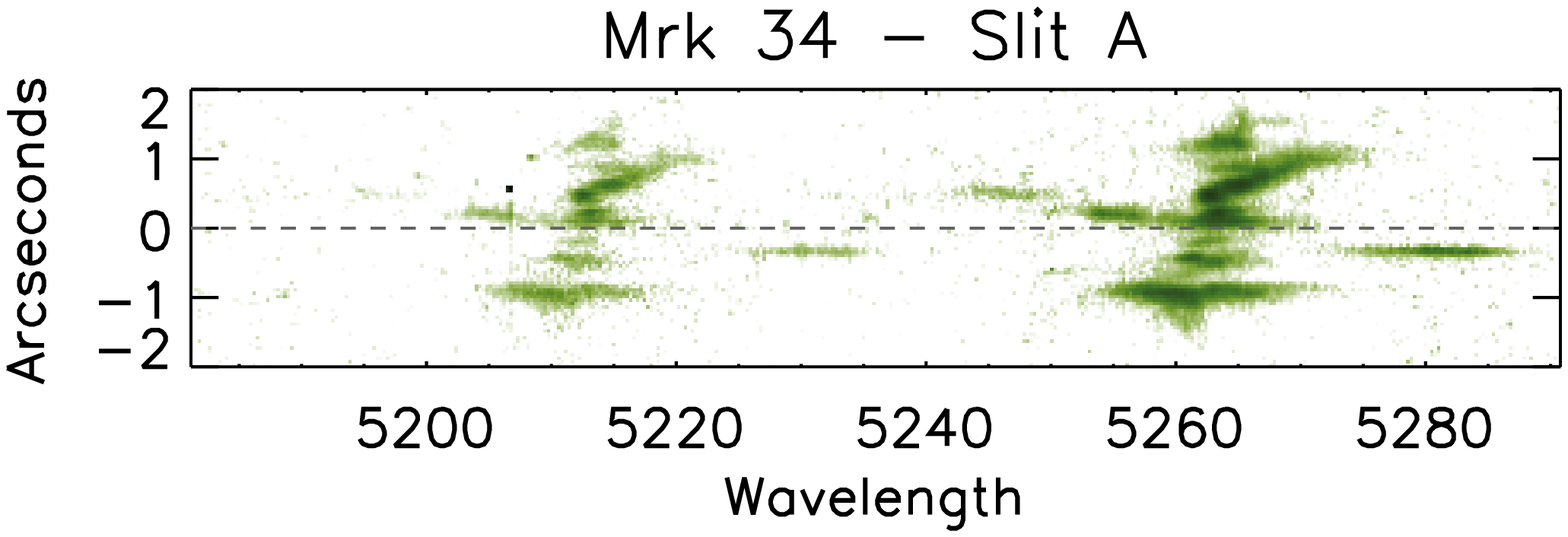} \\
\includegraphics[width=0.49\textwidth]{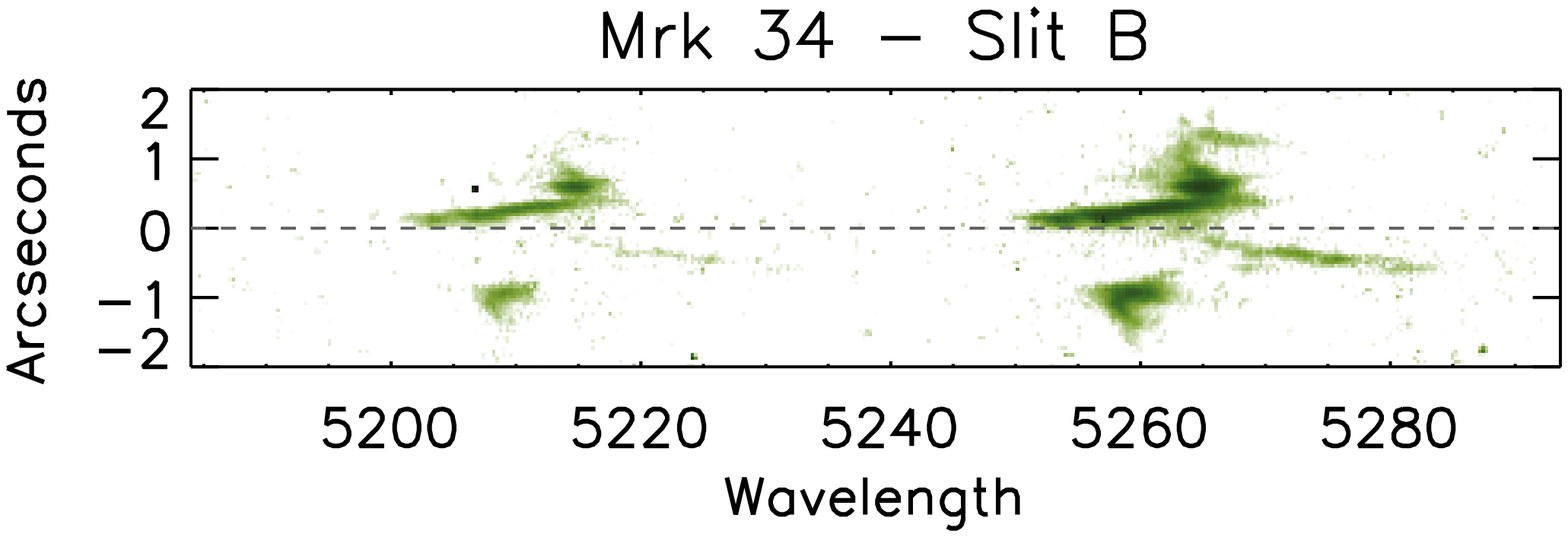}
\includegraphics[width=0.49\textwidth]{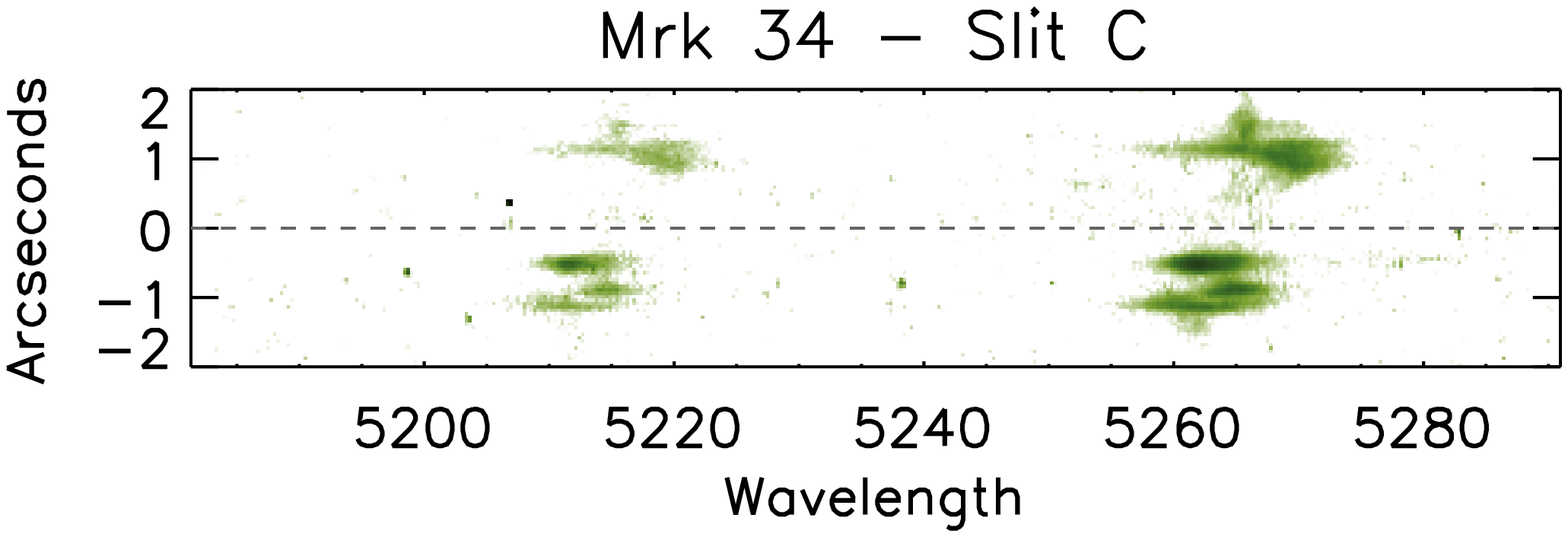} \\
\includegraphics[width=0.49\textwidth]{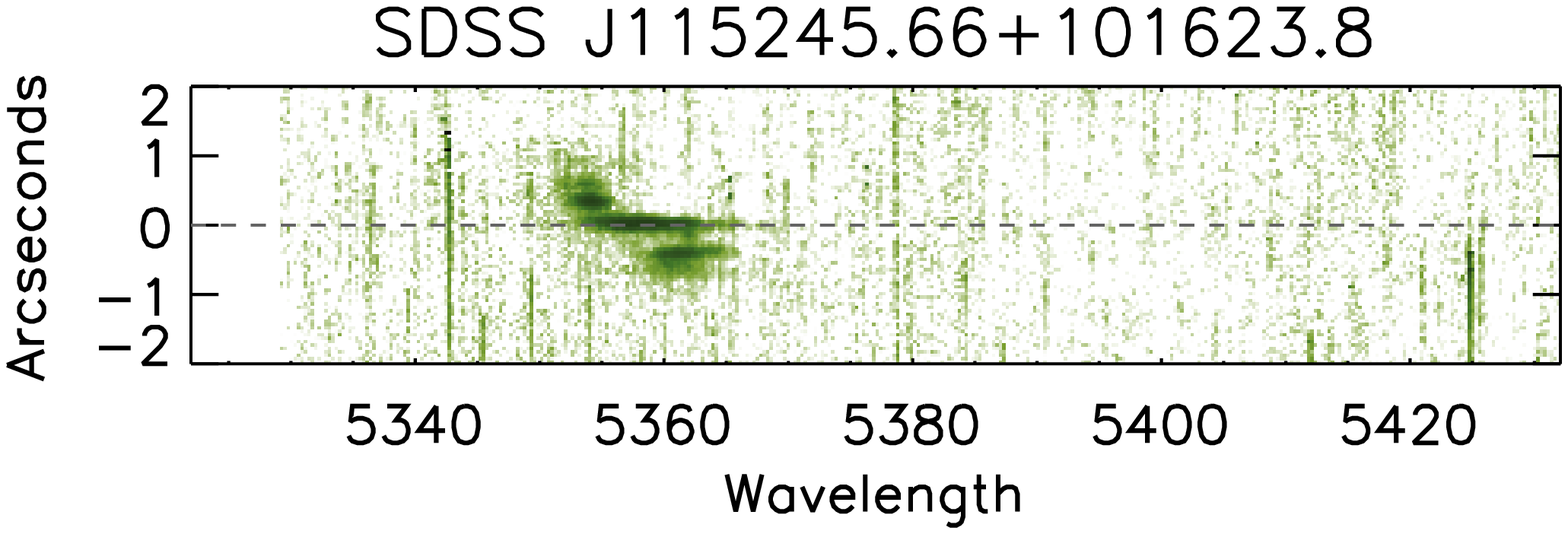}
\includegraphics[width=0.49\textwidth]{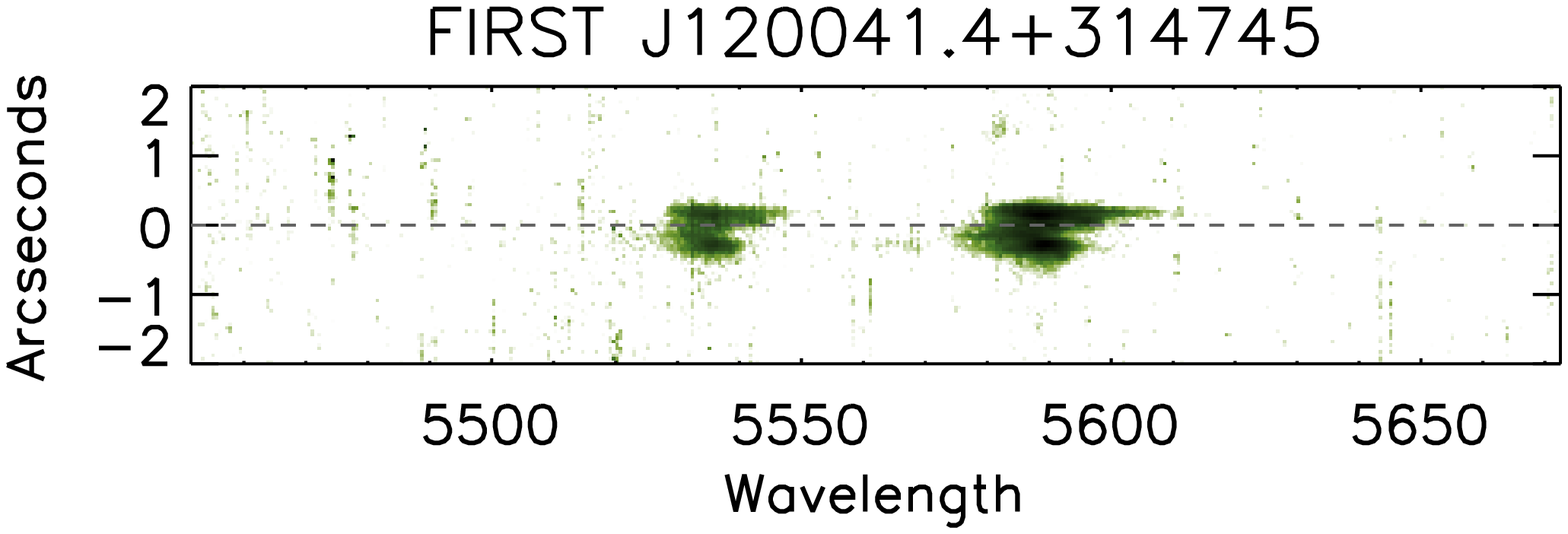}\\
\includegraphics[width=0.49\textwidth]{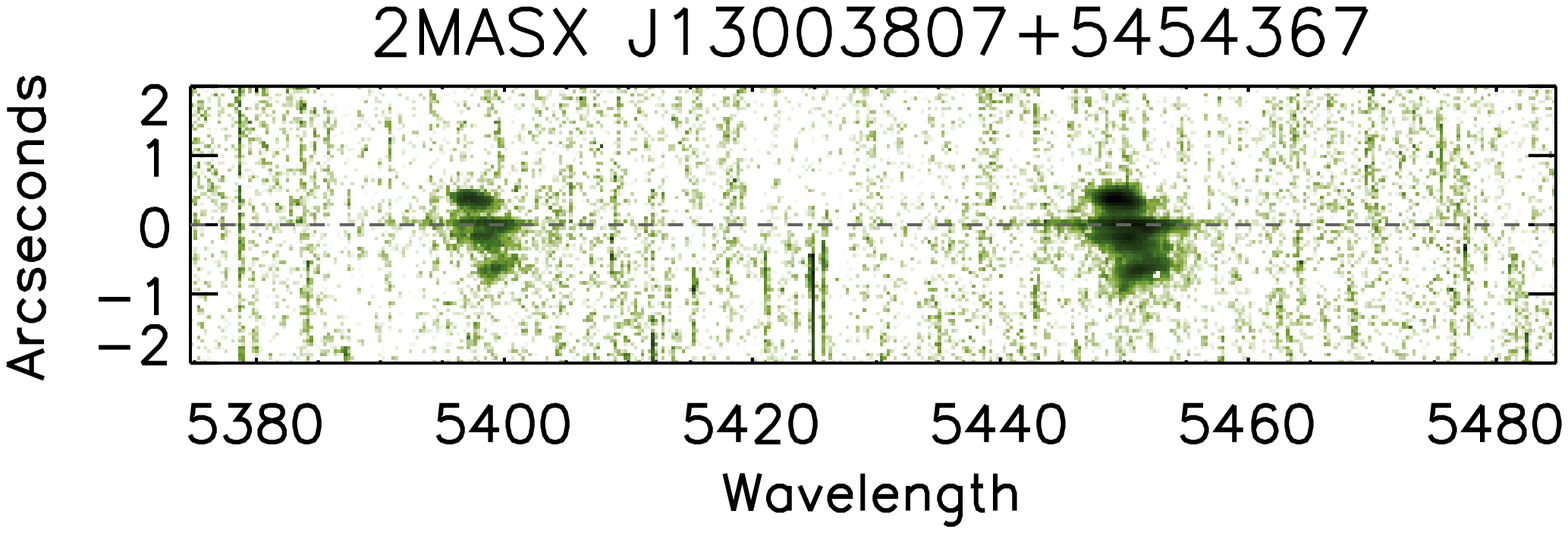}
\includegraphics[width=0.49\textwidth]{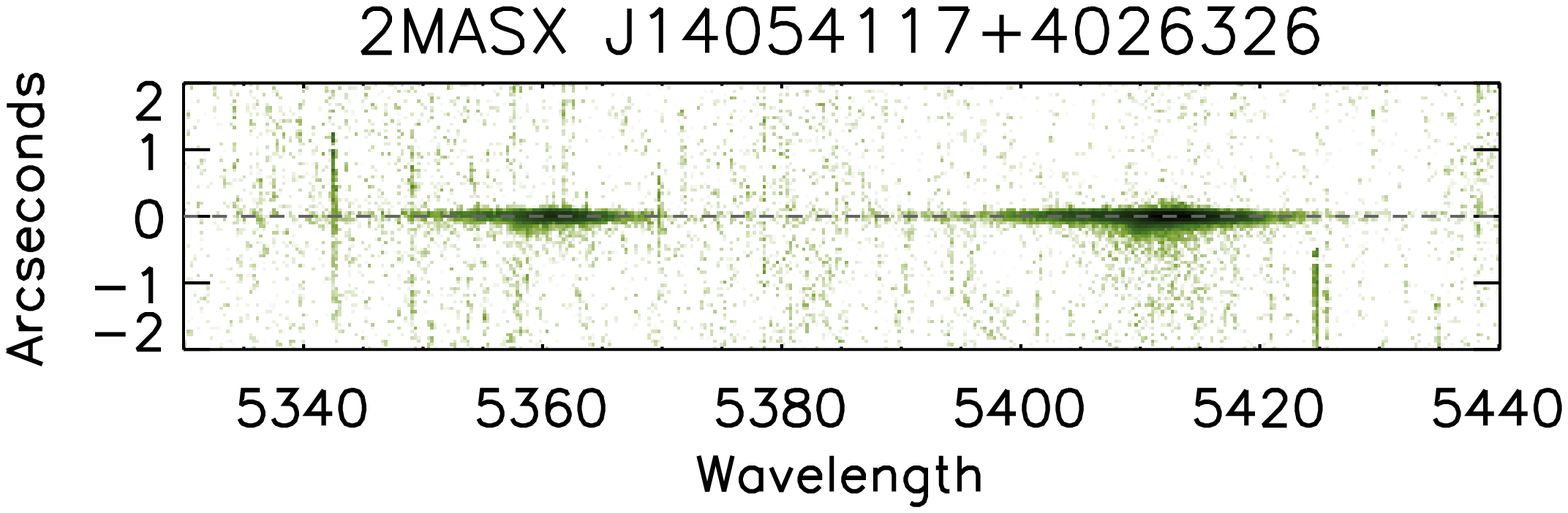}\\
\includegraphics[width=0.49\textwidth]{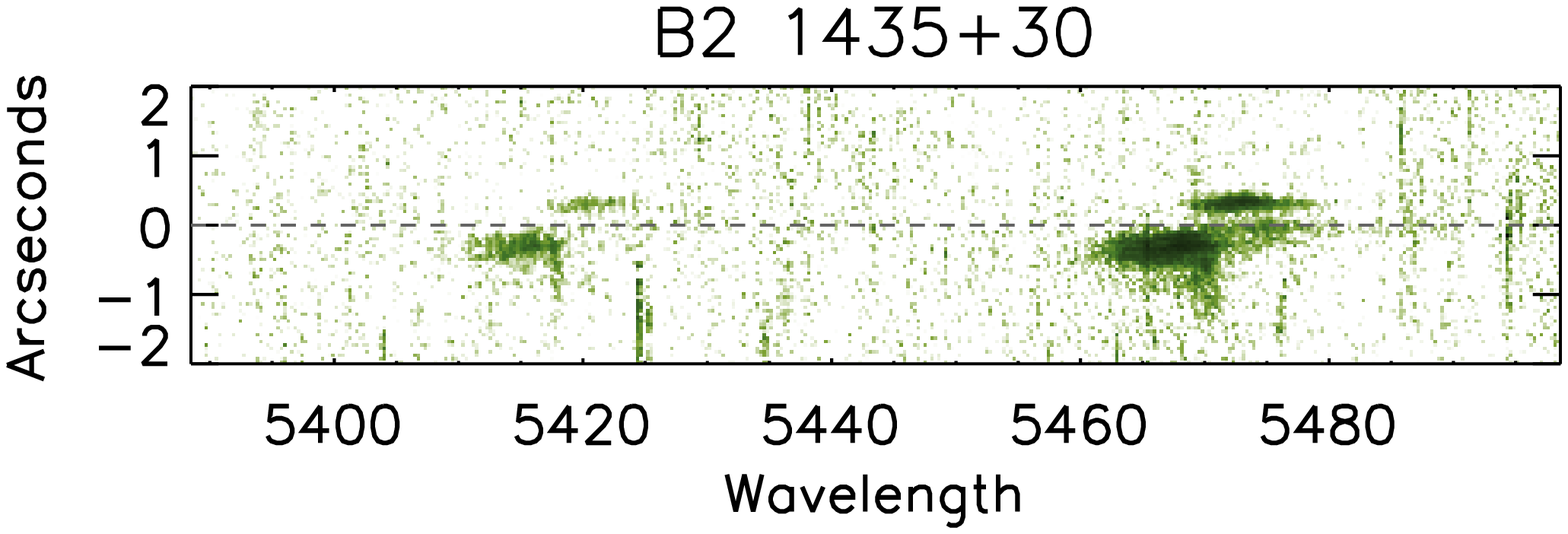}
\includegraphics[width=0.49\textwidth]{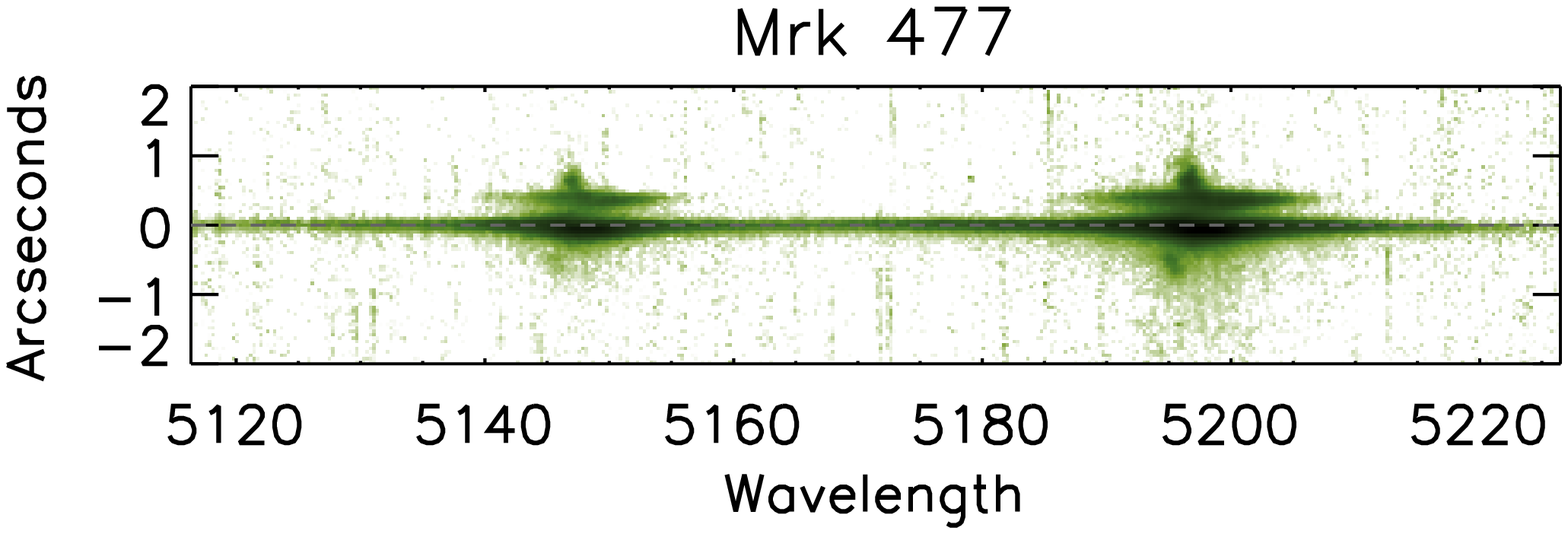}\\
\includegraphics[width=0.49\textwidth]{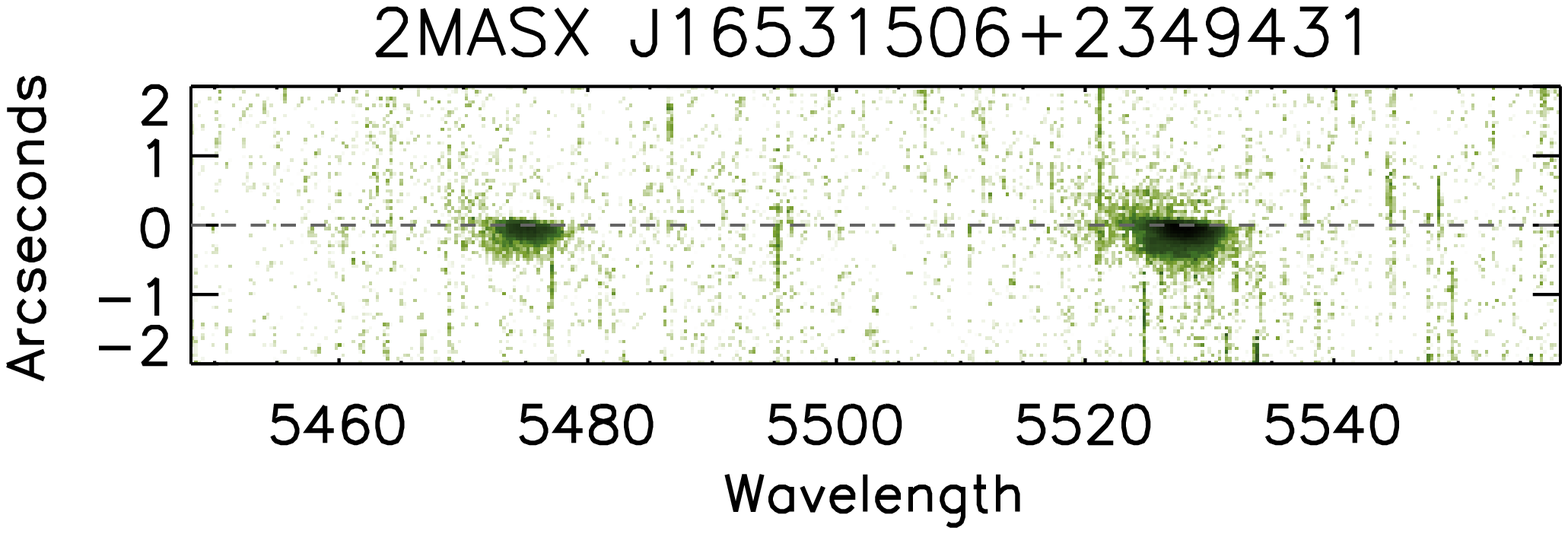}
\includegraphics[width=0.49\textwidth]{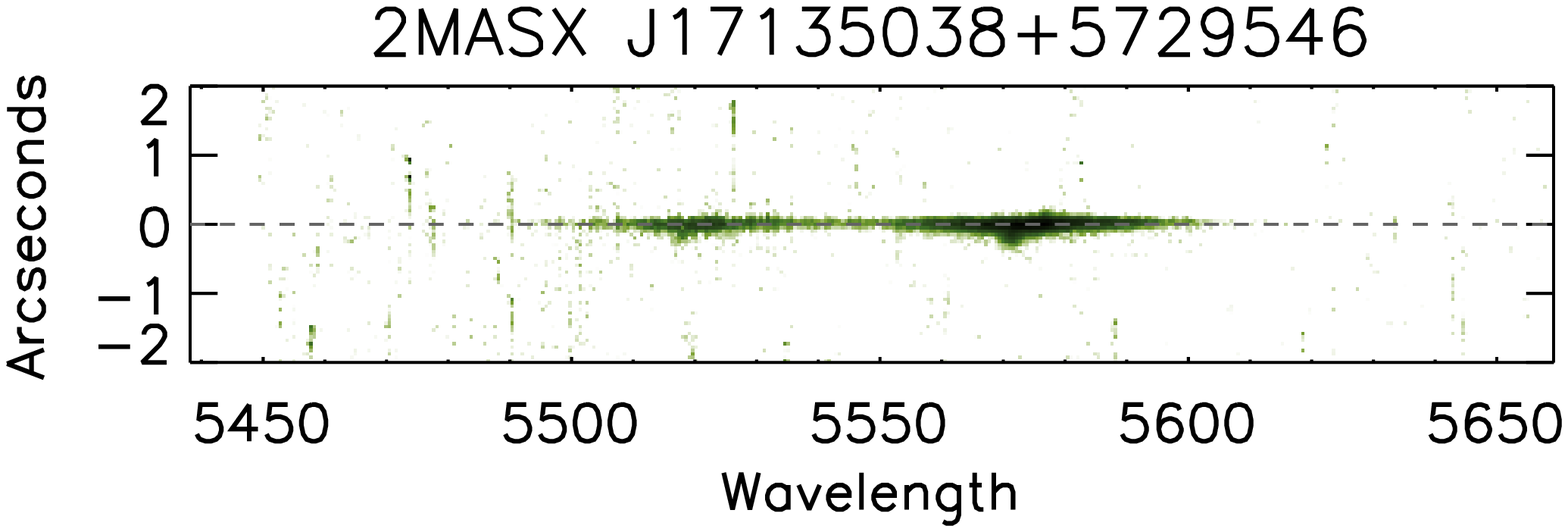}\\

\caption{STIS G430M and G750M long-slit spectral images of the [O~III] $\lambda\lambda$ 4959,5007 lines of our QSO2 sample. Each row 
of a spectral image represents a 0.05$"$ step along the STIS long-slit, with the top of each 2-D spectral image corresponding 
to the end of the long-slit at the position angle of the observations, as listed in Table \ref{tab:obs}}. Images are displayed with a minimum 
flux level of $3\times10^{-17}$ ergs cm$^{-2}$ s$^{-1}$ \AA$^{-1}$. Observations for SDSS J115245.66+101623.8 do not contain [O III] 
$\lambda$4959 line emission as the line is outside the observed G430M wave-band coverage. 
\label{fig:STISccds}

\end{figure*}

Data reduction for continuum and on-band images followed the standard {\it HST} pipeline procedures. On-band and continuum 
images were observed sequentially, not requiring realignment between images. The flux calibration of the FR647M and F467M 
continuum images and the FR505N, FR551N, and F547M on-band images was done in the standard way, using the information 
available on the image headers. Photometrically calibrated, 2D-rectified images of each long-slit observation were processed 
using Interactive Data Language software where multiple, dithered observations of each target were aligned and combined into 
a single, averaged dataset. Extracted, spatially resolved [O~III]$\lambda\lambda$4959,5007 emission lines from each combined 
dataset are shown in Figure \ref{fig:STISccds}. 

The final [O III] images of the galaxies are presented in Figures \ref{fig:images1} - \ref{fig:images7}. The lowest contour 
value of these images corresponds to the 3$\sigma$ level above the background in surface brightness, with interior contours 
increasing in powers of 2 $\times$ 3$\sigma$. 

\subsection{[O III] Spectroscopic Fitting}

\begin{figure}[h]
\centering

\includegraphics[width=0.48\textwidth]{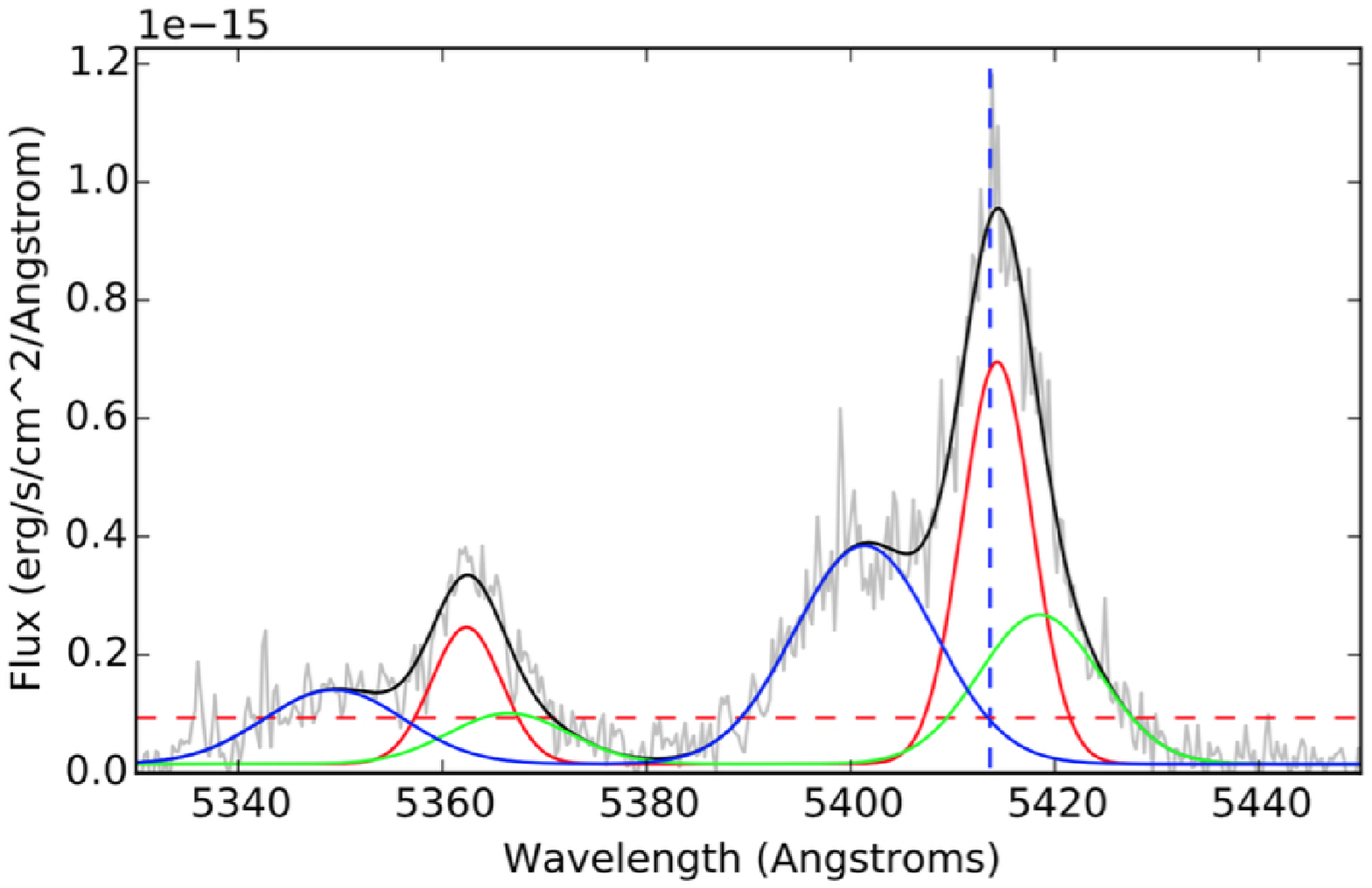}

\caption{[O III] $\lambda\lambda$4959,5007 emission-line component fitting example over the continuum peak (0.0$"$ offset)
in 2MASX J08025293+2552551. Gaussian fits to [O~III] $\lambda$4959 used centroid and width parameters 
fixed to be identical to those of [O~III] $\lambda$5007, with height fixed to be 1/3 of the 
[O~III] $\lambda$5007 Gaussian. Grey line represents STIS spectral data. Solid black line represents the 
total model. Blue, green, and red lines represent individual Gaussian components with 
first-, second-, and third-widest FWHM measurements, respectively. Vertical dashed blue line 
represents the [O III] $\lambda$5007 wavelength at systemic velocity. Horizontal dashed red line 
represents the 3$\sigma$ continuum-flux lower limit for [O~III] $\lambda$5007 Gaussian height in our fitting.}
\label{fig:linefit}

\end{figure}

[O~III] $\lambda$5007 kinematics and fluxes in our STIS observations were measured by fitting the 
emission line in each row of the 2D STIS CCD spectral image with Gaussians in an automated routine 
(Figure \ref{fig:linefit}). Our fitting process, previously discussed in depth in \citet{Fis17} uses the 
Importance Nested Sampling algorithm as implemented in the MultiNest library \citep{Fer08,Fer09,Fer13,Buc14} 
to compute the logarithm of the evidence, $lnZ$, for models containing a continuum plus zero, one, two, 
and three Gaussian components per emission line. When 
comparing two models, i.e. a model with zero Gaussians ($M_{0}$) and a model with one Gaussian ($M_{1}$), 
the simpler model is chosen unless the more complex model, $M_{1}$, has a 
significantly better evidence value, $|ln(Z_{1}/Z_{0})| > 5$ (99\% more likely). Models for measuring [O~III] 
$\lambda$5007 simultaneously fit a second set of components to [O~III] $\lambda$4959 in order to properly 
account for flux contributions from wing emission between both lines. Gaussian centroid and width parameters 
of the [O~III] $\lambda$4959 line were fixed to be identical to parameters used in fitting the [O~III] 
$\lambda$5007 line and flux was fixed to be 1/3 that of the [O~III] $\lambda$5007 flux parameter. Models for 
SDSS J115245.66+101623.8 did not fit the [O~III] $\lambda$4959 line as the redshift of this target placed 
the line outside the observed G430M waveband coverage. 

Initial input parameters in our models are selected based on physical considerations. The centroid position 
for each Gaussian was limited to a 40\AA~range around the systemic [O~III] wavelength that contained the 
entirety of the line profile throughout each long-slit data set. Gaussian standard deviation ranged from the 
spectral resolution of the G430M / G750M gratings, to an artificial limit of 15\AA~([O~III] FWHM $\sim$2000 km s$^{-1}$). 
Gaussian height was permitted to range from a minimum height of 3 times the standard deviation of the adjacent 
continuum (3$\sigma$) to a virtually unbound maximum height restriction of 3$\sigma \times$ 10$^8$.

\subsection{Host Disk Elliptical Fitting}

As illustrated in \citet{Fis17}, spatially-resolved, AGN-ionized gas can be explained as illuminated material located 
near the plane of the host disk, where rotating molecular gas outside the AGN area of influence in Mrk 573 connects 
to the ionized-gas structures inside the AGN area of influence. Therefore, if we assume that a majority of AGN ionized-gas 
structure in other Type 2 AGN also resides in their host disks, we can employ information on the host disk orientation of 
each target to deproject {\it HST} measurements for our QSO2 sample and determine true physical distances.  
Continuum imaging of our sample was obtained in our {\it HST} dataset, however these observations were intended for 
continuum subtraction of [O III] imaging and are too shallow for independent orientation analysis. Measurements of host disk 
ellipticities ($e$) and major axis position angles (PA$_{MA}$) were therefore obtained by fitting ellipses to isophotes in {\it gri} 
Sloan Digital Sky Survey (SDSS; \citealt{Yor00,Eis11}) imaging of each galaxy using the {\tt ellipse} routine in the STSDAS 
package of IRAF. Ellipses were fit from the inner $\sim$ 0.2$''$ of each image, out to where flux levels reached a 
3$\sigma$ level above the background in all bands. The background level and its standard deviation were determined 
from blank regions within the SDSS CCD field of each galaxy. Targets with interacting morphologies, 2MASX 
J08025293+2552551, B2 1435+30, and Mrk 477, were fit to a 5$\sigma$ level and FIRST J120041.4+314745, which 
is immediately adjacent to a neighboring galaxy, was fit to a 9$\sigma$ level to avoid fit disturbances in low flux 
isophotes. Ellipse parameters obtained from the fit of each SDSS image are shown in Appendix Figures \ref{fig:images8} 
- \ref{fig:images13}. Final values of PA$_{MA}$ and $e$ used for our calculations, as listed in Table \ref{tab:sdssobs}, are the 
average of values taken from each band at the greatest distance where the minimum S/N measurements exist in 
all bands. The difference in host major axis and STIS observation position angles is also provided in Table \ref{tab:sdssobs} 
as PA$_{diff}$.

\begin{table}[h!]
  \centering
  \ra{1.3}
  \caption{QSO2 Sample SDSS Imaging Measurements}
  \label{tab:sdssobs}
  \begin{tabular}{lrrrlr}
    \toprule	
    Target									&PA$_{MA}$		    & PA$_{diff}$   &R$_{MA}$		    & $e$			    \\
  										    &(degrees) 		    & (degrees)     &(kpc)			    &	             	\\                                                                                     
    J07594101                      	        & 152.0 $\pm$ 11.4 	& 49.3          & 6.5				& 0.137 $\pm$ 0.044 \\
    J08025293	                            & 111.0 $\pm$ 2.2	& 87.4          & 6.9				& 0.254 $\pm$ 0.023 \\
    MRK 34					        		& 29.7 $\pm$ 3.1	& 57.4          &13.0 	    		& 0.246 $\pm$ 0.023 \\
    J11001238	                        	& 61.3 $\pm$ 1.6	& 80.7          &15.3 		        & 0.339 $\pm$ 0.016 \\ 
    J115245.66 	                        	& 43.4 $\pm$ 5.1	& 33.3          &7.2				& 0.230 $\pm$ 0.040 \\
    J120041.4		                		& 57.2  $\pm$ 7.4	& 33.1          &5.5				& 0.204 $\pm$ 0.047 \\
    J13003807	                        	& 115.8 $\pm$ 26.7	& 48.6          &6.9				& 0.046 $\pm$ 0.060 \\
    J14054117	                        	& 91.7  $\pm$ 10.7 	& 45.4          &7.0				& 0.128 $\pm$ 0.056 \\
    B2 1435+30					            & 171.5 $\pm$ 7.1 	& 88.6          &8.7				& 0.134 $\pm$ 0.035 \\
    MRK 477					        		& 112.0 $\pm$ 9.8	& 82.2          &5.6				& 0.073 $\pm$ 0.023 \\
    J16531506	                        	& 153.0 $\pm$ 5.9 	& 77.9          &10.6		        & 0.363 $\pm$ 0.058 \\
    J17135038	                        	& 191.5 $\pm$ 4.6 	& 16.3          &9.4				& 0.238 $\pm$ 0.032 \\
    \bottomrule
  \end{tabular}
  \raggedright Column 2 lists the position angle of the host disk major axis used in our calculations. Column 3 lists the difference between the major 
  axis and STIS long-slit position angles. Columns 4 and 5 list the radius of the major axis and ellipticity of host disk used in our measurements, respectively.

  \vspace{.5cm}
\end{table}

\begin{table*}[h!]
  \centering
  \ra{1.3}
  \caption{QSO2 [O III] Imaging Measurements}
  \label{tab:hstphotobs}
  \begin{tabular}{lrrrrlr}
    \toprule	
    Target						 			&  PA$_{[OIII]}$  & Proj. R$_{max}$ 	    & Proj. R$_{90\%}$ 	&  3$\sigma$ F$_{\lambda}$ 	&Proj. 	 	  & Deproj. R$_{max}$     	    \\
  							 			    & (degrees)      	  &(kpc)		 	    & (kpc)				& (erg/cm$^{2}$/s/\AA/pixel)		&Scale		  & (kpc)			      	     \\    
										\midrule                                                                                                                                                 
    2MASX J07594101+5050245	                &   105	   	  & 2.98 			    & 0.42 				& 8.8e-20					& 0.919		  & 3.24			           \\
    2MASX J08025293+2552551	                &	0	  	  & 2.76			    & 1.35				& 8.2e-20					& 0.768		  & 3.61			          \\
    MRK 34					         		&	150	  	  & 1.75			    & 1.34				& 5.3e-20					& 0.799		  & 2.19			          \\
    2MASX J11001238+0846157	                &	---	  	  & 1.98			    & 0.75				& 7.2e-20					& 0.661\footnote{\label{mynote}Maximum deprojection along minor axis.}             & 3.00    \\ 
    SDSS J115245.66+101623.8 	            &	10	  	  & 3.25 			    & 1.33				& 7.4e-20					& 0.909		  & 3.57			          \\
    FIRST J120041.4+314745		            &	75	  	  & 5.92			    & 2.60				& 5.6e-20					& 0.974 	  	  & 6.07			           \\
    2MASX J13003807+5454367	                &	165	  	  & 4.53			    & 1.58				& 5.6e-20					& 0.972 	 	  & 4.66			            \\
    2MASX J14054117+4026326	                &	---	   	  & 0.88 			    & 0.39				& 6.3e-20					& 0.872\footref{mynote}             & 1.01	  \\
    B2 1435+30					         	&	90	   	  & 3.32 			    & 1.46				& 5.9e-20					& 0.868 	  	  & 3.82			           \\
    MRK 477					         		&	35	  	  & 2.54			    & 0.83				& 2.1e-19					& 0.930		  & 2.74			           \\
    2MASX J16531506+2349431	                & 	77	  	  & 4.21			    & 1.76				& 6.6e-20					& 0.648		  & 6.50			         \\
    2MASX J17135038+5729546	                &	---	  	  & 1.38			    & 0.52				& 5.0e-20					& 0.762\footref{mynote}             & 1.81	  \\
    \bottomrule
  \end{tabular}
  
  \raggedright Column 4 is the projected radius that encircles 90\% of the [O III] flux inside Proj. R$_{max}$. Column 6 lists the scaling factor used to deproject measurements to their true extents, assuming the 
  observed emission is near the plane of the host galaxy. 
  
\end{table*}

\begin{table*}[h!]
  \centering
  \ra{1.3}
  \caption{QSO2 [O III] Spectroscopic Measurements}
  \label{tab:hstspecobs}
   \tabcolsep=0.2cm
  \begin{tabular}{lrrrrrrrr}
    \toprule
    Target						& Proj.  	    & Proj. 	    & Proj. 	    & Nuclear 		& Deproj.       & Deproj.	      & R$_{out} /$ \\
                                & R$_{out}$     & R$_{dist}$    & V$_{out}$     & FWHM          &  R$_{out}$    &  R$_{dist}$     & R$_{max}$   \\
							    & (kpc)		    & (kpc)			& (km/s)		& (km/s)		&	(kpc)		&	(kpc)	      &             \\
    \midrule
    2MASX J07594101+5050245	    & 0.62			& 0.62			&315			& 1680			& 0.67 			& 0.67			  & 0.21		\\
    2MASX J08025293+2552551	    & 0.44			& 0.68			& -700			& 870			& 0.57			& 0.89			  & 0.08		\\
    MRK 34					    & 1.50  	    & 1.50      	& 1100			& 580			& 1.89       	& 1.89      	  & 0.86    	\\
    2MASX J11001238+0846157 	& 0.45			& $>$1.00 		& -350			& 1780			& 0.68          & $>$1.51		  &	0.23		\\ 
    SDSS J115245.66+101623.8 	& 0.13			& $>$1.12		& -300 			& 360			& 0.15 			& 1.23			  &	0.04		\\
    FIRST J120041.4+314745 		& 1.04 			& $>$1.55		& 450			& 720			& 1.07	 		& $>$1.59	      &	0.18		\\
    2MASX J13003807+5454367	    & 0.16			& 0.16			& 100			& 580			& 0.16			& 0.16			  &	0.04		\\
    2MASX J14054117+4026326	    & 0.29			& $>$0.82		& 100			& 760			& 0.33	 		& $>$0.94		  &	0.33		\\
    B2 1435+30					& 0.17			& $>$1.51      	& 200			& 630			& 0.20	 		& $>$1.74		  &	0.05		\\
    MRK 477					    & 0.50       	& 0.84			& -500			& 2040			&  0.54     	& 0.90			  &	0.20    	\\
    2MASX J16531506+2349431	    & 0.37			& $>$1.14 		& 110			& 510			& 0.57			& 1.23 			  &	0.09		\\
    2MASX J17135038+5729546	    & 0.49      	& $>$0.70		& -160			& 1660			& 0.65      	& $>$0.92 		  & 0.36    	\\
    \bottomrule
  \end{tabular}
  
  \raggedright Column 2 lists the maximum projected distance of outflowing kinematics. Column 3 lists the maximum projected distance of 
  disturbed kinematics, gas exhibiting FWHM $>$ 250 km/s that is at systemic or follows rotation. Column 4 lists the maximum velocity 
  measured inside R$_{out}$, as the maximum outflow velocity. Column 9 lists the ratio of maximum outflow radius to maximum [O III] extent 
  in each target.
  
\end{table*}

\begin{figure*}
\centering

\includegraphics[width=0.95\textwidth]{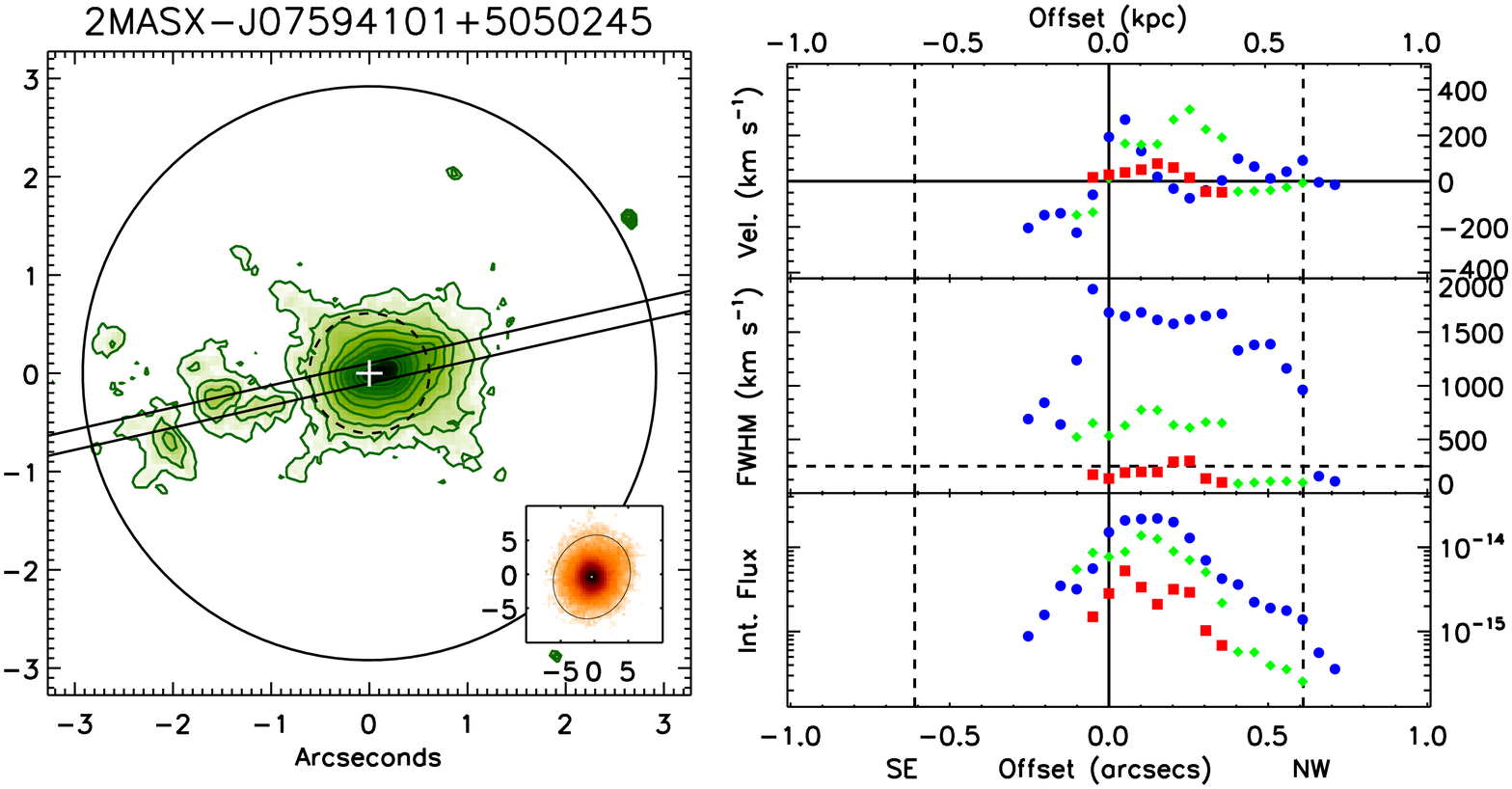}\\
\includegraphics[width=0.95\textwidth]{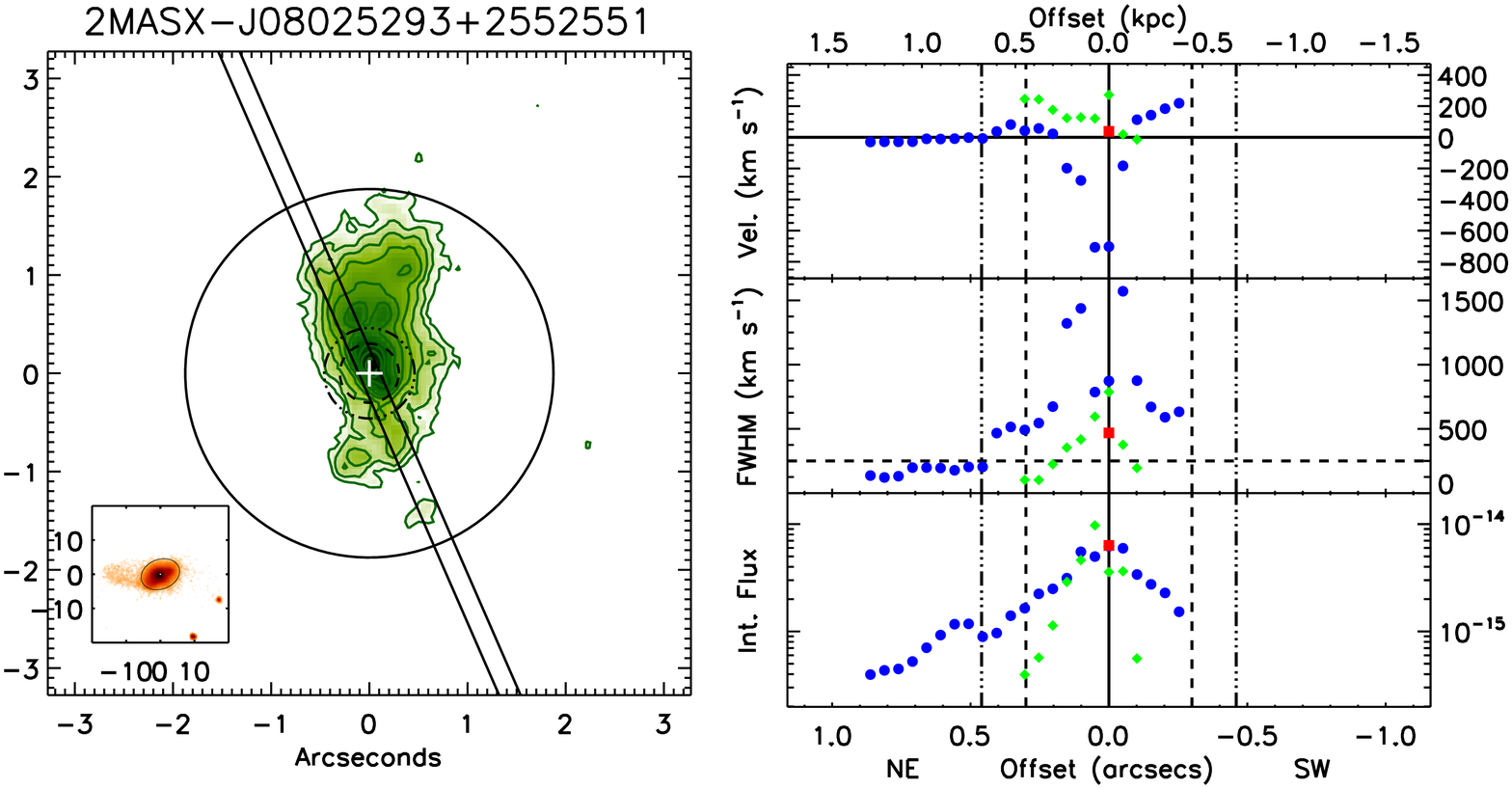}\\

\caption{{\it HST} and SDSS datasets for 2MASX J07594101+505245 and 2MASX J08025293+2552551. 
Left: $5''\times5''$ Continuum-subtracted [O~III] $\lambda5007$ images of 
each target. [O~III] flux contours start at 3$\sigma$ over background and increase in powers of 2 times 3$\sigma$ 
(i.e. 3$\sigma \times 2^{n}$). The position of the nucleus, measured at the continuum-image flux peak of each target, 
is plotted as a white cross. Dashed, dotted-dashed, and solid circles represent R$_{out}$, R$_{dist}$ (when different 
than R$_{out}$), and R$_{max}$ radii, respectively. Slits depict the position of STIS long-slit positions. 
Inset images display $20''\times20''$ and $40''\times40''$ $r$-band SDSS images of 2MASX J07594101+505245 and 
2MASX J08025293+2552551, respectively, with the host disk elliptical fit used in our analysis overplotted. North is up 
and east is left in all images.
Right: Velocities, FWHMs, and integrated fluxes (ergs s$^{-1}$ cm$^{-2}$) for each emission-line component 
of our kinematic fits to the [O~III] $\lambda$5007 emission-line along our {\it HST}/STIS medium-resolution observations. 
Kinematic data points are marked as blue circles, green diamonds, and red squares corresponding to components with 
first-, second-, and third-widest FWHM measurements, respectively. The horizontal, dashed line in FWHM plots 
signifies the approximate limit for non-disturbed kinematics of 250 km s$^{-1}$. Vertical dashed and dotted-dashed 
lines represent the maximum radius of observed outflows (R$_{out}$) and disturbed kinematics (R$_{dist}$; when 
different than R$_{out}$), respectively. R$_{dist}$ radii for 2MASX J07594101+505245 are equal in size to R$_{out}$. 
R$_{out}$ and R$_{dist}$ radii in each target are mirrored across the nucleus to illustrate the maximum extent of each 
region and associate the kinematics to what is observed in imaging, even if no data points exist out to those radii 
on a given side of the nucleus.}
\label{fig:images1}
\end{figure*}

\begin{figure*}
\centering

\includegraphics[width=0.95\textwidth]{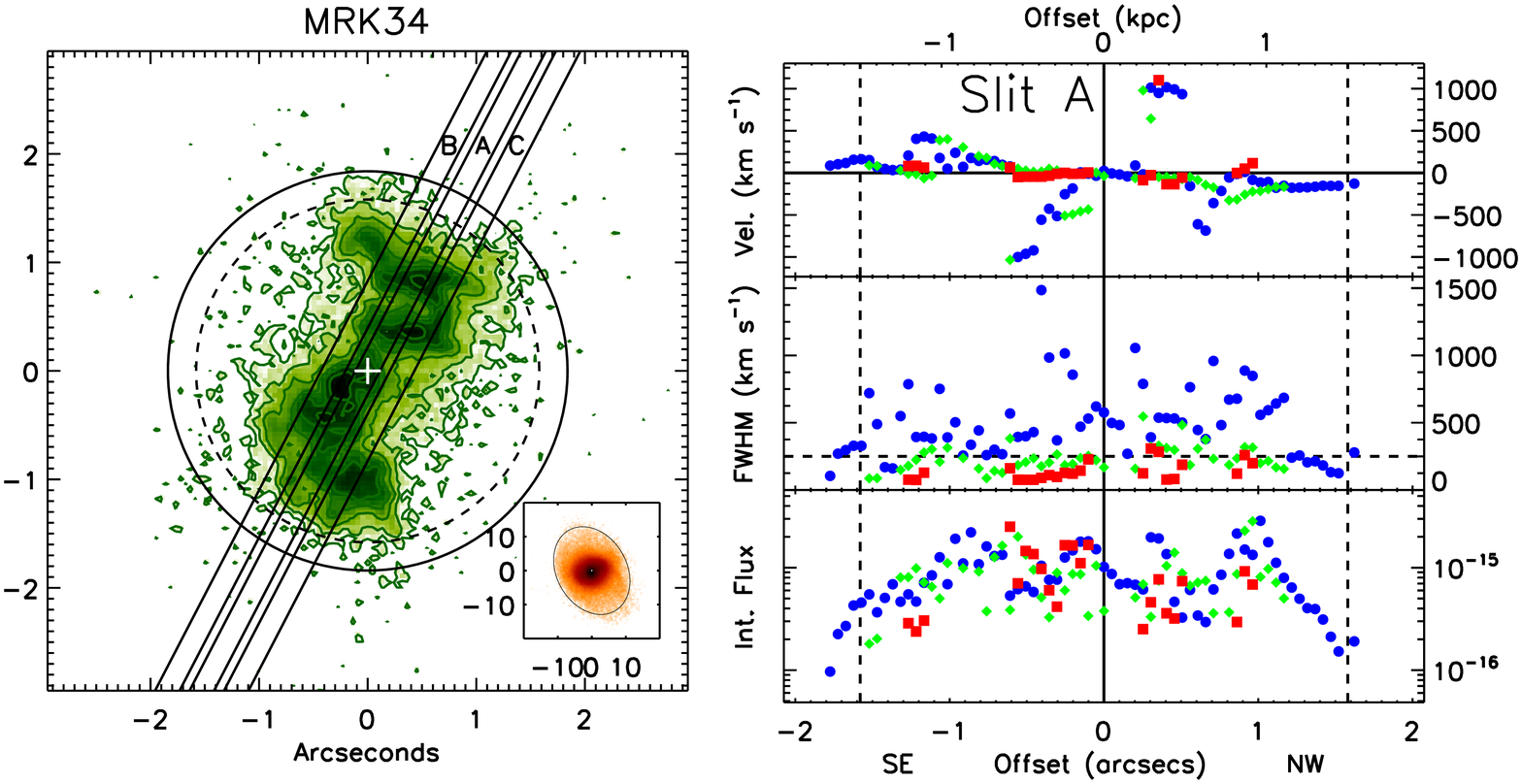}\\
\includegraphics[width=0.95\textwidth]{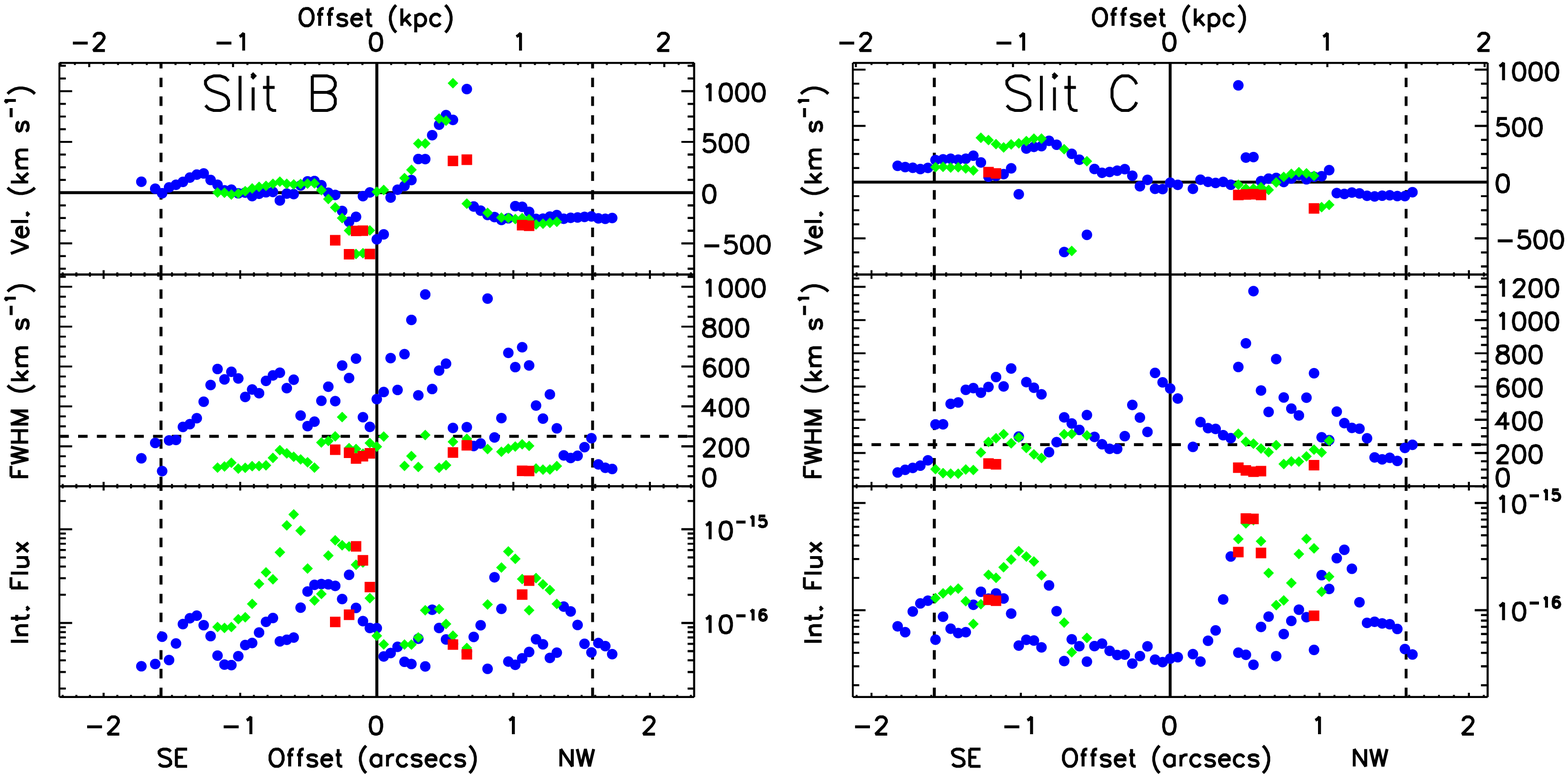}\\
\caption{Same as Figure \ref{fig:images1} for Mrk 34. Inset image $r$-band SDSS image is $40''\times40''$, with the host disk ellipticals fit to 3$\sigma$. R$_{dist}$ radii for Mrk 34 are equal in size to R$_{out}$. R$_{out}$ is measured as the maximum radius measurement across all three long-slit observations.}
\label{fig:images2}

\end{figure*}

\begin{figure*}
\centering
\includegraphics[width=0.95\textwidth]{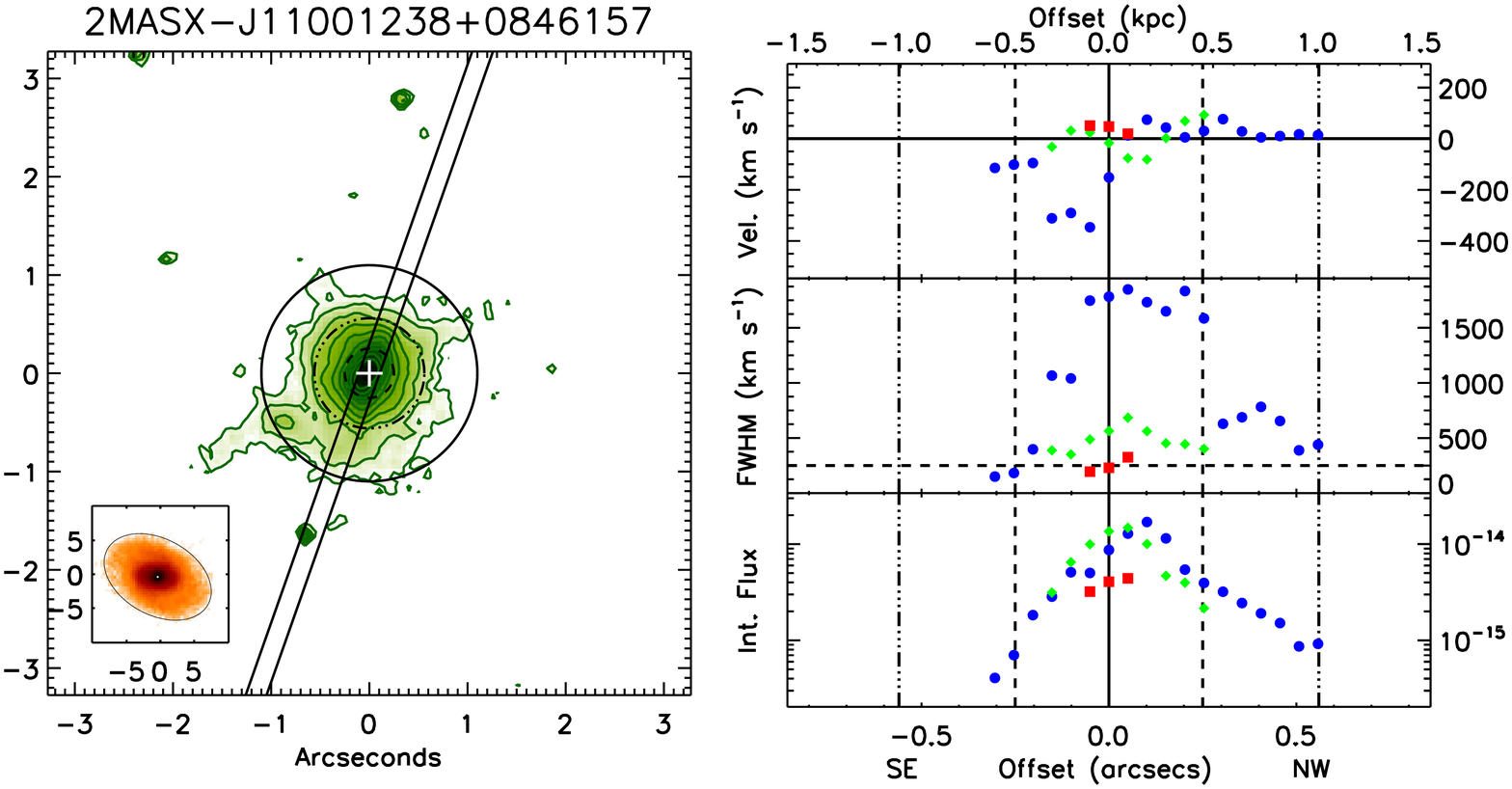}\\
\includegraphics[width=0.95\textwidth]{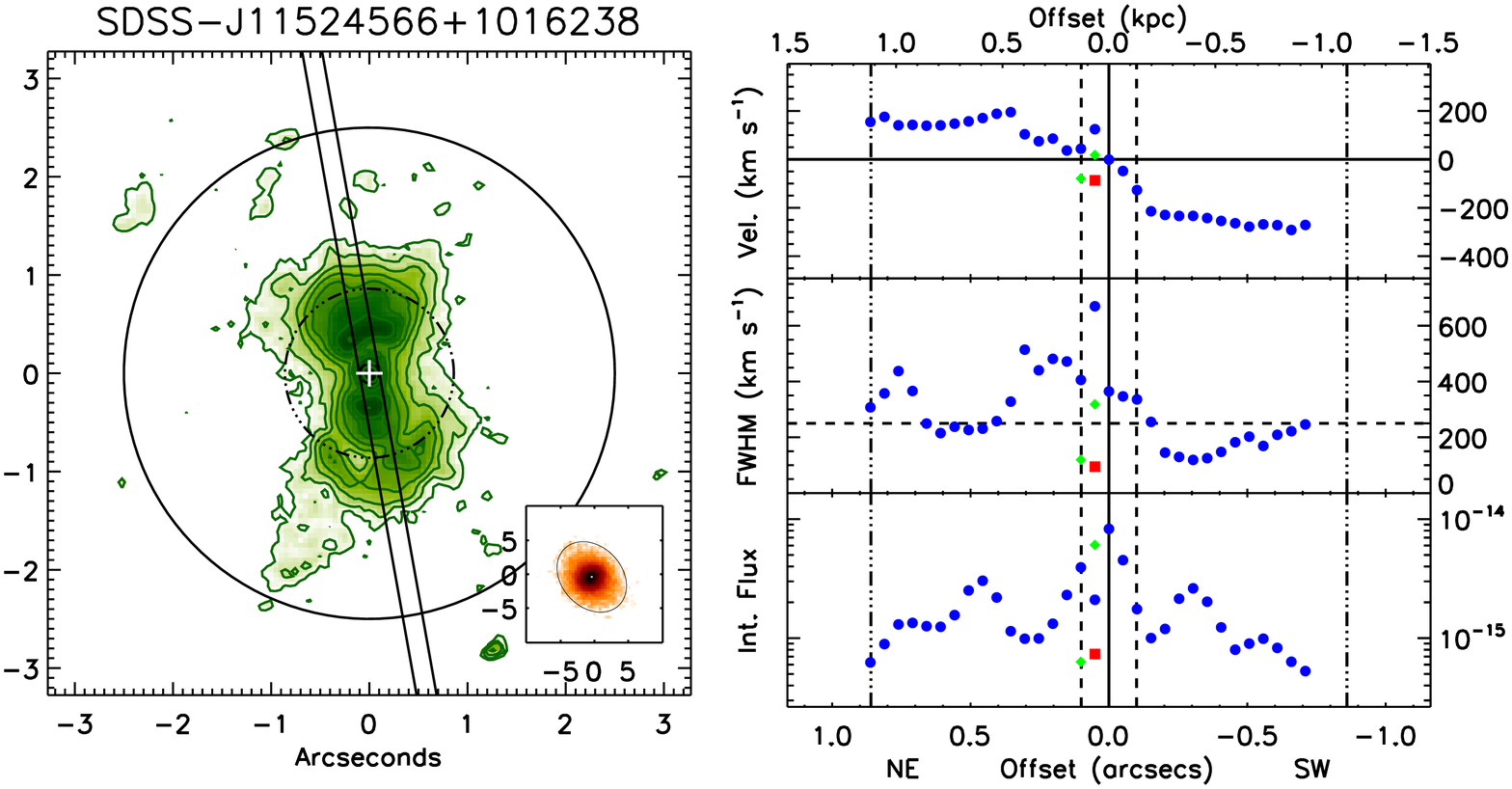}\\

\caption{Same as Figure \ref{fig:images1} for 2MASX J11001238+0846157 and SDSS J115245.66+101623.8. Inset image $r$-band SDSS images are $20''\times20''$, with host disk ellipticals both fit to 3$\sigma$.}
\label{fig:images3}

\end{figure*}

\begin{figure*}
\centering
\includegraphics[width=0.95\textwidth]{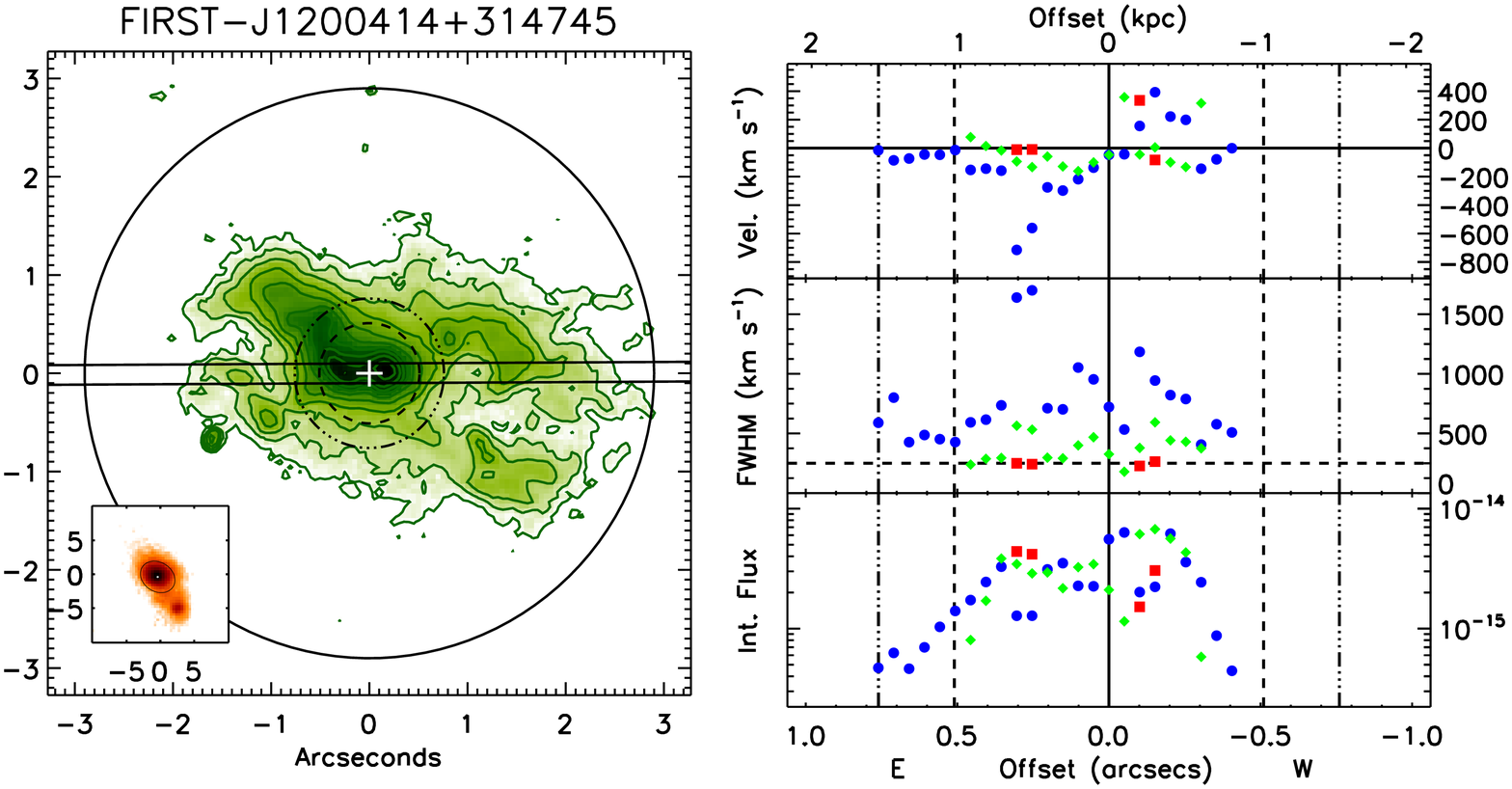}\\
\includegraphics[width=0.95\textwidth]{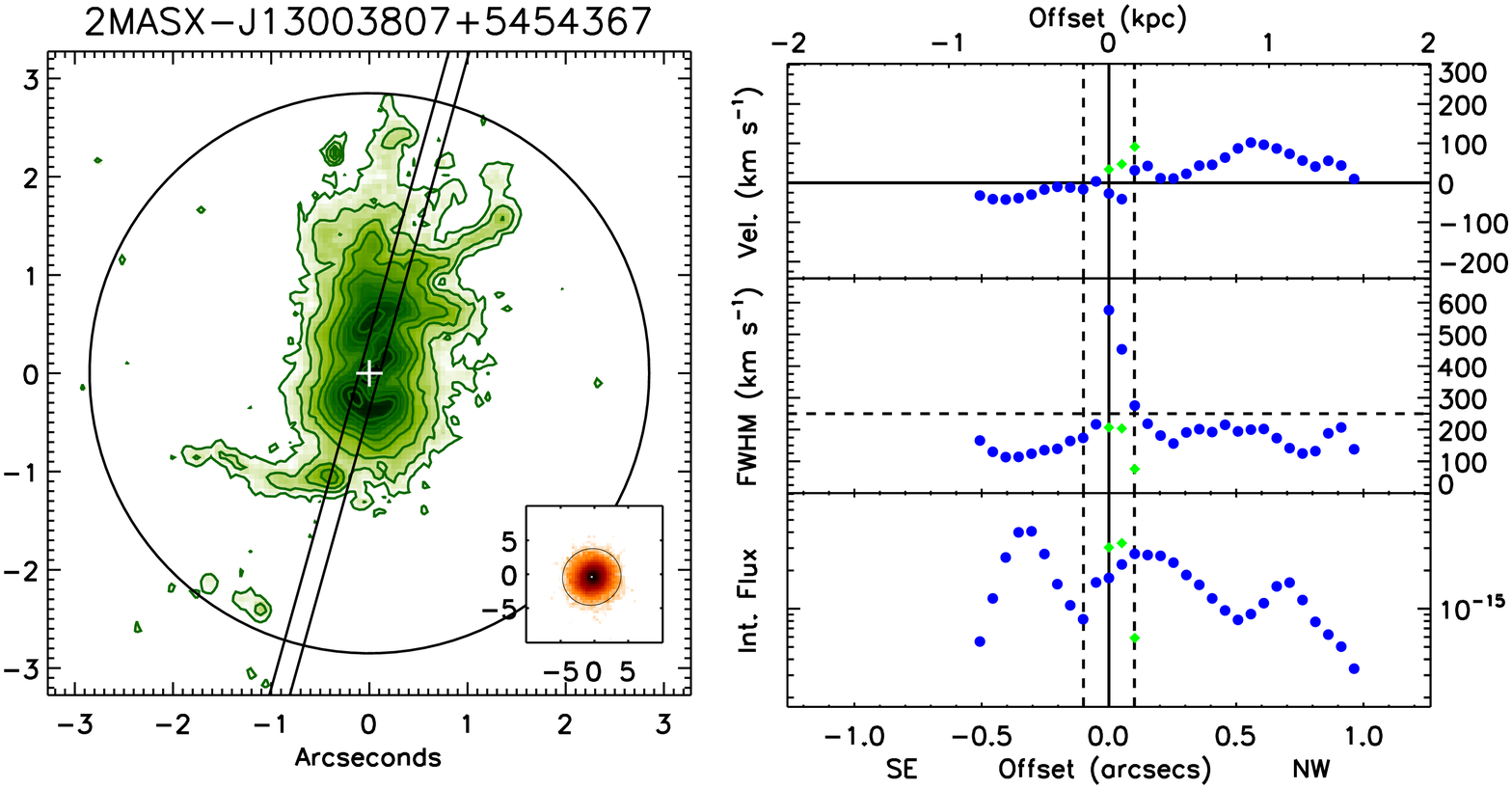}\\	

\caption{Same as Figure \ref{fig:images1} for FIRST J120041.4+314745 and 2MASX J13003807+5454367. R$_{dist}$ radii for 2MASX J13003807+5454367 are equal in size to R$_{out}$. Inset image $r$-band SDSS images are $20''\times20''$, with host disk ellipticals fit to 9$\sigma$ and 3$\sigma$, respectively.}
\label{fig:images4}

\end{figure*}

\begin{figure*}
\centering
\includegraphics[width=0.95\textwidth]{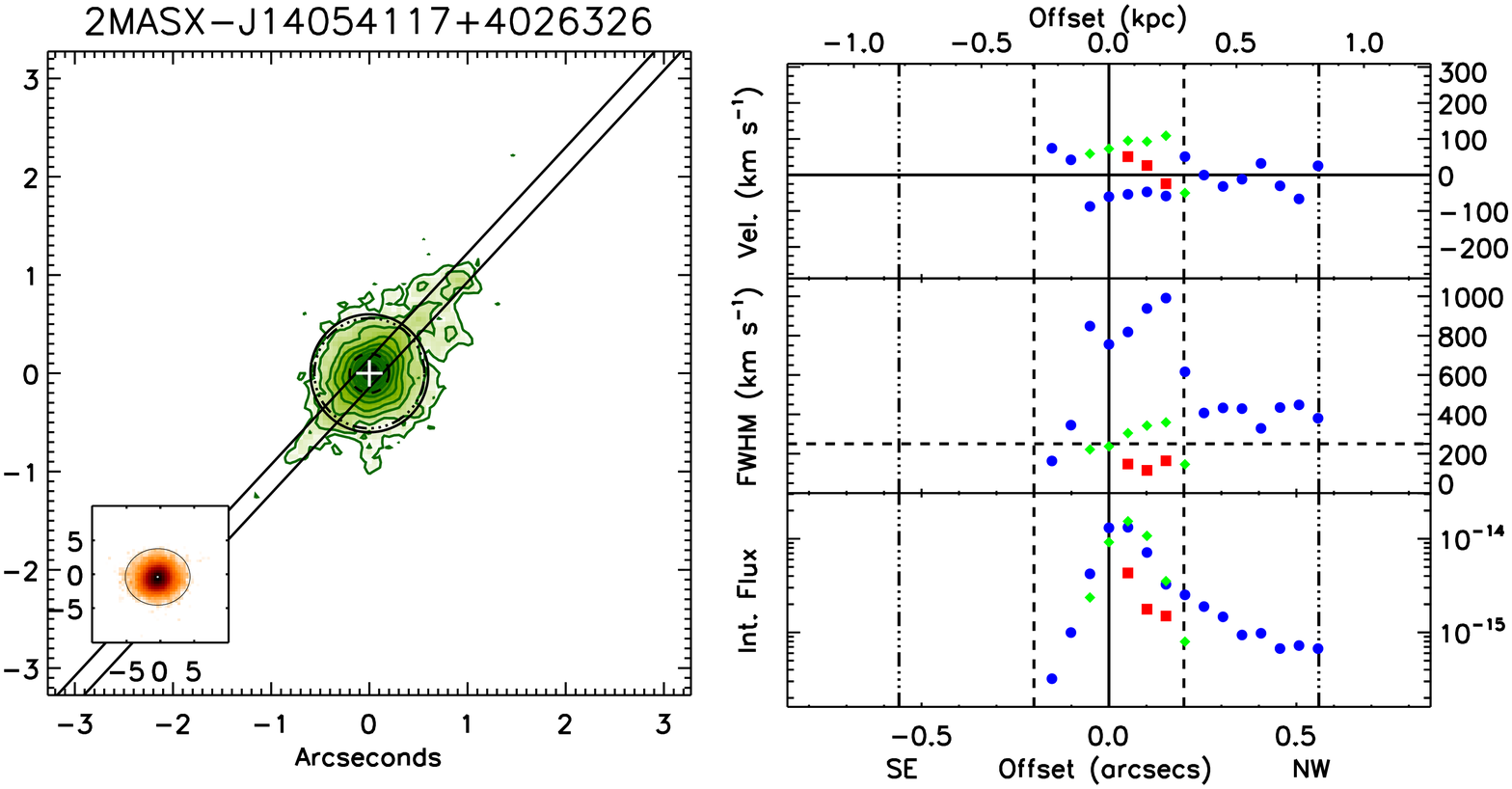}\\
\includegraphics[width=0.95\textwidth]{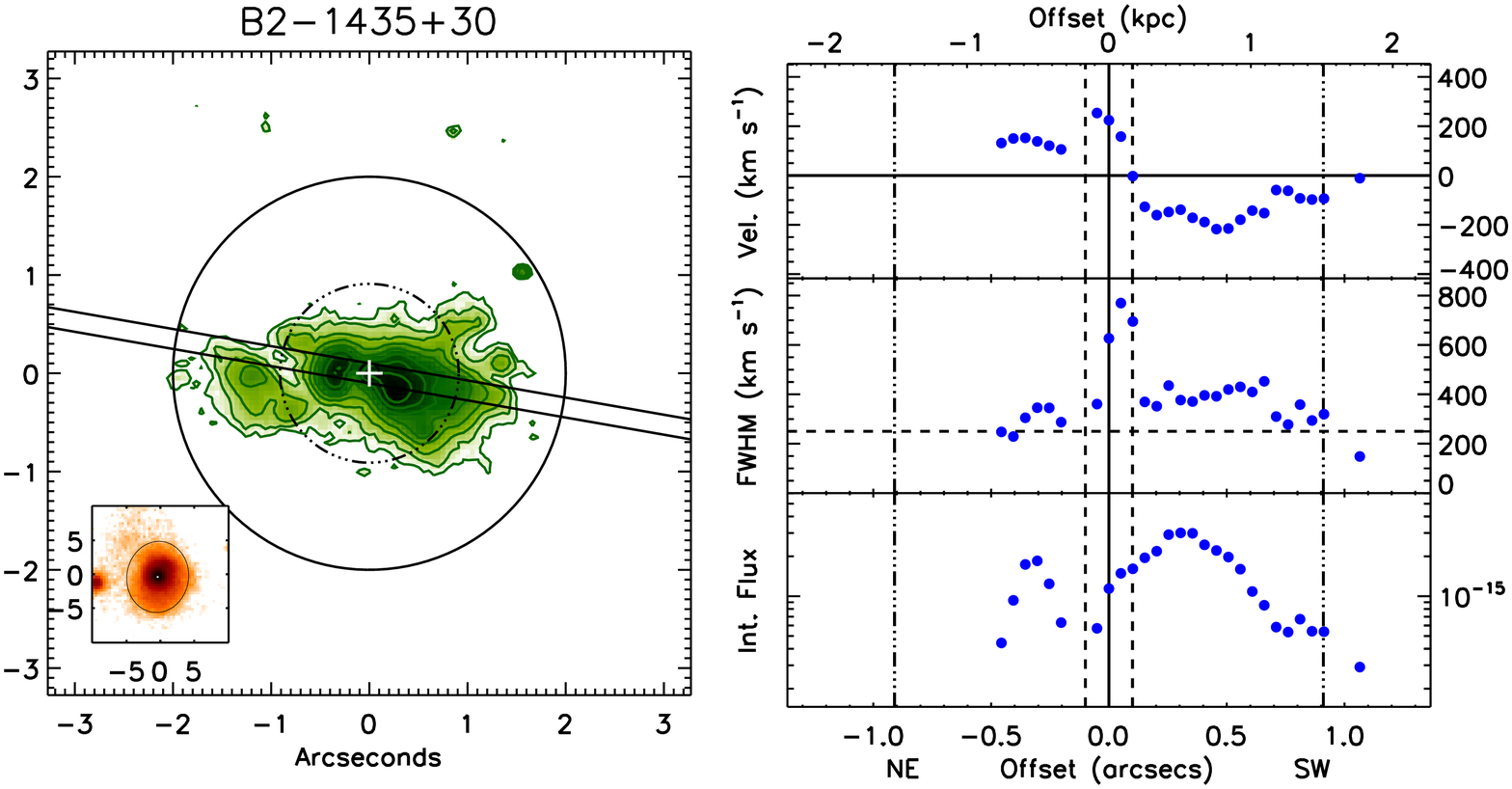}\\

\caption{Same as Figure \ref{fig:images1} for 2MASX J14054117+4026326, B2 1435+30. Inset image $r$-band SDSS images are $20''\times20''$, with host disk ellipticals fit to 3$\sigma$ and 5$\sigma$, respectively.}
\label{fig:images5}

\end{figure*}

\begin{figure*}
\centering

\includegraphics[width=0.95\textwidth]{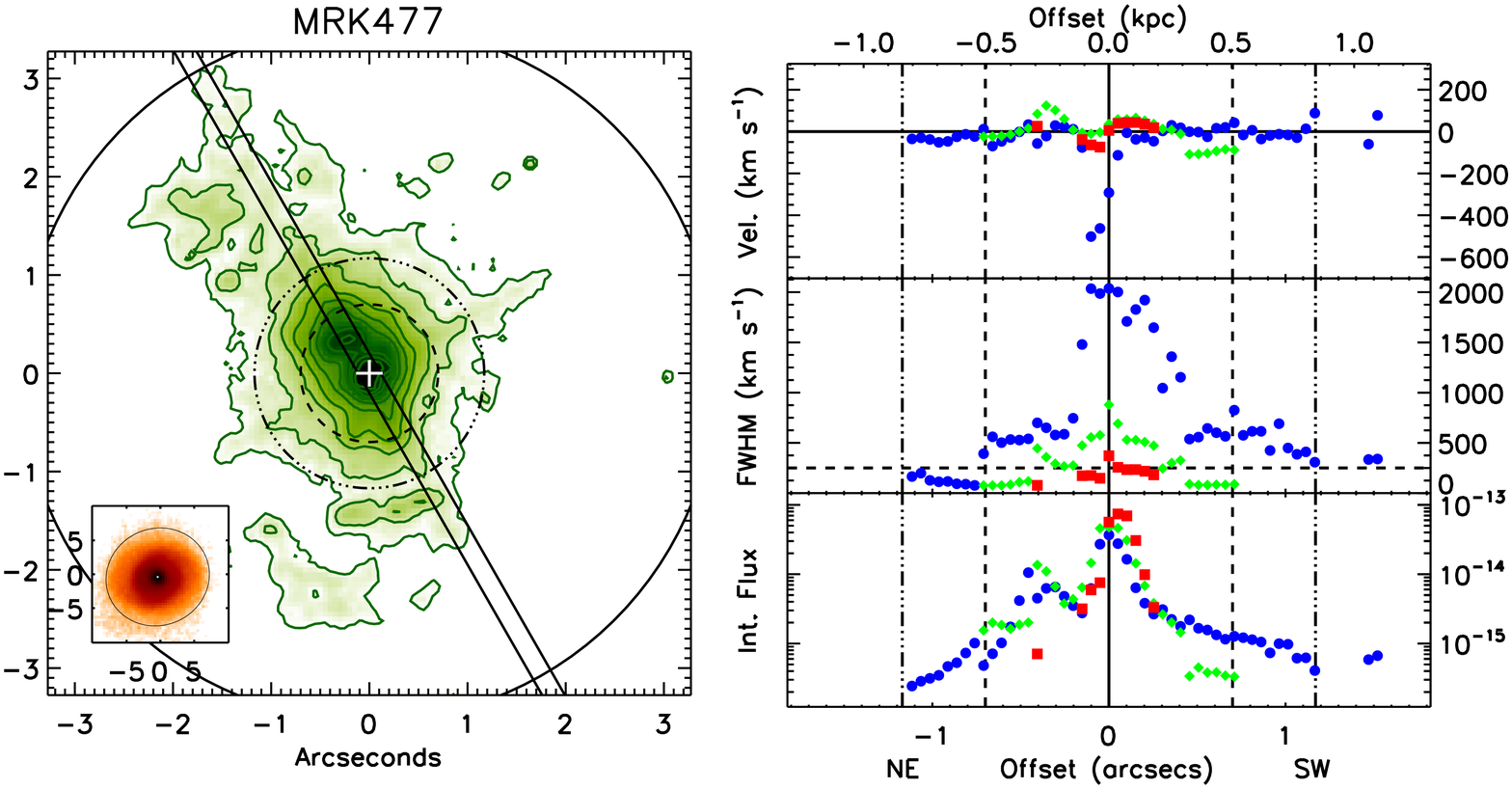}\\
\includegraphics[width=0.95\textwidth]{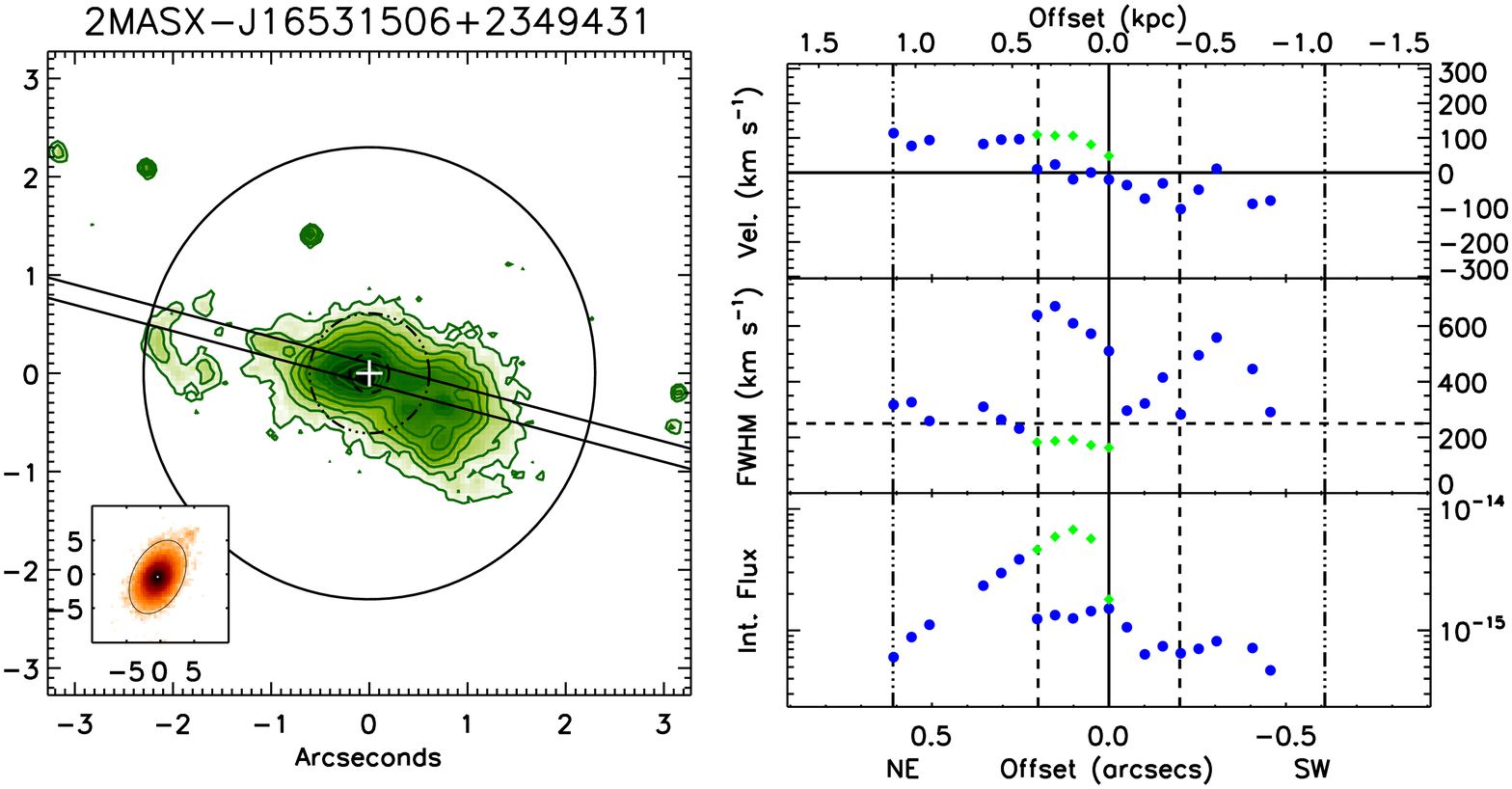}\\	

\caption{Same as Figure \ref{fig:images1} for Mrk 477 and 2MASX J16531506+2349431. 
Inset image $r$-band SDSS images are $20''\times20''$, with host disk ellipticals fit 
to 5$\sigma$ and 3$\sigma$, respectively.}
\label{fig:images6}

\end{figure*}

\begin{figure*}
\centering

\includegraphics[width=0.95\textwidth]{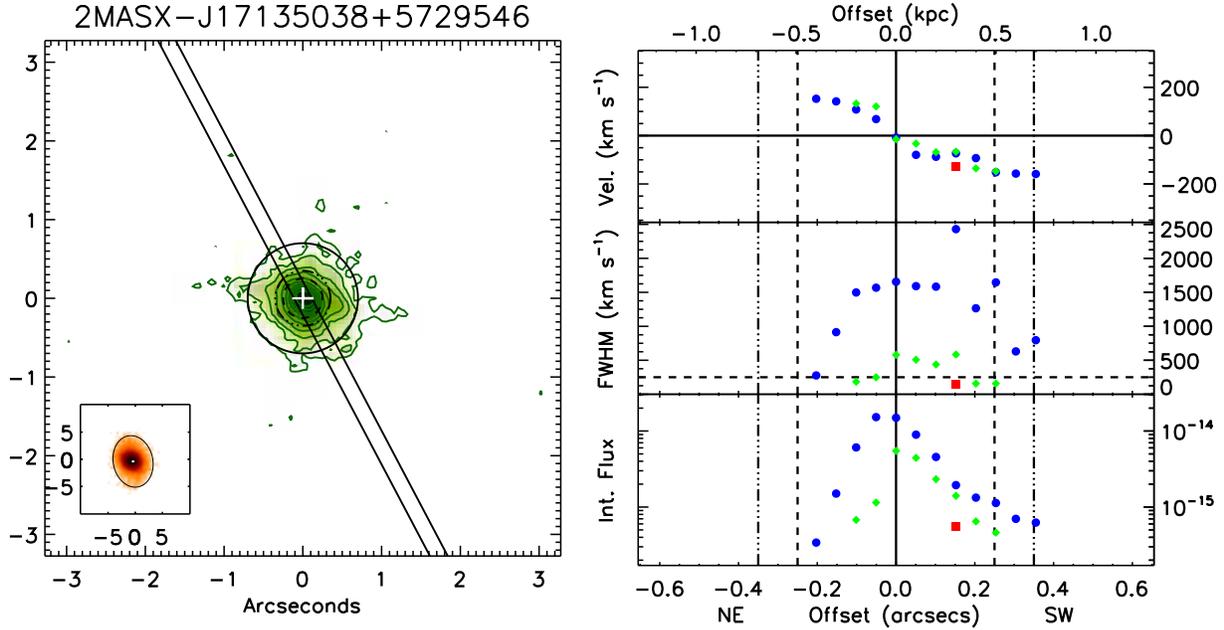}\\

\caption{Same as Figure \ref{fig:images1} for 2MASX J17135038+5729546. Inset image $r$-band 
SDSS image is $20''\times20''$, with a host disk elliptical fit to 3$\sigma$.}
\label{fig:images7}

\end{figure*}

\section{Results}
\label{analysis_sec}

\subsection{Morphologies}
Continuum-subtracted [O~III] images were used to measure the projected extension of AGN ionization in these galaxies. Using the 
continuum centroid as the nucleus location in each galaxy, we measured the length between the continuum centroid and the furthest 
radial extent in each [O~III] image (R$_{max}$), as determined by the 3$\sigma$ surface brightness contour in each image. 
R$_{max}$ measurements are illustrated in the left side of Figures \ref{fig:images1} - \ref{fig:images4} as solid, black circles. 
To distinguish targets that have similar [O III] extents, but different flux distributions, we also provide the radius that encompasses 
90\% of the flux inside R$_{max}$, R$_{90\%}$, which is smaller for centrally peaked [O III] sources and larger for sources with 
bright, extended structures. Position angles (PAs) of the [O~III] emission axis in each target were measured by eye. [O III] emission 
in three galaxies, 2MASX J11001238+0846157, 2MASX J14054117+4026326, and 2MASX J17135038+5729546, was not extended enough to 
accurately determine a PA. Table \ref{tab:hstphotobs} gives the measured values of R$_{max}$, R$_{90\%}$, PA, and the 3$\sigma$ 
flux density used to trace the extent of the [O~III] gas, with angular radii converted to true radii using scales listed in Table 
\ref{tab:obs} derived from the luminosity distance of each target. 

\subsection{Kinematics}

We mapped the observed ionized-gas kinematics in each galaxy by extracting Gaussian parameters from the best 
fit to the [O~III] $\lambda\lambda$4959,5007 emission-line for each row of the 2D spectral images of our 
long-slit data, as seen in Figure \ref{fig:STISccds}. Resultant velocities (given in the rest frame of the galaxy 
and using a vacuum rest wavelength of 5008.24\AA.), FWHMs, and integrated fluxes (integrated under the Gaussian) for individual Gaussian 
components along each QSO2 long-slit spectrum in our sample are shown in the right side of Figures \ref{fig:images1} - \ref{fig:images4}, with 
left to right progression representing 0.05$"$ steps along the STIS long-slit and a 0.0$"$ offset located over the continuum 
centroid observed in imaging. Spectra containing multiple Gaussian components, as seen in Figure \ref{fig:linefit}, are sorted 
by measured FWHM, with blue circles, green diamonds, and red squares corresponding to first-, second-, and third-widest FWHM 
components respectively. Specifically, Gaussian parameters extracted from the emission-line fits in Figure \ref{fig:linefit} 
are seen reproduced at the 0.0$"$ offset in the kinematics plots of 2MASX J08025293+2552551 in Figure \ref{fig:images1}.

Kinematic analysis of the [O III] gas was performed based on our findings from \citet{Fis17}, where 
we found the influence of the central AGN to be stratified into two regions extending from the nucleus.
At small radii, outflows are visible as emission-lines deviating from the overall rotation pattern of the system 
or emission-lines that possess multiple components that travel at different velocities from one another in the NLR. 
At greater radii, gas is ionized by AGN radiation, but possess velocities and FWHMs consistent with gas in rotation 
with the host galaxy, as derived by stellar kinematics, in the extended NLR (ENLR). Applying these qualifications to 
the kinematics of our current sample, we can measure the extent of the outflowing gas within the observed [OIII] emission 
of each target. In doing so, we also find that several targets exhibit a third, composite set of AGN kinematics 
where the gas appears to be in rotation but is experiencing some influence by the AGN as FWHM measurements are 
found to be consistently $>$ 250 km s$^{-1}$, indicative of some kinematic disturbance \citep{Bel13,Ram17}. While 
the gas sampled in these regions is not outflowing, it is being disturbed by the AGN, therefore we reference these 
regions as 'disturbed' kinematics. We specifically classify the stratification of outflow, disturbed, and 
non-disturbed rotation kinematics below, from which we measure the extent of outflowing and AGN disturbed 
gas in each system. 

\medskip

1) AGN driven outflows: Emission-lines exhibit high centroid velocities ($>$ 300 km s$^{-1}$) from systemic and/or emission-lines 
with multiple components. Velocities do not follow velocity pattern of adjacent regions. The maximum radial extent of these 
kinematics is measured as R$_{out}$, designated by dashed lines in both the STIS measurements and the ACS or WFPC2 imaging 
of Figures \ref{fig:images1} - \ref{fig:images7}.

2) Disturbed rotation: Single component emission-lines with FWHMs $>$ 250 km s$^{-1}$ that exhibit low centroid velocities 
($<$ 300 km s$^{-1}$) that follow the orderly rotation pattern observed in similar radial distances. The maximum radial 
extent of these kinematics is measured as R$_{dist}$, mark with dotted-dashed lines in Figures \ref{fig:images1} - \ref{fig:images7}. 
Targets without R$_{dist}$ dotted-dashed lines have equal R$_{out}$ and R$_{dist}$ measurements.

3) Non-disturbed rotation: Emission lines that exhibit low centroid velocities ($<$ 300 km s$^{-1}$) with low FWHM 
($\lesssim$ 250 km s$^{-1}$). Consistent with rotation, where measurable non-systemic extents exhibit blueshifts on one 
side of nucleus, redshifts on other side.

\medskip

Maximum projected R$_{out}$ and R$_{dist}$ distances in each target are marked in the kinematics plots of 
Figures \ref{fig:images1} - \ref{fig:images4} and mirrored across the nucleus to illustrate the extent of each 
region and aid in associating the kinematics to what is observed in imaging, even if no spectroscopic data points exist 
out to those radii on a given side of the nucleus. Descriptions of the individual R$_{out}$ and R$_{dist}$ measurements for each 
target are provided in Appendix \ref{sec:targets}, with resultant distances listed in Table \ref{tab:hstspecobs}.

We deproject our maximum distance measurements of the outflowing and disturbed gas by dividing our maximum 
projected distances by a scaling factor ($s$) calculated from the orientation of the [O~III] morphology 
and the orientation and ellipticity of the host galaxy, as determined from our elliptical fits to the SDSS host images:

\begin{equation}
s = \frac{b}{\sqrt{(b\cos(\theta))^{2} + \sin(\theta)^{2}}}
\end{equation}

\noindent where $b$ is the fractional size of the minor axis relative to the major axis (i.e. $a = 1$), 
as determined from the ellipticity of the host ($e = 1 - \frac{b}{a}$) and $\theta$ is the difference in position 
angles of the host major axis and the [O~III] morphology. Scales in targets without a clear ionized-gas position 
angle, 2MASX J11001238+0846157, 2MASX J14054117+4026326, and 2MASX J17135038+5729546, are the maximum deprojection 
of each targets measurements assuming that the difference in position angle of the host major axis and the [O~III] 
morphology in each target is 90$^{\degree}$. After deprojecting our initial measurements, we find that the maximum 
outflow radii range from 150 to 1890 pc, with a mean value of $\sim$625 pc, and the observed maximum 
disturbed radii similarly range from 160 to 1890 pc, with a mean value of $\sim$ 1130 pc. We note that measurements 
of disturbed gas radii extend to the full length of our observations in several targets.

\section{Discussion}
\label{dis_sec}

\subsection{[O III] Extent - Luminosity Relation}

\begin{figure}
\centering

\includegraphics[width=0.45\textwidth]{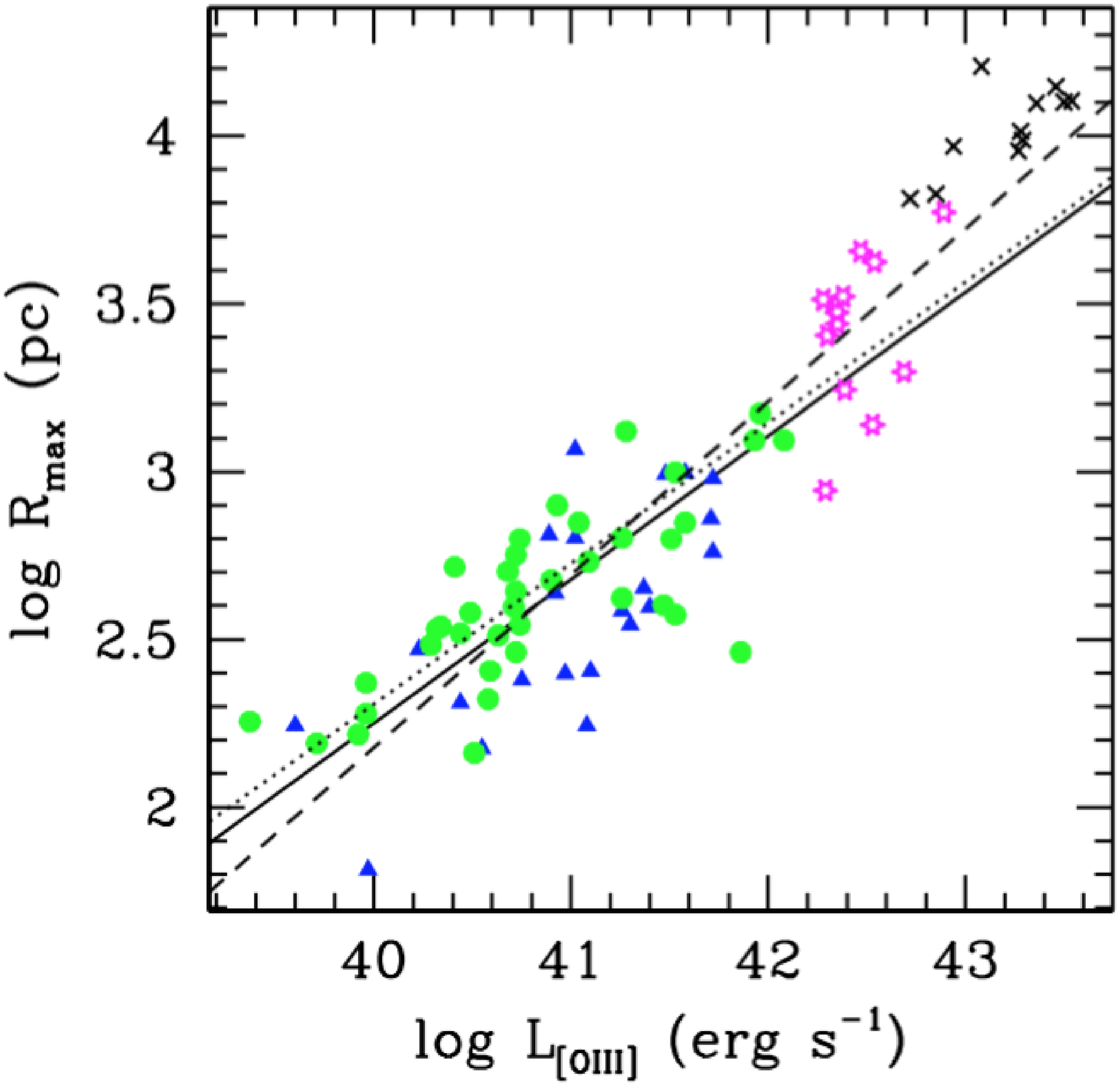}\\

\caption{Comparison between projected log R$_{max}$ and log L$_{[OIII]}$ for Seyfert 1s 
(blue triangles) and Seyfert 2s (green circles) from \citet{Sch03}, QSO2s (black Xs) from 
\citet{Liu13a}, and our QSO2s (magenta stars). The best fitting linear relations obtained 
using all data (slope = 0.52), only {\it HST} observations (0.43), or only Seyfert 2 and QSO2 {\it HST} observations (0.42), 
are shown as dashed, solid, and dotted lines, respectively.}

\label{fig:lvr}
\end{figure}

Comparing our projected QSO2 [O III] extents (R$_{max}$), as presented in Table 3, with similar 
measurements (i.e. measured to similar sensitivities of $\sim$ 10$^{-15}$ erg cm$^{-2}$ s$^{-1}$ arcsecond$^{-1}$) 
for non-merging Seyfert galaxies \citep{Sch03b} and higher redshift QSO2s \citep{Liu13a} we find a 
continuous trend between [O III] extent and [O III] luminosity, as shown in Figure \ref{fig:lvr}. 
The trend suggests higher luminosity targets produce [O III] out to greater distances. This figure 
contains three linear fits. The first fit (dashed line), obtained using all observed galaxies has 
the expression

$$ log R_{max} = -18.41 + 0.52 log L_{[OIII]}$$

\noindent 
with a Spearman rank of $\rho=0.92$ and a slope uncertainty of 0.024. The second fit (solid line) 
was obtained only using galaxies observed with {\it HST}, our targets and those from \citet{Sch03}, 
and has the expression

$$ log R_{max} = - 14.85 + 0.43 log L_{[OIII]}$$

\noindent
with a Spearman rank $\rho=0.82$ and a slope uncertainty of 0.031. The third fit (dotted line)
was obtained using only {\it HST} observed Seyfert 2s and QSO2s. The resulting expression is

$$ log R_{max} = -14.45 + 0.42 log L_{[OIII]}$$

\noindent
with a Spearman rank $\rho=0.88$ and a slope uncertainty of 0.031.

We find that fits from {\it HST} observed targets have similar slopes $\sim$0.42, with the slight
offset from including Seyfert 1s likely due to projection effects. The measured slopes are steeper than the 
value of 0.33 found by \cite{Sch03b}, but shallower than the value of 0.52 observed by \cite{Ben02}. A slope of 0.5, 
similar to what is found fitting all targets, is consistent with a $1/r^{2}$ dilution of the ionizing radiation, 
assuming similar gas densities and covering factors in the ENLRs.  A smaller value for the slope may simply be due to the absence of 
gas at large radial distances. 
 
\subsection{Extent of non-rotating kinematics}
In measuring the extent of non-rotating kinematics in our QSO2 sample, we find that the [O III] kinematics observed 
in Mrk 34 remain exceptionally large, such that the extent of the outflowing gas in this object, R$_{out}$ = 1.89 kpc, is 
$\sim$75\% larger than the next largest outflow radius in our sample, FIRST J120041.4 at R$_{out}$ = 1.07 kpc, and 
on average, the observed [O III] gas contains radial outflows extending to a distance of $\sim$0.6 kpc. 
As Type 2 AGN have been shown to have systemically lower Eddington ratios than Type 1 AGN \citep{Ric17}, 
one explanation for the lack of strong, kpc-scale winds in a majority of our sample may be due to low accretion rates.
However, our sample contains the most luminous Type 2 AGN from \citealt{Rey08} within a redshift of
z $<$ 0.12, with [OIII] luminosities 3 to 10 times larger than that of Mrk 573. Therefore, it is likely that a 
majority of our sample is radiating at or near the Eddington limit.

We were not able to determine the bulge radii for the galaxies in our sample to compare to the 
measured outflow radii as our {\it HST} continuum images were too shallow to detect the extent of the bulge and the 
SDSS images are of such scales that the bulges could not be resolved. However, considering that in the case of 
Mrk 573 the bulge has an effective radius of $\sim$1 kpc and that most of its mass is enclosed in a $\sim$2 kpc radius, even if 
one makes the simplifying assumption that the QSO2s were located in a similar host galaxy, their winds 
would most likely not be capable of clearing material from the bulge. Similar to our previous findings in 
Mrk 573, the [O III] outflow radii in our sample are typically a small fraction of the radii encompassing 
the entire [O III] structures in their host disks. We note that when comparing the radial extent of [O~III] emission 
between imaging and spectroscopy for these QSO2s, spectroscopic measurements only sample the brightest knots 
of emission observed in imaging for a majority of our targets. However, it is likely that we are observing the 
majority of radially outflowing kinematics in all of our targets, as outflows are often observed as 
the brightest knots of AGN-ionized gas and we observe the kinematics returning from very high velocities or 
FWHMs by the furthest radial extents of our observations.

As we do not have information on the kinematics of the stellar disks or surrounding molecular hydrogen for 
these targets to confirm the influence of rotation in our sample, distance measurements reported here are made using the 
assumption that the observed, extended [O~III] emission is created via an intersection between ionizing radiation 
from the central engine and preexisting, rotating gas in the host disk. Qualitatively, evidence that the 
extended, low velocity gas in these systems is in rotation exists in the morphologies of the [O~III] regions, as the arc 
structures observed in several systems resemble spiral arms one would expect to reside in their hosts. Quantitatively, 
if we compare the difference in position angle between the host major axis and the STIS long-slit observation (PA$_{diff}$; 
Table \ref{tab:sdssobs}), we find that targets with apparent high amplitude, pronounced rotation curves (i.e. SDSS 
J115245.66+101623.8 and 2MASX J17135038+5729546) have a smaller PA$_{diff}$, while targets with higher PA$_{diff}$ 
(i.e. 2MASX J08025293+2552551 and 2MASX J11001238+0846157) exhibit extended velocities near systemic, suggesting we are 
sampling rotation in the host disk along the major and minor axes, respectively.


Our findings also agree with similar, recent kinematic studies in nearby, non-merging QSOs 
\citep{Liu13b,Har14,McE15,Kar16,Vil16,Kee17,Ram17}, where maximum outflow distances are measured to be between 
1 - 2 kpc, and disrupted gas kinematics are observed out to several kpcs in a majority of targets. These findings 
suggest that AGN may still be disrupting gas that would form stars in their host galaxies on bulge-size scales 
without requiring complete evacuation via outflows. 

\subsection{[OIII] extent vs [OIII] FWHM}

\begin{figure}
\centering

\includegraphics[width=0.47\textwidth]{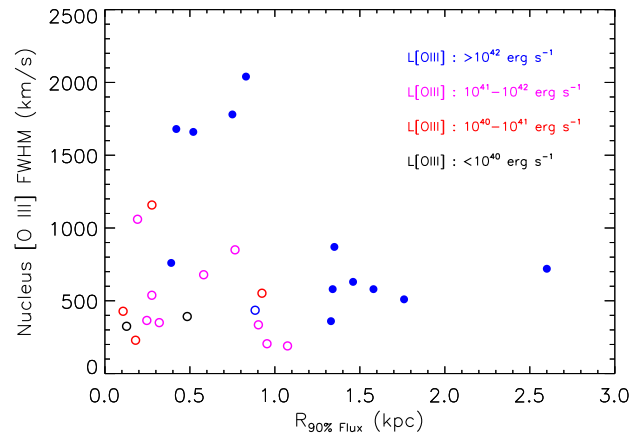}\\

\caption{A comparison between the [O III] flux distribution and the maximum nuclear FWHM of each 
QSO2. R$_{90\%}$ represents the radius that encompasses 90\% of the observed [O III] flux in each target 
inside R$_{max}$. Solid circles represent measurements from this work. Open circles represent measurements 
for nearby Type 2 Seyferts from \citet{Nel95} and \citet{Sch03}.}

\label{fig:fwhm_vs_r}
\end{figure}

The combination of the [O~III] images and spectra for the QSO2s in our sample suggest that they can generally be 
divided into two, distinct categories, as shown in Figure \ref{fig:fwhm_vs_r}. Specifically,
targets with compact [O III] morphologies 
tend to exhibit broad nuclear FWHMs ($>$1500 km s$^{-1}$), while targets with more extended [O~III] 
morphologies have narrower nuclear FWHMs ($<$ 900 km s$^{-1}$). 2MASX J14054117+4026326 overlaps 
with both catagories, as it has a compact [O III] morphology, but also possesses a comparitively small 
nuclear FWHM of 720 km s$^{-1}$. We note that in targets with compact morphologies (2MASX J07594101+505245, 
2MASX J11001238+0846157, 2MASX J14054117+4026326, 2MASX J17135038+572954, and Mrk 477), the [O~III] extent 
is so compact that we observe the diffraction spikes typical of a point source in the emission line images. 
If this is due to an orientation effect, the compact QSO2s would be viewed roughly face-on, similar to Type 1 AGN. 
However, SDSS spectra of these sources show no evidence of hydrogen emission-lines being broader 
than [O~III] and the continuum images of these targets (Appendix Figures \ref{fig:images8} 
- \ref{fig:images13}) show only the diffuse emission due to the host galaxy and do not 
show a corresponding nuclear point source that would be expected if these targets were seen face-on. Additionally, other 
than the broad, central components, the [O III] velocity profiles of the compact sources are generally similar 
to the extended sources, e.g., 2MASX-J07594101 (Figure \ref{fig:images1}) compared to FIRST-J12004141 
(Figure \ref{fig:images2}). Therefore, as the compact sources are true Type 2 AGN, the simplest
explanation for the broad [O~III] FWHMs is the presence of more material closer to the AGN, with large 
velocity gradients along our line of sight being produced by high velocity gas that is outflowing 
more isotropically. The presence of high velocity [O~III] gas close to the AGN does not necessarily 
account for the compactness of these sources, as a smaller [O III] morphology in the compact sources could be due to a 
weaker AGN. However, assuming that [O~III] is an isotropic quantity \citep{Hec04}, the five compact 
sources have roughly the same bolometric luminosities as the remaining sources. 

\begin{figure*}
\centering

\includegraphics[width=0.98\textwidth]{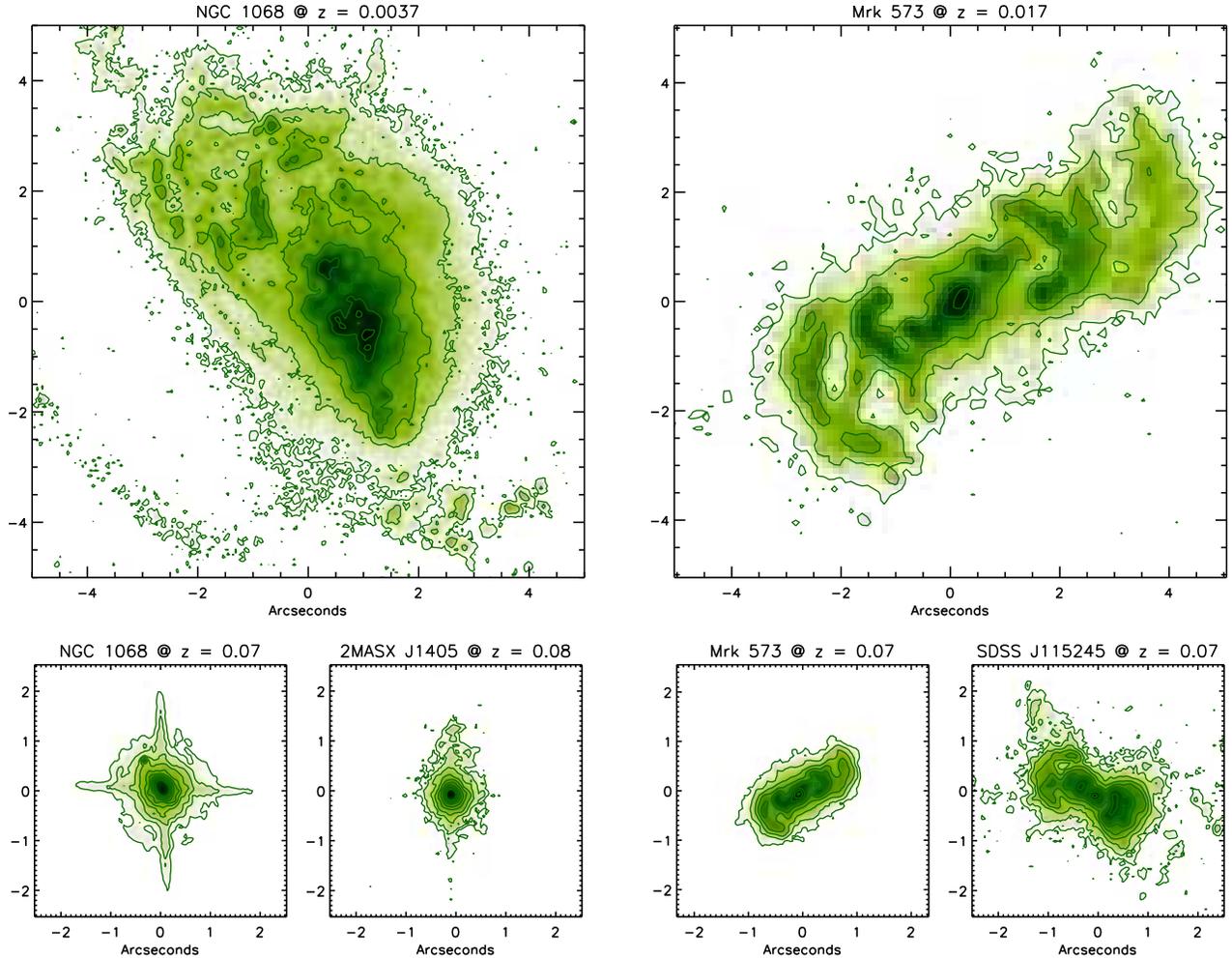}\\

\caption{Top: Archival {\it HST}/WFPC2 [O III] images of NGC 1068 and Mrk 573. Bottom: Comparisons between 
archival {\it HST} [O III] images of NGC 1068 and Mrk 573, resampled to simulate being located at z $\sim$ 
0.07 and convolved with the PSF of ACS using the F502N filter, and {\it HST}/ACS images of 2MASX J140541 and 
SDSS J115245 at comparable redshifts. High concentrations of [O III] flux in the nucleus of NGC 1068 at 
z = 0.07 produce diffraction spikes, similar to the compact sources in our QSO2 sample.}
\label{fig:ngc1068}
\end{figure*}

The morphological dichotomy that we observe in our QSO2 sample is analogous to what is seen in nearby,
high accretion-rate Seyfert galaxies. For example, Figure \ref{fig:ngc1068} shows {\it HST} imaging of [O III] morphologies 
in the Seyfert 2 galaxies NGC 1068 and Mrk 573 at their observed redshifts, which represent galaxies with
broad and narrow [O III] FWHMs, respectively. This figure also shows simulated imaging of what we would
expect to observe for the same targets at a redshift of z = 0.07, typical of our QSO2 sample. These 
images were created by resampling the [OIII] images of NGC 1068 and Mrk 573 to the expected resolution 
at z = 0.07, and convolving these images with the point spread function of a corresponding narrow band
filter, to simulate the passage of the image through the telescope optics. The results from this
experiment show that in the case of NGC 1068, its [O III] structure is compact enough that if it was
placed at a redshift typical of our QSO2s, the [O III] emission would be detected as a point source 
with fainter line emission surrounding it, similar to what we observe in our compact, high-FWHM targets. 
Alternatively, in the case of Mrk 573, the resampling and convolution of the image does not cause a 
significant change to the structure of the [O III] morphology, other than reduced resolution. This 
is similar to the case of SDSS J115245, a QSO2 with narrow nuclear [O III] FWHM measurements.

A possible explanation for the difference in morphology is that there is more gas close to the AGN in the compact targets, 
which results in the attenuation of a fraction of ionizing radiation. STIS observations of NGC 1068 support this idea, 
as they reveal bright, nuclear [O~III] emission with FWHM $>$ 1000 km s$^{-1}$ 
\citep{Cre00b,Das06}. Photoionization analysis of these observations, performed by \citet{Kra00b}, suggested that the observed 
ionizing radiation is attenuated by gas close to the AGN. Additionally, the scattered optical continuum is centrally-peaked
\citep{Cre00b} and, based on {\it Chandra}/ACIS imaging, the X-ray source is $\lesssim$ 165 pc in extent \citep{You01}.
Therefore, all of this together indicates there is a large concentration of ionized gas at small radii (100 - 200 pc). 

It is unlikely that the optical emission-line gas is attenuating much of the ionizing radiation, since, based on our 
studies of Seyfert galaxies \citep{Kra00b,Col09}, the [O~III] gas has a small covering factor. On the 
other hand, X-ray emission-line gas in the NLR can possess covering factors $\sim$ unity 
(e.g., \citealt{Kra15}). As the medium responsible for the scattering/polarization of the hidden 
broad-line emission lines and continuum in Type 2 AGN must possess large covering factors and column 
densities \citep{Ant93}, it is possible that the ionizing radiation has been attenuated by 
absorption/scattering in high ionization gas. Note, however, that there would be strong soft X-ray emission lines 
formed in the attenuating medium, assuming that elements such as O, N, C and Ne are not fully ionized.  

Given the evidence for strong outflows in NGC 1068 \citep{Das05} and that it is likely radiating near Eddington \citep{Kra15}, 
this gas could eventually be driven away from the inner nucleus. This would reduce the amount of emission-line gas at small 
radial distances, which would weaken or eliminate the broad [O~III]. This also could drive out the X-ray emission-line gas, 
which would reduce the attenuation of the ionizing radiation, resulting in a more extended [O III] structure. As such, it 
may be that the compact QSO2s are in a similar state, and that a transition occurs from compact AGN, with higher central 
concentrations of gas, to the more extended sources observed in the remainder of our sample, which resemble Mrk 573 with
extended morphologies and nuclear FWHMs $\sim$ 100s km s$^{-1}$ \citep{Fis10,Fis17}.

\section{Conclusions}
\label{conc_sec}

We have analyzed the [O~III] morphology and emission-line kinematics of a sample of 12 QSO2s via high-resolution 
imaging and spectroscopy with {\it HST}. Our major findings are as follows:

1) Our targets consist of several of the most luminous Type 2 AGN within a redshift of z = 0.12, yet 
[O~III] morphologies in our sample vary from compact, core-like structures to extended structures,
all several kpc in length. Overall, the size of [O~III] regions scale with luminosity in comparison to 
non-merging Seyfert galaxies and QSO2s observed to similar sensitivities.

2) Radially outflowing kinematics exist in all of our targets, with maximum outflow distances ranging 
between 150 and 1890 pc and a mean R$_{out}$ of $\sim$600 pc. The extent of these outflows are relatively 
small compared to the overall extent of the [O III] morphology, with an average R$_{out}$/R$_{max}$ = 0.22. 
As such, our findings suggest that a majority of observed extended [O III] emission 
is often in rotation and not driven radially by AGN winds, questioning the effectiveness of AGN being 
capable of clearing material from their host bulge in the nearby universe. 

3) A majority of our sample also show signatures of gas at systemic velocity or following rotation that 
is disturbed by the central AGN (FWHM $>$ 250 km s$^{-1}$) at distances outside the maximum measured radial 
outflow radii, with maximum radii ranging from 160 to 1860 pc, with a mean R$_{dist}$ of $\sim$ 1130 pc. 
These findings suggest that AGN activity may be disrupting gas that forms stars without 
requiring complete evacuation.

4) We find our targets fall into two classes when comparing [O~III] radial extent and nuclear FWHM, such that 
QSO2s with a more compact [O~III] morphology have broader nuclear emission-lines. We hypothesize that this 
could be due to a transitional effect, where QSO2s with compact morphologies possess comparatively 
large amounts of gas near the AGN, which may be driven out at later states.

\acknowledgments

This research has made use of the NASA/IPAC Extragalactic Database (NED), which is operated 
by the Jet Propulsion Laboratory, California Institute of Technology, under contract with the National Aeronautics 
and Space Administration. 

Funding for the Sloan Digital Sky Survey IV has been provided by the Alfred P. Sloan Foundation, 
the U.S. Department of Energy Office of Science, and the Participating Institutions. SDSS-IV acknowledges
support and resources from the Center for High-Performance Computing at
the University of Utah. The SDSS web site is www.sdss.org.

SDSS-IV is managed by the Astrophysical Research Consortium for the 
Participating Institutions of the SDSS Collaboration including the 
Brazilian Participation Group, the Carnegie Institution for Science, 
Carnegie Mellon University, the Chilean Participation Group, the French Participation Group, Harvard-Smithsonian Center for Astrophysics, 
Instituto de Astrof\'isica de Canarias, The Johns Hopkins University, 
Kavli Institute for the Physics and Mathematics of the Universe (IPMU) / 
University of Tokyo, Lawrence Berkeley National Laboratory, 
Leibniz Institut f\"ur Astrophysik Potsdam (AIP),  
Max-Planck-Institut f\"ur Astronomie (MPIA Heidelberg), 
Max-Planck-Institut f\"ur Astrophysik (MPA Garching), 
Max-Planck-Institut f\"ur Extraterrestrische Physik (MPE), 
National Astronomical Observatories of China, New Mexico State University, 
New York University, University of Notre Dame, 
Observat\'ario Nacional / MCTI, The Ohio State University, 
Pennsylvania State University, Shanghai Astronomical Observatory, 
United Kingdom Participation Group,
Universidad Nacional Aut\'onoma de M\'exico, University of Arizona, 
University of Colorado Boulder, University of Oxford, University of Portsmouth, 
University of Utah, University of Virginia, University of Washington, University of Wisconsin, 
Vanderbilt University, and Yale University.

The authors thank the anonymous referee for their helpful comments that improved the clarity
of this work. T.C.F. was supported by an appointment to the NASA Postdoctoral Program at the NASA Goddard Space Flight Center, 
administered by Universities Space Research Association under contract with NASA. Basic research at the Naval 
Research Laboratory is funded by 6.1 base funding. M.R. gratefully acknowledges support from the National Science 
Foundation through the Graduate Research Fellowship Program (DGE-1550139). L.C.H. was supported by the National 
Key R\&D Program of China (2016YFA0400702) and the National Science Foundation of China (11473002, 11721303). 
M.V. gratefully acknowledges financial support from the Danish Council for Independent Research via grant no. 
DFF 4002-00275. 


\bibliographystyle{apj}             
\bibliography{apj-jour,hst_qso2}       

\appendix
\section{Individual Target Descriptions}
\label{sec:targets}

\begin{figure*}{t}
\centering

\includegraphics[width=0.95\textwidth]{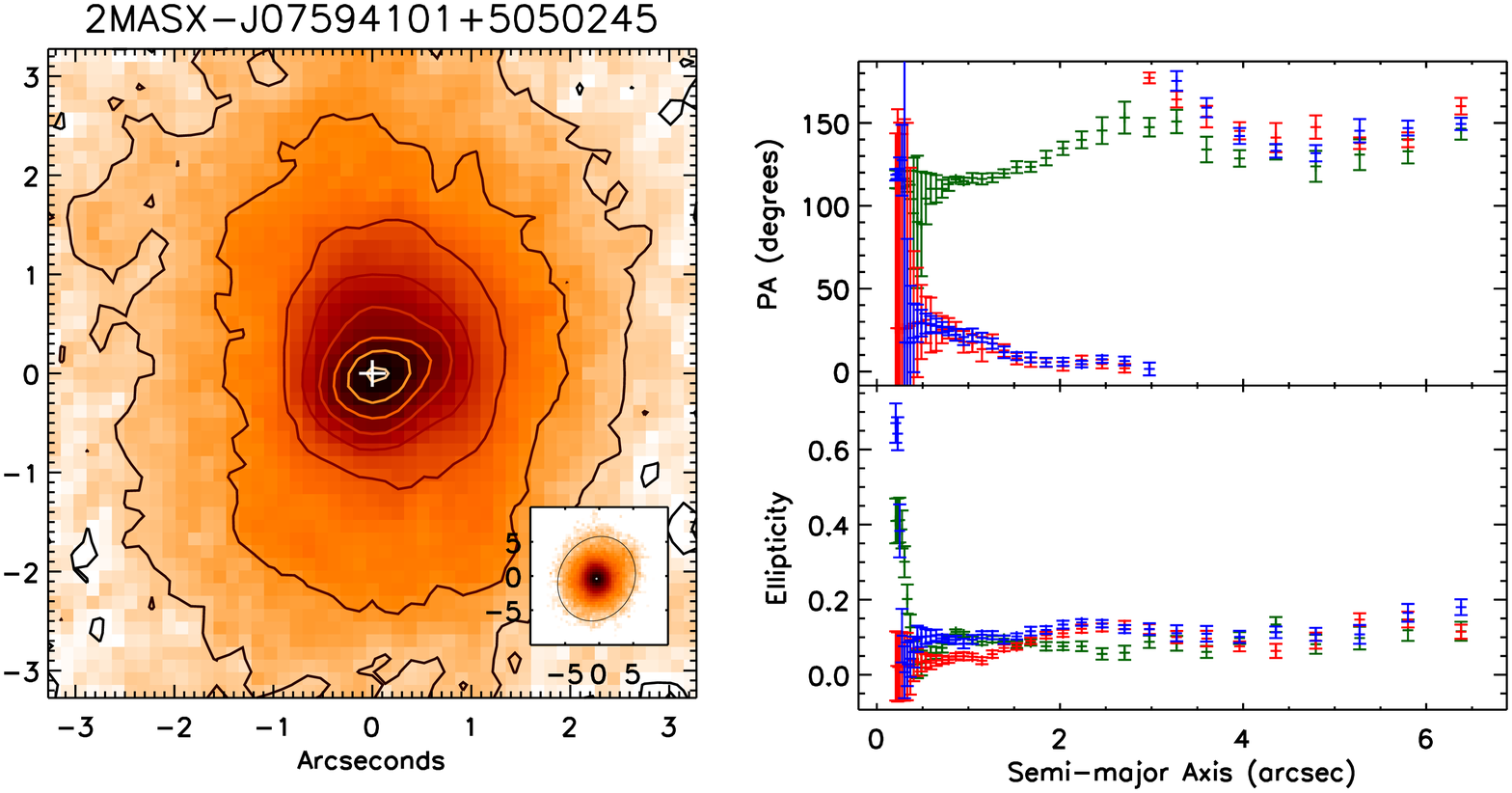}\\
\includegraphics[width=0.95\textwidth]{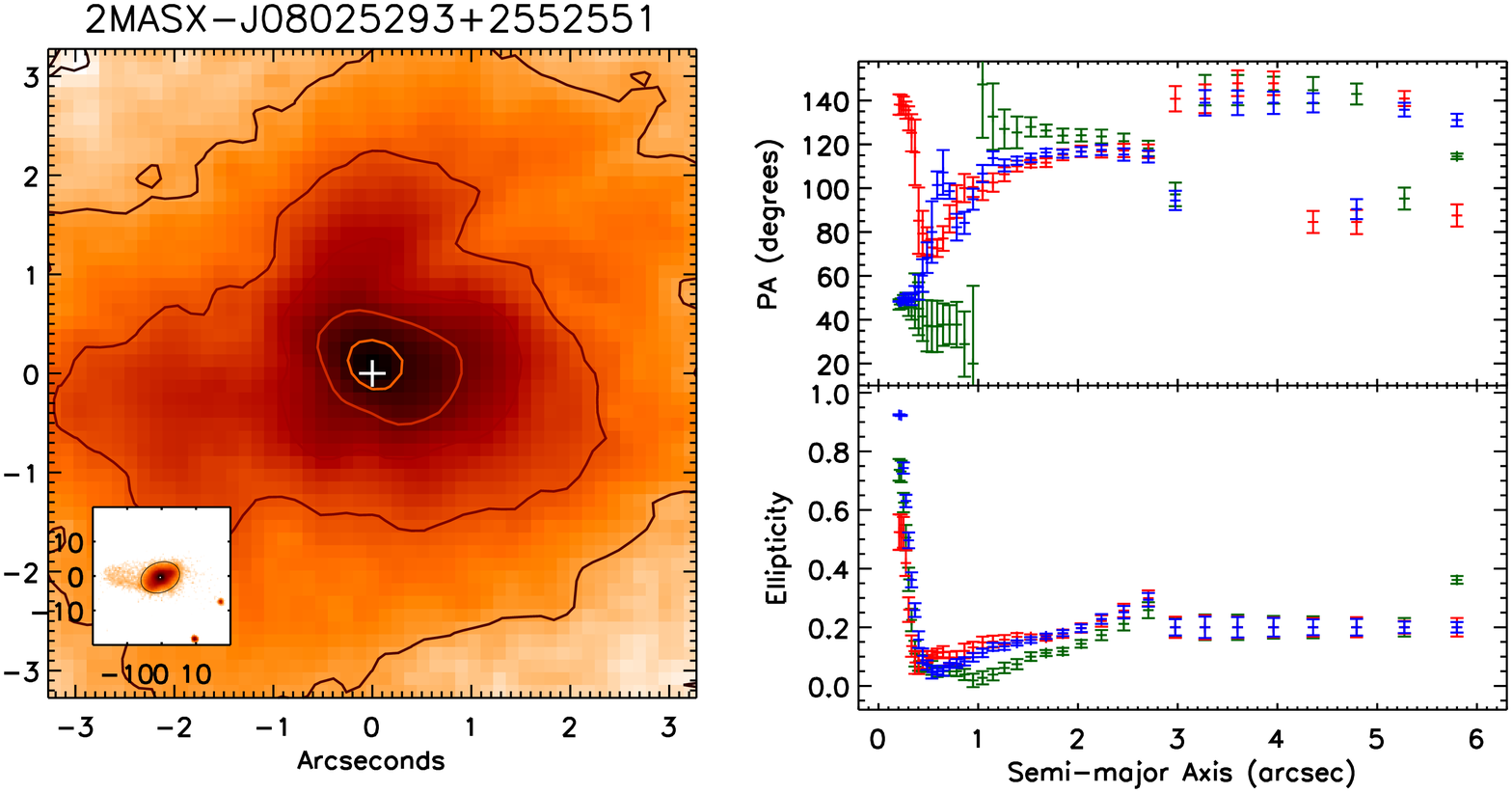}\\
\caption{{\it HST} and SDSS datasets for 2MASX J07594101+505245 and 2MASX J08025293+2552551. 
Left: The large window presents $5''\times5''$ Continuum images of 
each target. Flux contours start at 3$\sigma$ over background and increase in powers of 2 times 3$\sigma$ 
(i.e. 3$\sigma \times 2^{n}$). The position of the nucleus, measured at the flux peak of each target, 
is plotted as a white cross. The inset image displays a zoomed-out continuum $r$-band SDSS image of the 
target host galaxy with the outer-most elliptical fit overplotted. Right: Position angle and ellipticity 
outputs of elliptical fits to SDSS $g-$(green), $r-$(red), and $i-$band (blue) imaging out to a 3$\sigma$ 
detection for 2MASX J07594101+505245 and 5$\sigma$ for 2MASX J08025293+2552551.}
\label{fig:images8}
\end{figure*}

\subsection{2MASX J075941}
The [O III] structure in this target is highly asymmetric, with faint, extended 
knots of emission to the east of the nucleus. Kinematics are obtained near the 
nucleus (total extent $<$1''), with a two-component emission line (FWHMs of 
$\sim$1500 and 100 km s$^{-1}$) observed near systemic throughout, and an 
intermediate width component (FWHM $\sim$ 500 km s$^{-1}$) traveling in different 
directions on either side of the nucleus inside 500 pc. We measure the 
maximum extent of these kinematics as the observed outflow radius R$_{out}$ = 620 
pc. At greater distances, we measure single component emission-lines at systemic 
with FWHMs $<$ 250 km s$^{-1}$, which we define as undisturbed rotation. As such, 
we do not measure a distinct disturbed kinematics maximum radius, R$_{dist}$ in 
this target. This QSO resides in a relatively face-on system, that does not have 
any companions or tidal tails, suggesting no recent galaxy interaction activity.

\subsection{2MASX J080252}
This QSO has an asymmetric, conical [O III] morphology pointing north of the 
nucleus. Kinematics are obtained from the nucleus to distances $>$ 1 kpc north, 
with emission-line fluxes rapidly becoming fainter south of the nucleus, 
consistent with the opposite cone being extinguished by the host disk. Rapid 
differences in FWHM and velocity between nucleus and immediately off-nucleus 
measurements are due to blending of the blueshifted, offset component, as seen in 
Figure \ref{fig:linefit}, at off-nuclear positions. R$_{out}$ is measured as the 
extent of the multicomponent emission lines, which extend along the STIS slit to 
the northeast approximately 500 pc. Single-component, broad FWHM ($>$250 km s$^{-1}$) 
emission-lines near systemic are observed to extend to an additional 200 pc, for 
an R$_{dist}$ of $\sim$700 pc. Emission-lines at greater radii to the northeast 
have a single, narrow ($<$ 250 km s$^{-1}$) component that suggests that the 
gas is kinematically undisrupted. This target resides in a host that has recently 
experienced some galaxy interaction, with a long tidal tail extending 15'' east 
of the system. Elliptical fits to this system result in a highly inclined disk, 
which may be affected by the resultant morphology.

\begin{figure*}
\centering
\includegraphics[width=0.95\textwidth]{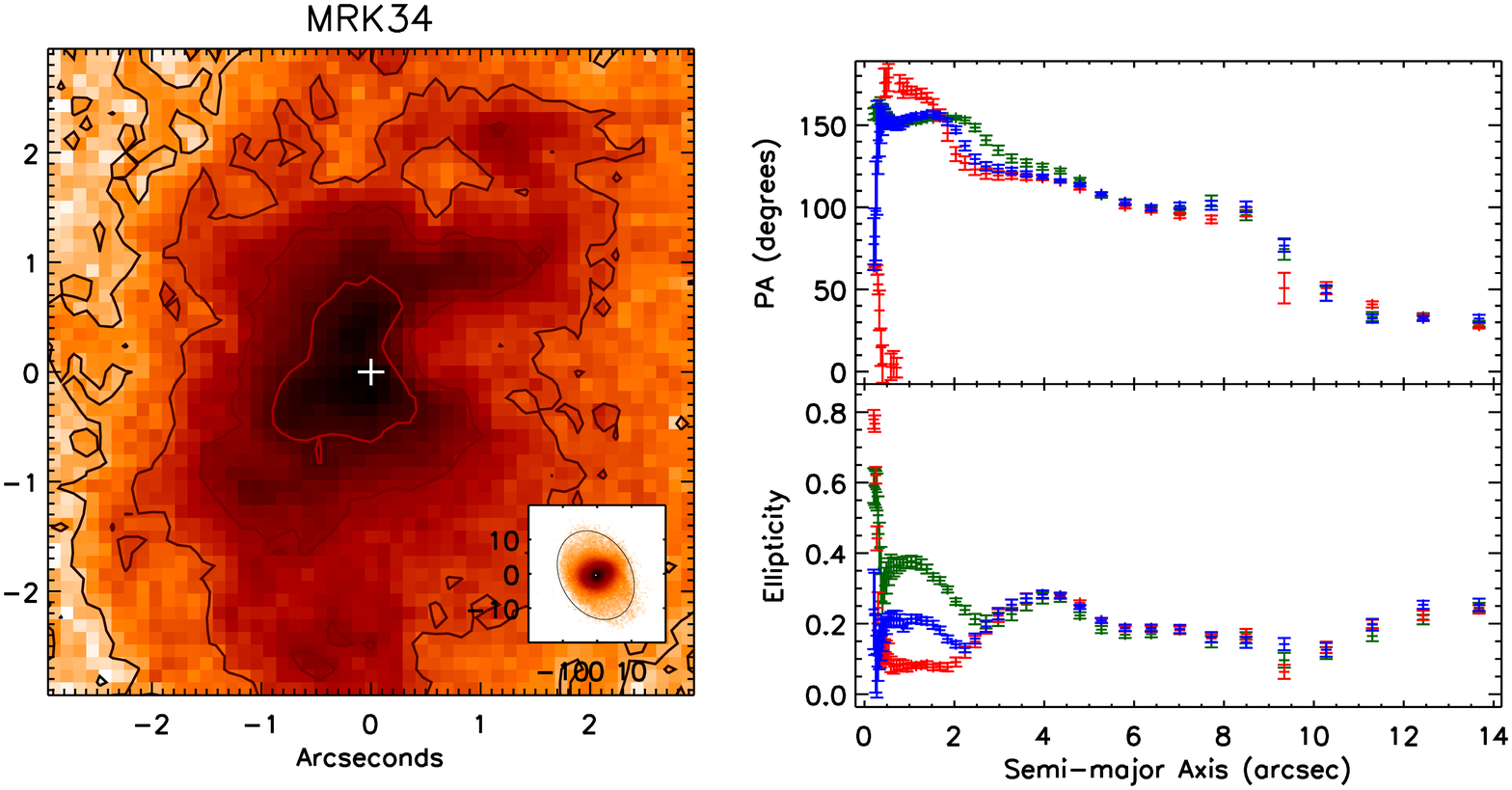}\\
\includegraphics[width=0.95\textwidth]{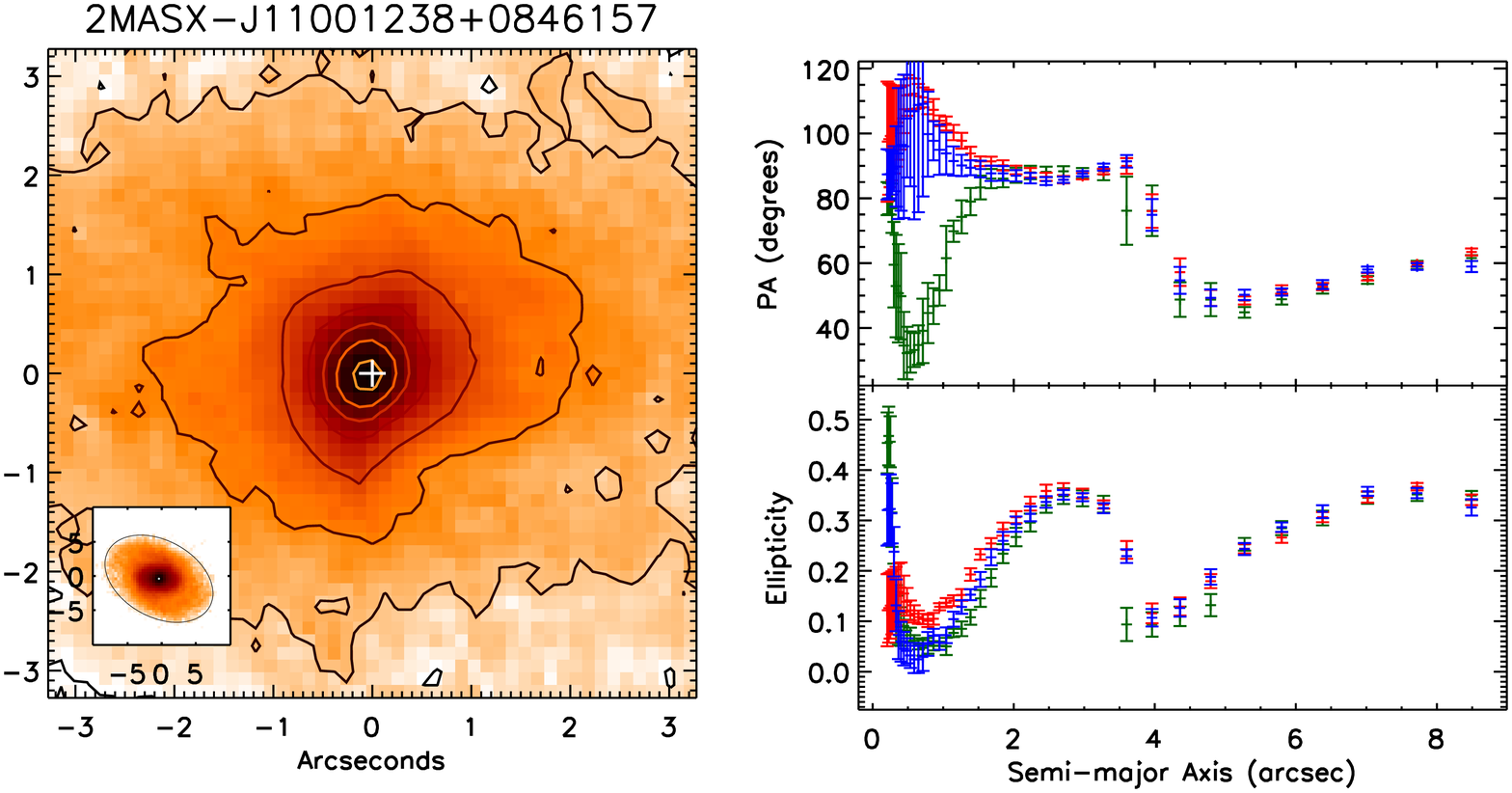}\\

\caption{Same as Figure \ref{fig:images8} for Mrk 34 and SDSS J115245.66+101623.8.
Host disk ellipticals are bot fit to 3$\sigma$.}
\label{fig:images9}

\end{figure*}

\subsection{Mrk 34}
The [O III] morphology in Mrk 34 is a clear, biconical structure, with a 'Z' shape 
likely due to the illumination of spiral arms in the host disk, similar to previously 
observed targets like Mrk 3 \citep{Cre10b} and Mrk 573 \citep{Fis10,Fis17}. [O III] 
kinematics across the three separate slit positions show high-velocity ($\sim$1000 km 
s$^{-1}$) outflows existing inside projected distances of 500 pc. Extended kinematics 
for the rest of the system appear to be largely rotational, however large deviations 
from rotation and emission-line splitting are present on either end of the nucleus out 
to $\sim$ 1.5 kpc. Here, we observe line splitting with both lines deviating from the 
rotation curve in opposite directions, to an approximate maximum difference of 500 km 
s$^{-1}$. Regions that display line splitting are observed as the outer arcs being 
illuminated in the NLR, suggesting that these kinematics are due to material in the 
spiral arms being driven or ablated in directions perpendicular to the radial outflows. 
We define R$_{out}$ to encompass the entirety of the observed kinematics with 
multiple components for a radial distance of 1.50 kpc. This distance also marks the 
extent of R$_{dist}$, as kinematics at greater radii follow a likely rotation pattern 
and have FWHM measurements of approximately 250 km s$^{-1}$. The host galaxy of Mrk 34 
is a well-resolved spiral galaxy that is moderately inclined with no evidence of 
merger activity.

Mrk 34 has been studied significantly more than a majority of our targets. VLA radio 
observations \citep{Fal98} show a radio jet structure well-aligned with the optical 
ionized-gas morphology, where radio hot spots coincide with regions of low excitation. 
{\it NuSTAR} 3-40 keV observations by \citet{Gan14} found that Mrk 34 is the nearest 
non-merging Compton-thick QSO2, with soft X-ray emission being driven by AGN 
photoionization versus star formation. This follows the findings that Mrk 34 lacks 
nuclear star formation \citep{Gon01,Sto09}, which may be suppressed due to AGN 
feedback \citep{Wan07}.

\subsection{2MASX J110012}
This target has a compact [O III] morphology, which is nearly circular with diffraction 
spikes and a slight extension to the southeast. High FWHM emission ($>$ 1000 
km s$^{-1}$) is measured across nucleus that is blueshifted to the southeast. We 
measure R$_{out}$ to the maximum extent of these high FWHM 
measurements at 450 pc northwest of the nucleus. Disturbed kinematics extend to the 
furthest observations northwest of the nucleus, R$_{dist}$ = 1 kpc, where we measure 
single-component emission-lines at systemic with widths of $\sim$ 500 km s$^{-1}$, 
which may be a product of blending multiple components as observed over the nucleus. 
This QSO2 resides in a well-resolved barred spiral galaxy that is moderately inclined 
with no evidence of recent interactions or ongoing mergers.

\begin{figure*}
\centering
\includegraphics[width=0.95\textwidth]{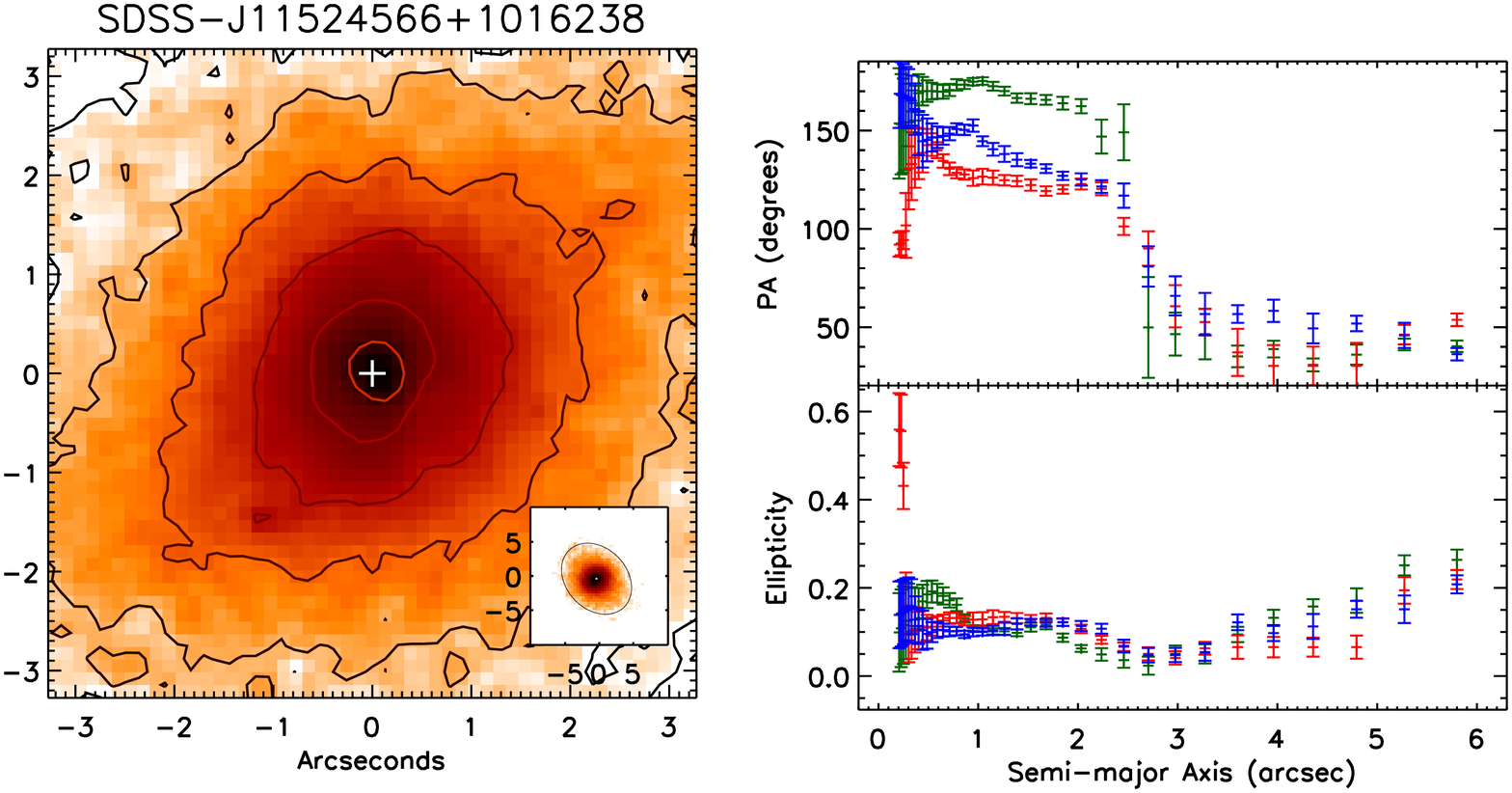}\\
\includegraphics[width=0.95\textwidth]{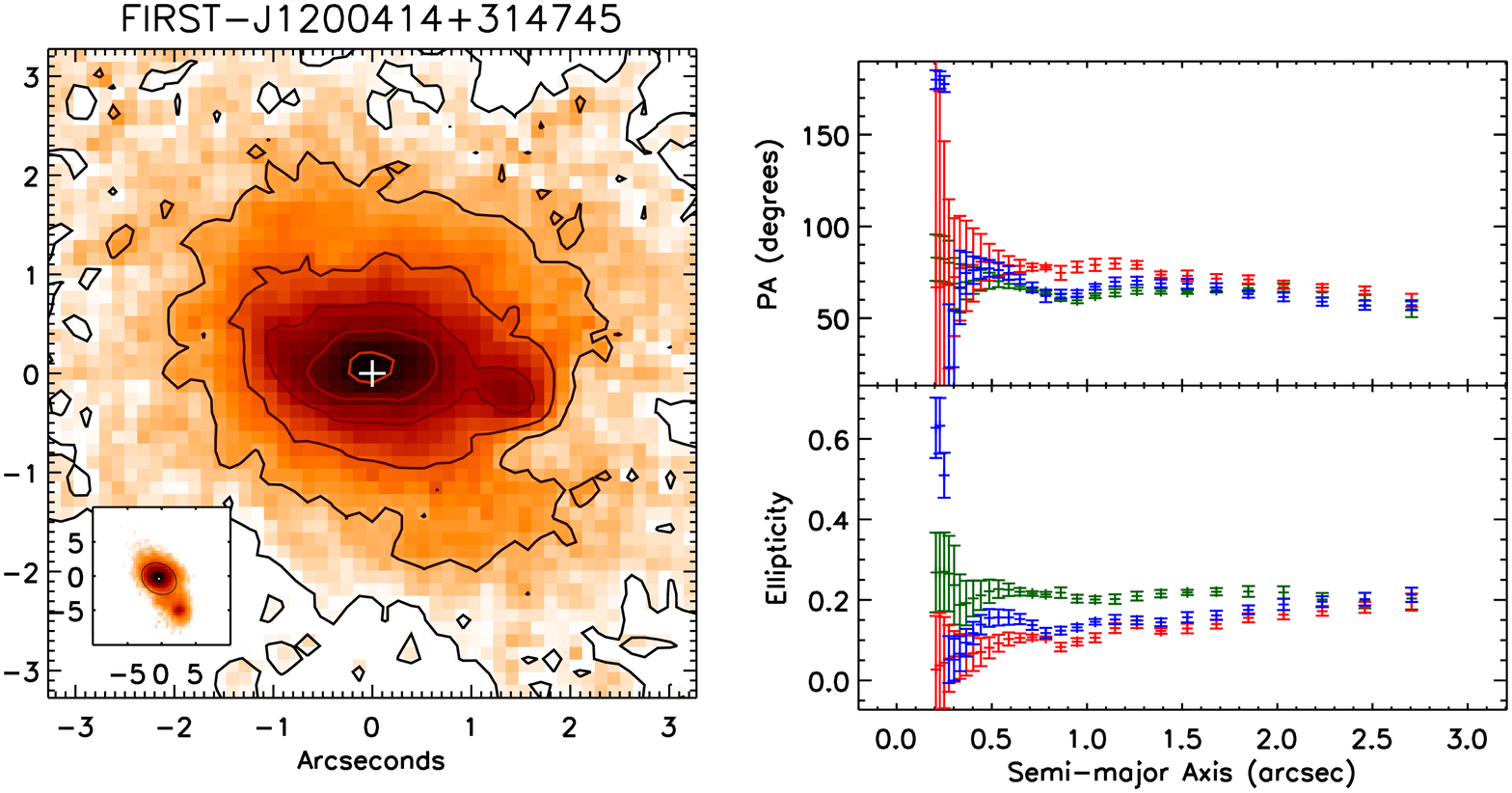}\\

\caption{Same as Figure \ref{fig:images8} for SDSS J115245.66+101623.8 and FIRST J120041.4+314745, and . 
Host disk elliptical fits are to 3$\sigma$ and 9$\sigma$, respectively.}
\label{fig:images10}

\end{figure*}

\subsection{SDSS J115245}
This target has a clear biconical [O III] morphology, with arcs of emission that suggest 
that a majority of the gas exists in illuminated spiral arms. The observed kinematics 
appear to be largely rotational, with redshifted rotation to the north and blueshifted 
rotation to the south. High peak rotation velocities ($\sim$200 km s$^{-1}$) are consistent 
with the alignment between the modeled host major axis and STIS long-slit position angles 
(PA$_{diff}$ = 33.3$^{\degree}$). Narrow emission line components are measured going in the 
opposite direction of rotation immediately northeast of the nucleus, which represent the 
extent of outflows with R$_{out}$ = 130 pc. $\sim$ 400 km s$^{-1}$ FWHM gas is observed 
in rotation at radii outside the first northern arc of emission, for an R$_{dist}$ of 
$\sim$1.1 kpc. The host galaxy for this target is moderately inclined with no signs of recent 
interaction activity.

\subsection{FIRST J120041}
FIRST J120041 has the largest projected [O III] morphology extent in our sample, with an 
R$_{max}$ of $\sim$6 kpc, which is on the same scale as the extent of the observable host 
galaxy in SDSS imaging. [O III] velocities are measured on much smaller scales, 
extending to distances between 0.8 - 1.5 kpc from the nucleus, which can be mapped to 
the two bright lobes on either side of the nucleus along the STIS slit. Multi-component 
and high-velocity emission lines are observed in the inner 0.5'', for an R$_{out}$ of $\sim$1 kpc, with high 
FWHM, single-component emission lines detected to the furthest extent of our measurements 
(R$_{dist}$ = 1.12 kpc). Outflows and disturbed kinematics may reach to further 
distances outside the area sampled by the STIS slit, as the bright knots where we 
observe the high-velocity gas extend to a radius of $\sim$0.76$''$, or $\sim$1.55 kpc. 
This target appears to be part of an ongoing or recent interaction with a smaller, 
satellite galaxy to the southwest, as filamentary structure can be observed linking 
the two systems.

\begin{figure*}
\centering
\includegraphics[width=0.95\textwidth]{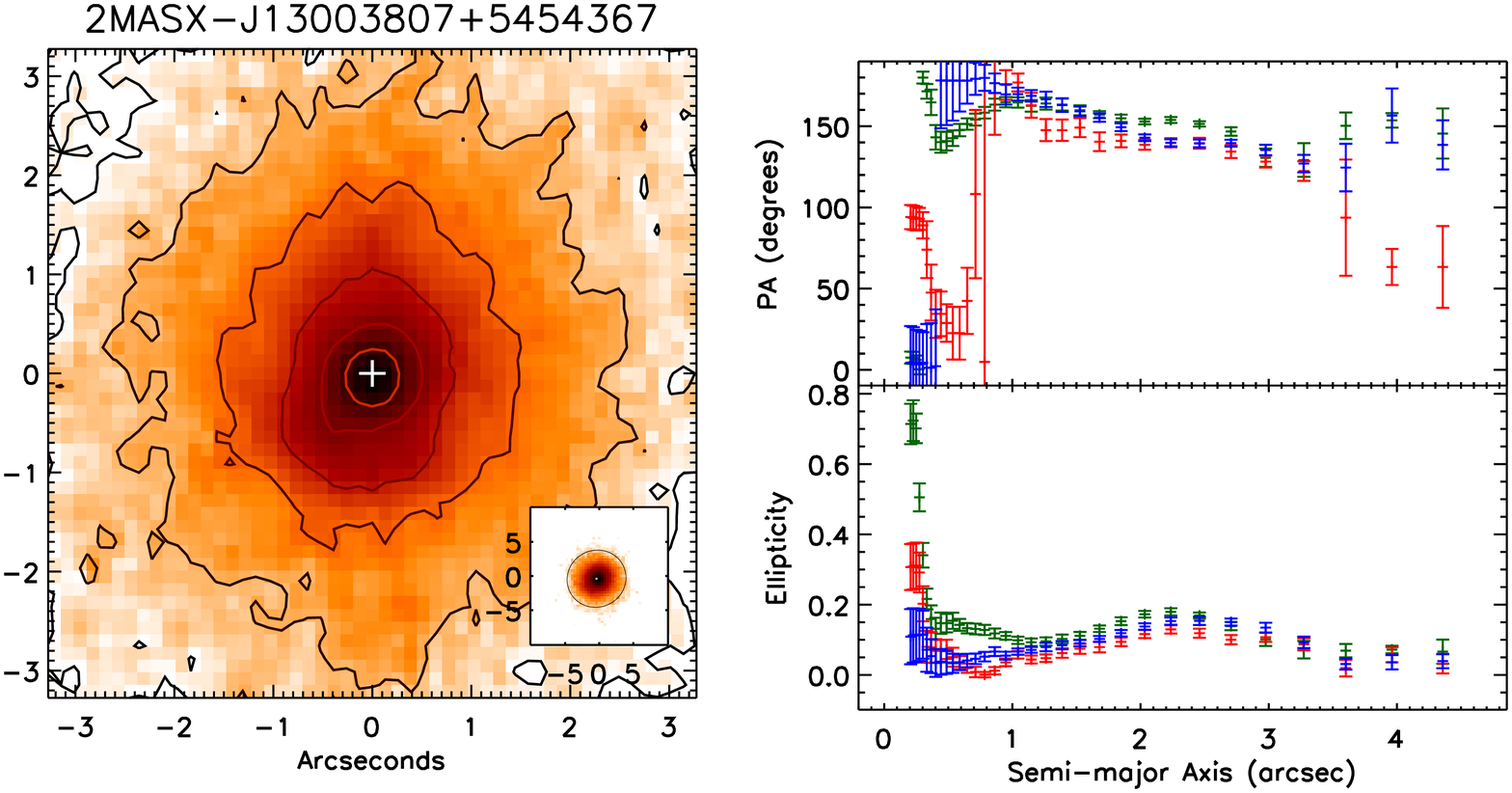}\\
\includegraphics[width=0.95\textwidth]{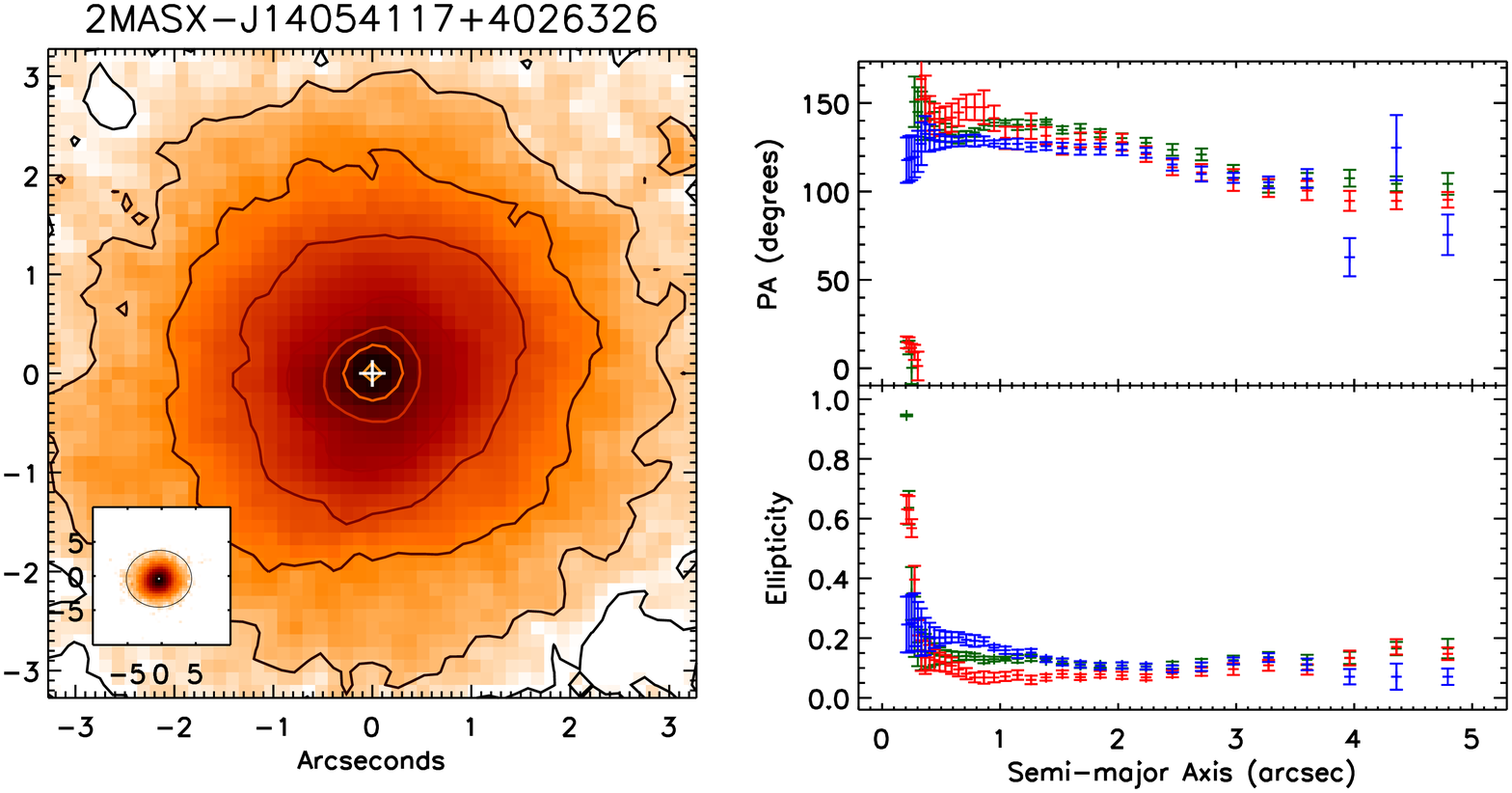}\\

\caption{Same as Figure \ref{fig:images8} for 2MASX J13003807+5454367 and 2MASX J14054117+4026326. Host disk elliptical 
fits are to both to 3$\sigma$.}
\label{fig:images11}

\end{figure*}

\subsection{2MASX J130038}
The [O III] structure of this target is extended, with two arcs of bright emission 
and a linear feature near the nucleus, suggesting that the observed morphology 
is due to spiral arms being illuminated by the central engine. The kinematics 
sampled by our STIS observations agree with this interpretation, as they appear 
to be largely in kinematically un-disturbed rotation. Velocities are blueshifted 
to the southeast and redshifted to the northwest, with FWHMs ranging between 
100 - 200 km s$^{-1}$. A compact, broad FWHM component is observed immediately 
northwest of the nucleus, with R$_{out}$ $\sim$160 pc, suggesting outflows are 
being driven off the linear filament near the nucleus observed in imaging, 
similar to Mrk 573. As single-component emission lines at greater radii possess 
FWHMs $<$ 250 km s$^{-1}$, R$_{out}$ and R$_{dist}$ are the same distance. The 
host of this QSO2 appears to be relatively face-on without any signs of recent 
interactions.

\subsection{2MASX J140541}
This source has a compact [O III] morphology. The measured R$_{max}$ is inside the observed 
[O III] morphology as it is artificially extended by diffraction spikes. Kinematics show the 
gas to be near systemic, with slightly blueshifted outflows over the nucleus possessing FWHMs 
of $\sim$800 km s$^{-1}$ out to an R$_{out} \sim$ 300 pc. FWHM measurements remain near 400 
km s$^{-1}$ for single-component emission lines at the furthest extent of our 
measurements, suggesting kinematically disturbed gas exist throughout (R$_{dist}$ 
$\sim$ 850 pc). The farther extension of gas kinematics to the northwest suggest that 
the host disk may be extinguishing emission to the southeast. The host is moderately 
inclined, without signs of recent interaction activity.

\begin{figure*}
\centering
\includegraphics[width=0.95\textwidth]{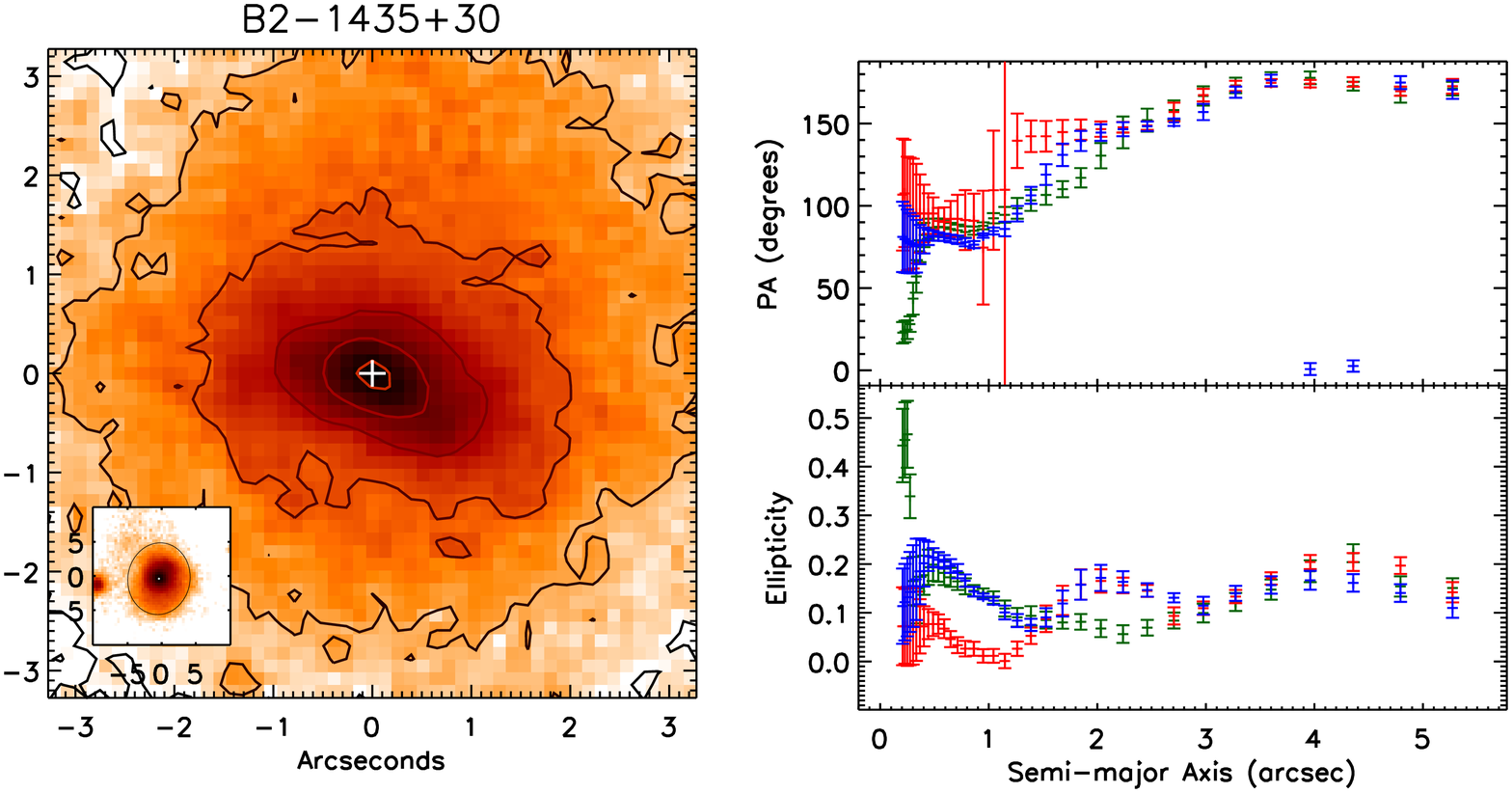}\\
\includegraphics[width=0.95\textwidth]{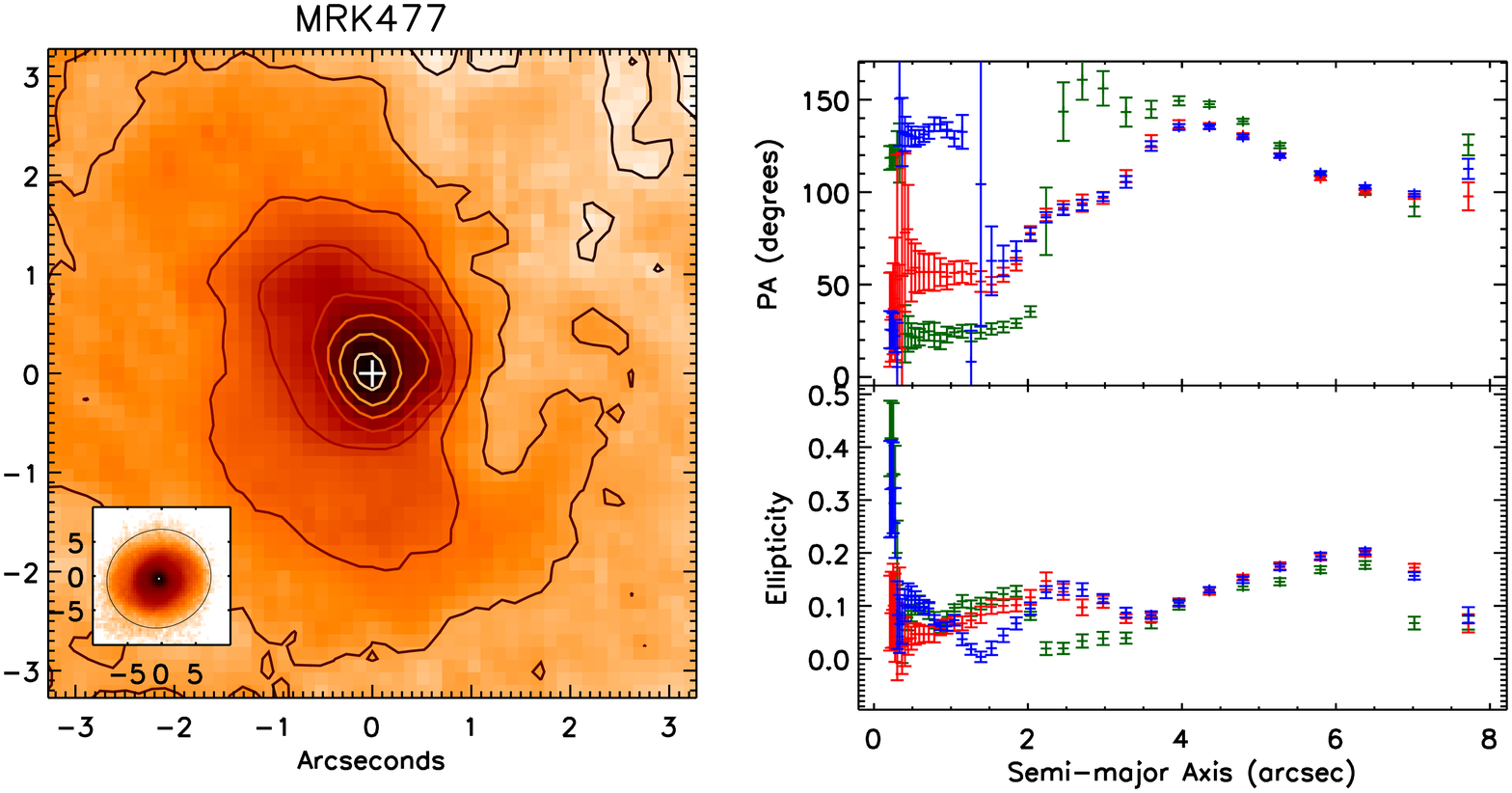}\\

\caption{Same as Figure \ref{fig:images8} for B2 1435+30 and Mrk 477. Host disk ellipticals are fit to 5$\sigma$ for both targets.}
\label{fig:images12}

\end{figure*}

\subsection{B2 1435+30}
The [O III] morphology of B2 1435+30 is biconical, with a compact arc to the east, and a 
bright, extended cone to the west, which are both traced kinematically by our STIS observations. 
This target does not possess significantly high velocities or multiple components in its 
kinematics, however, we find that the redshifted, high FWHM ($\sim$700 km s$^{-1}$) gas 
kinematics over the nucleus do not follow the overall rotation pattern of the surrounding 
gas at greater radii and deem these kinematics to be in outflow, with R$_{out}$ = 170 pc.
Kinematics across the features outside the nucleus resemble kinematically disturbed rotation, 
observed as 400 km s$^{-1}$ FWHM gas, extending throughout the southwest cone, with 
R$_{dist} >$ 1.5 kpc. This target appears to be part of a recent galaxy interaction, with 
extended structure in SDSS imaging linking it to an eastern satellite galaxy.

\subsection{Mrk 477}
The [O III] morphology of this target is largely comprised of two emission line knots inside 
500 pc, with fainter emission extending to radii of $\sim$2.5 kpc. [O III] kinematics suggest 
that the gas is largely in rotation near systemic, with additional emission over the nucleus 
extending to radii $\sim$500 pc that displays multiple emission-line components and excessively 
large FWHMs $\sim$ 2000 km/s. Relatively moderate FWHM ($>$ 500 km s$^{-1}$), single-component, 
near-systemic emission-line kinematics extend to further distances of R$_{dist}$ = 840 pc.
This target resides in a well-resolved spiral galaxy that shows signs of a recent interaction in tidal 
streams that bridge Mrk 477 to neighboring galaxies to the northeast \citep{DeR87}.

Similar to Mrk 34, Mrk 477 is also well-studied. UV through near-IR analysis by \citet{Hec97} show 
that a large portion of the energetics in this system is provided by an ongoing compact, nuclear 
starburst in this target. 8.4 GHz radio emission extends to a radius of $\sim$500 pc along a position 
angle $\sim 30\degree$, similar to our STIS long-slit position.

\begin{figure*}
\centering
\includegraphics[width=0.95\textwidth]{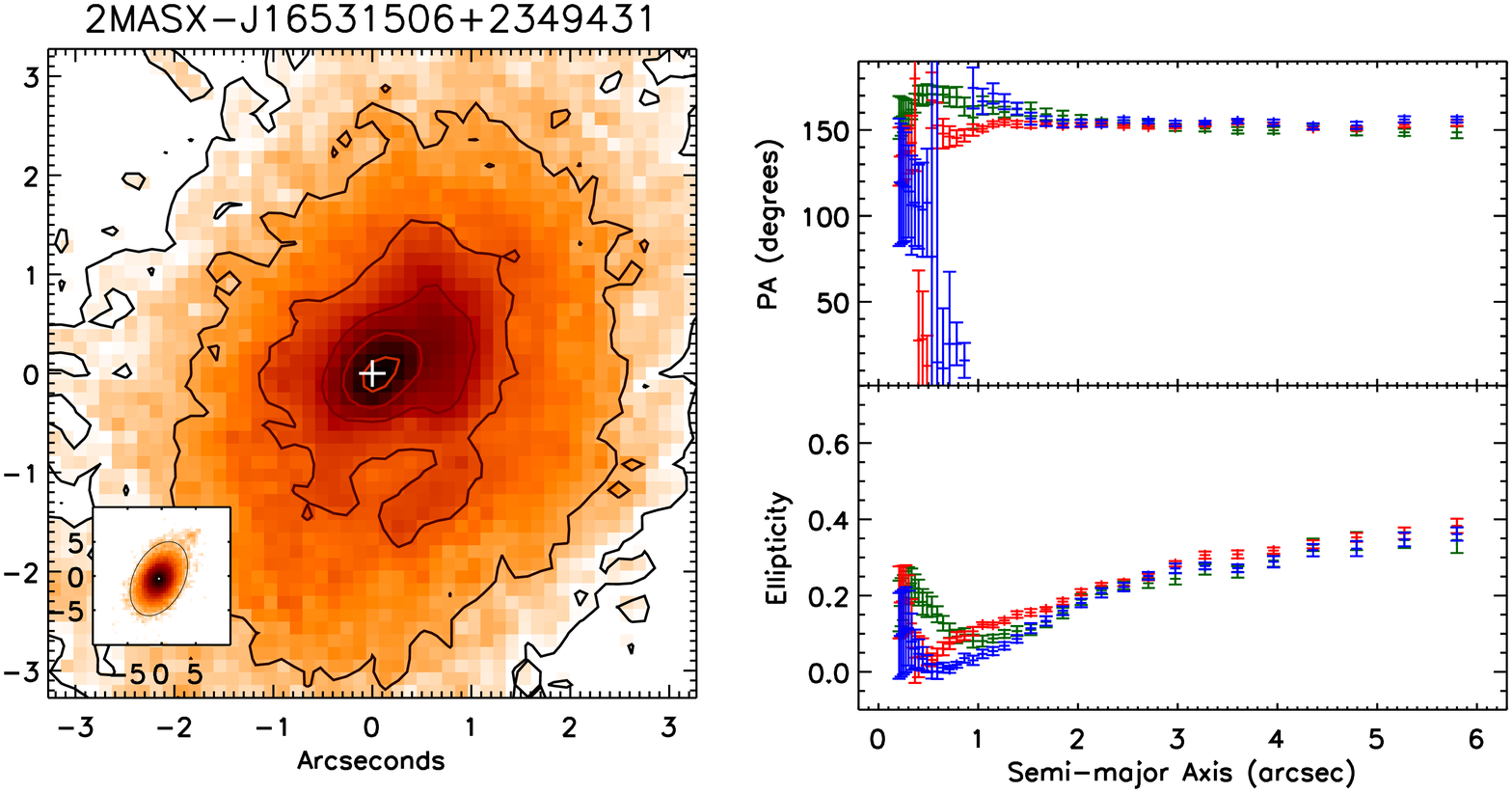}\\
\includegraphics[width=0.95\textwidth]{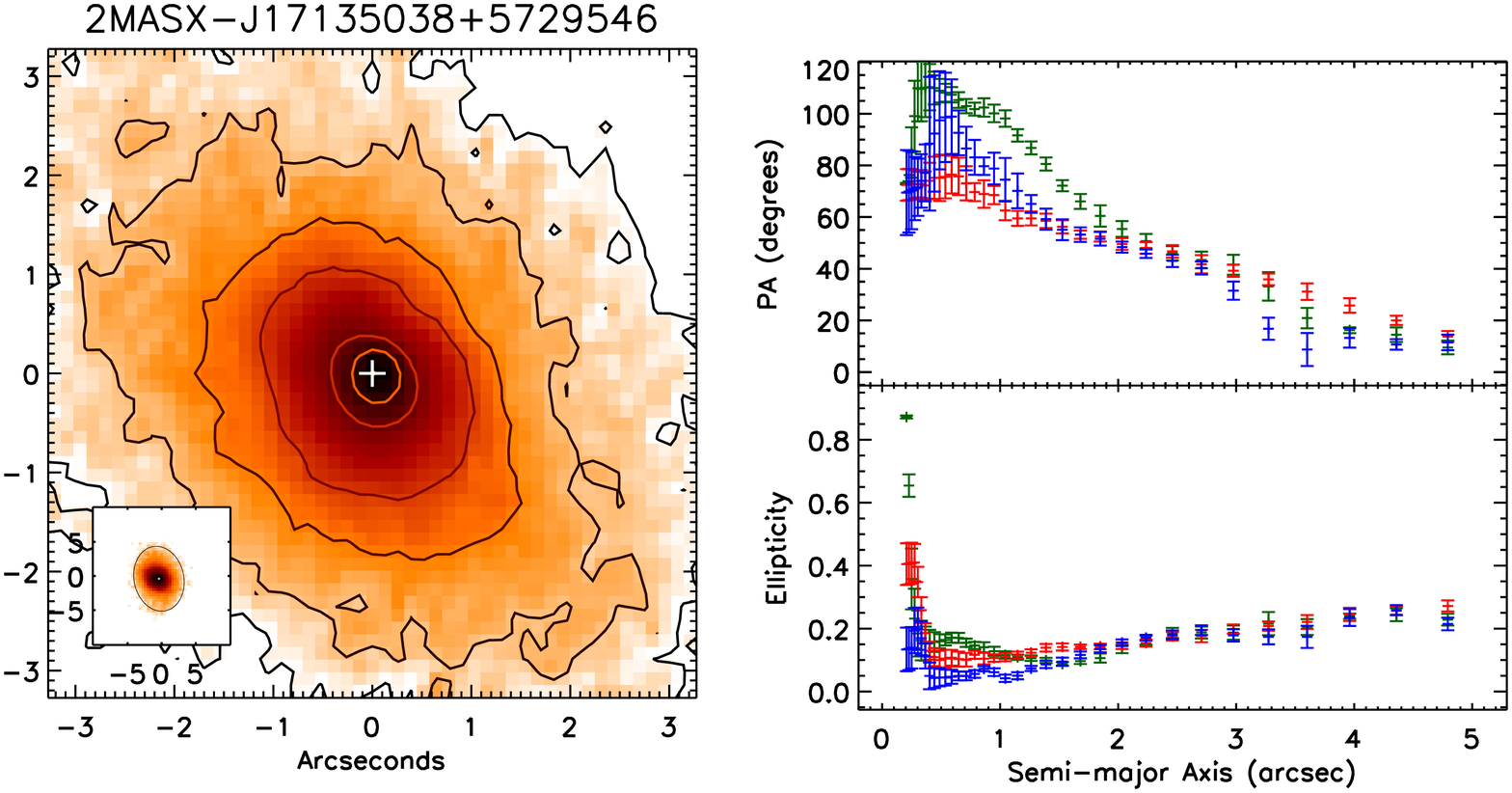}\\

\caption{Same as Figure \ref{fig:images8} for 2MASX J16531506+2349431 and 2MASX J17135038+5729546. Host disk ellipticals 
are fit to 3$\sigma$ for both targets.}
\label{fig:images13}

\end{figure*}

\subsection{2MASX J165315}
The [O III] morphology in this target is distributed in three features; a faint arc 
$\sim$ 4 kpc east of the nucleus, a central lobe extending east from the nucleus, and 
a fan or conical shape west of the nucleus. [O III] kinematics are only obtained for 
the central lobe and western fan, with a largely rotational pattern. The maximum outflow 
radius is defined by a relatively broad ($>$500 km s$^{-1}$), systemic, (or blueshifted 
with respect to the narrow component), secondary component observed east of the nucleus 
for an R$_{out}$ = 370 pc. Single-component line emission with FWHM $>$ 250 km s$^{-1}$ 
that follows a low amplitude rotation pattern is measured throughout the rest of the 
system, for an R$_{dist}$ of 1.1 kpc. The host galaxy of this target is moderately inclined, 
with no signs of a recent merger.

\subsection{2MASX J171350}
Similar to 2MASX J140541, this source has a compact [O III] morphology and the measured 
R$_{max}$ is inside the observed [O III] morphology as it is artificially extended by diffraction 
spikes. Kinematics in this system follow a rotation pattern, redshifted to the northeast and 
blueshifted to the southwest, with broad emission-line FWHMs frequently around 1500 km s$^{-1}$. 
High peak rotation velocities ($\sim$200 km s$^{-1}$) are consistent with the alignment between 
the modeled host major axis and STIS long-slit position angles (PA$_{diff}$ = 16.3$^{\degree}$). 
Multi-component emission lines are observed out to 490 pc, which we define as R$_{out}$. Blending 
of these components likely occurs in measurements near the ends of our STIS observations, with 
the very broad component becoming too faint to individually detect at a 3$\sigma$ level, creating a 
single-component emission line possessing disturbed kinematics that extends to an R$_{dist}$ of $\sim$700 pc. 
The host is moderately inclined, without signs of a recent merger.

\end{document}